\documentclass[format=acmsmall, review=false, screen=true]{acmart}
\usepackage{url}
\usepackage{verbatim}
\usepackage{float}
\usepackage{subfig}
\usepackage{lineno,hyperref}
\usepackage{comment}
\usepackage{booktabs}
\usepackage{amssymb,amsmath,epsfig,psfrag}
\setcopyright{usgovmixed}
\usepackage{amsfonts,color,graphics,graphicx}
\usepackage{array}
\usepackage{makecell}
\usepackage{xcolor, soul}
\usepackage{multirow}
\usepackage{rotating}
\setcitestyle{square}
\usepackage[normalem]{ulem}
\acmJournal{CSUR}

\usepackage{wrapfig}
\usepackage{colortbl}
\usepackage{lscape}
\usepackage{ltablex}
\usepackage{pifont}
\usepackage{graphicx}
\usepackage{tikz}
\usetikzlibrary{shapes,snakes,patterns}
\usetikzlibrary{shapes.gates.logic.US,trees,positioning,arrows}
\usepackage{pgfplots}
\pgfplotsset{width=18cm, height=6cm}
\usepackage{pgf-pie}
\usepackage[ruled]{algorithm2e}
\SetAlFnt{\small}
\SetAlCapFnt{\small}
\SetAlCapNameFnt{\small}
\SetAlCapHSkip{0pt}
\IncMargin{-\parindent}
\newcolumntype{P}[1]{>{\centering\arraybackslash}p{#1}}

\usepackage{enumitem}
\setlist[itemize]{leftmargin=*}
\newcommand{\xmark}{\ding{55}}%

\begin{document}

\title{Online Social Deception and Its Countermeasures for Trustworthy Cyberspace: A Survey}


\author{Zhen Guo}
\author{Jin-Hee Cho}
\author{Ing-Ray Chen}
\affiliation{%
  \institution{Computer Science, Virginia Tech}
\streetaddress{7054 Haycock Road}
  \city{Falls Church}
  \state{VA}
  \postcode{22043}
  \country{USA}
  }
\author{Srijan Sengupta}
\affiliation{%
  \institution{Statistics, Virginia Tech}
  \streetaddress{250 Drillfield Drive}
  \city{Blacksburg}
  \state{VA}
  \postcode{24061}
  \country{USA}
  }
\author{Michin Hong}
\affiliation{%
  \institution{School of Social Work, Indiana University}
  \streetaddress{902 West New York Street}
  \city{Indianapolis}
  \state{IN}
  \postcode{46202}
  \country{USA}
  }
\author{Tanushree Mitra}
\affiliation{%
  \institution{Computer Science, Virginia Tech}
\streetaddress{2202 Kraft Drive}
  \city{Blacksburg}
  \state{VA}
  \postcode{24060}
  \country{USA}
  }
\begin{abstract}
We are living in an era when online communication over social network services (SNSs) have become an indispensable part of people's everyday lives.   As a consequence, online social deception (OSD) in SNSs has emerged as a serious threat in cyberspace, particularly for users vulnerable to such cyberattacks.  Cyber attackers have exploited the sophisticated features of SNSs to carry out harmful OSD activities, such as financial fraud, privacy threat, or sexual/labor exploitation.  Therefore, it is critical to understand OSD and develop effective countermeasures against OSD for building a trustworthy SNSs.  In this paper, we conducted an extensive survey, covering (i) the multidisciplinary concepts of social deception; (ii) types of OSD attacks and their unique characteristics compared to other social network attacks and cybercrimes; (iii) comprehensive defense mechanisms embracing prevention, detection, and response (or mitigation) against OSD attacks along with their pros and cons; (iv) datasets/metrics used for validation and verification; and (v) legal and ethical concerns related to OSD research.  Based on this survey, we provide insights into the effectiveness of countermeasures and the lessons from existing literature.  We conclude this survey paper with an in-depth discussions on the limitations of the state-of-the-art and recommend future research directions in this area.  

\end{abstract}



\maketitle

\section{Introduction} \label{sec:intro}

\subsection{Motivation} \label{subsec:motivation}

Social media and social network services (SNSs) have become an indispensable part of people's everyday lives.
In 2018, approximately 70\% of Americans reported using social media~\cite{Rainie18}.  
This surge in the popularity of SNSs is due to various benefits that users enjoy, such as easy communications with others, engagement in civic and political activities, searching jobs, marketing, and/or exchanging/sharing information or emotional support.  Even with these significant benefits, many people have ambivalent feelings about social media due to privacy concerns and/or deceptive activities aiming to harm normal, legitimate users~\cite{Rainie18}. The proliferation of highly advanced social media technologies has been exploited by perpetrators as convenient tools for deceiving users~\cite{Anderson17}.
The widespread damages due to {\em online social deception} (OSD) attacks have increased significantly in recent times, with about 25\% of people experiencing some types of social deception, such as identity theft, cyberbullying, fraud, or phishing in 2018~\cite{Reinhart18}.  The serious consequences have led to such OSD attacks being defined as {\em cybercrimes}~\cite{Nykodym05} since early 2000's.  
The advanced features of SNS technologies further have facilitated the significant increase of serious, sophisticated cybercrimes, beyond simple phishing or spamming, such as human trafficking, online consumer fraud, identity cloning, hacking, child pornography, and online stalking~\cite{van2018determinants}. 

It is therefore the need of the hour to understand OSD and develop effective countermeasures against OSD to develop a trustworthy cyberspace for SNSs.
The concept of `social deception' is highly multidisciplinary and has been extensively studied in various domains, such as psychology~\cite{Adair85, Nicks97, Riggo83}, sociology~\cite{Langer72, Meltzer03, Patwardhan09}, philosophy~\cite{Carson10, Gert15, Mahon16, Russow86}, behavioral science~\cite{Hauser97, Langleben02, Spence01}, public relations~\cite{Englehardt94, Ortmann10, Seiffert-Brockmann17, Stone01}, communications or linguistics~\cite{Buller96-testing, Buller96, Hancock08, Zhou04-linguistics}, and computing/engineering~\cite{Albladi18, Han18-deception-survey} (see the multidisciplinary concept of deception discussed in Section~\ref{subsec:deception-concept}). 

Deception is commonly understood as a planned action to mislead a potential victim in order to achieve a deceiver's goal, although more general notions of deception have been discussed based on their various goals and/or intent~\cite{Rowe16}.  Despite this common understanding of deception, different types of deception have been discussed based on their various goals and/or intent depending on a different context/domain.  The current countermeasures against OSD related cybercrimes have mainly focused on detecting them using data mining~\cite{lee2010uncovering}, text mining using machine learning (e.g., text mining for posts, tweets/retweets, or clicks)~\cite{Algaradi16, song2015crowdtarget, wang2013characterizing}, or user and network features analysis using data mining or machine/deep learning~\cite{lee2013crowdturfers, yao2017automated}.  In the cybersecurity domain, deception is heavily used by both attackers and defenders. Any online deception to achieve a deceiver's malicious goals, such as phishing, identity theft, spamming, cyber bullying, grooming, or stalking, is an act of deception by online social attackers.  Defenders also have taken various types of defensive deception techniques as strategic actions~\cite{Han18-deception-survey}.

In this survey paper, our goal is two-fold: (a) to provide an in-depth understanding of online social deception through the lens of cybersecurity, and (b) to describe and assess the state-of-the-art countermeasures against OSD as defense mechanisms for its prevention, detection, and response/mitigation.  Although several survey papers have been published on this topic (see Section~\ref{subsec:comparison-other-surveys}), there is still a lack of comprehensive survey that embraces the fundamental concepts and cues of social deception and the key susceptibility factors to the major defense strategies.  In addition, no prior work provided a comprehensive survey on defense strategies to OSD attacks in terms of prevention, defense, and response/mitigation, and evaluation methodologies discussing datasets and metrics used in the state-of-the-art literature. 

\subsection{Research Goal \& Questions} \label{subsec:goal-questions}

To fill the gap identified as above, this study aims to deliver a comprehensive, systematic survey for researchers 
to efficiently and effectively grasp a large volume of the state-of-the-art literature on OSD and its countermeasures in a broad sense.  To achieve this goal, {\bf the scope of this work} focuses on answering the following research questions:
\begin{itemize}[leftmargin=-.5mm]
\item[] {\bf RQ1}: {\em How is online social deception (OSD) affected by the fundamental concepts and characteristics of social deception which have been studied in multidisciplinary domains?}  
\item[] {\bf RQ2}: {\em What are new attack types based on the recent trends of OSD attacks observed in real online worlds and how are they related to common social network attacks, cybercrimes, and security breaches based on cybersecurity perspectives?} 
\item[] {\bf RQ3}: {\em How can the cues of social deception and/or susceptability traits to OSD affect the strategies by attackers and defenders in OSNs?}  
\item[] {\bf RQ4}: {\em What kinds of defense mechanisms and/or methodologies still need to be explored to develop better defense tools combating OSD attacks?}  
\item[] {\bf RQ5}: {\em What are the key limitations of validaverification methodologies, particularly in terms of datasets and metrics used in the state-of-the-art approaches?}  
\item[] {\bf RQ6}: {\em What are the key concerns associated with legal and/or ethical issues in conducting OSD research?}  
\end{itemize}
In Section~\ref{sec:conclusion}, we will discuss how the above research questions have been answered in this paper.

\begin{table}[t]
\scriptsize 
\centering
\caption{Comparison of the key contributions of our survey paper and other existing survey papers.}
\label{tab:comparison-survey-papers}
\begin{tabular}{|P{3cm}|P{1cm}|P{1cm}|P{1cm}|P{1cm}|P{1cm}|P{1.2cm}|P{1.5cm}|}
\hline
{\bf Criteria} & {\bf Our Survey} &\citet{Rathore17-is-survey} &\citet{novak2012survey} &\citet{gao2011security}  &\citet{fire2014online}  &\citet{kayes2017privacy}  &\citet{tsikerdekis2014online} \\
\hline
\multicolumn{8}{|c|}{\bf Concepts and Characteristics of Online Social Deception}\\
\hline
Discussion on multidisciplinary concepts &\cellcolor{green!40}\checkmark &\cellcolor{black!40}\xmark &\cellcolor{black!40}\xmark &\cellcolor{black!40}\xmark &\cellcolor{black!40}\xmark &\cellcolor{black!40}\xmark &\cellcolor{black!40}\xmark\\
Deception cues &\cellcolor{green!40}\checkmark &\cellcolor{black!40}\xmark &\cellcolor{black!40}\xmark &\cellcolor{black!40}\xmark &\cellcolor{black!40}\xmark &\cellcolor{black!40}\xmark &\cellcolor{yellow!60} Limited\\
Spectrum of deception with/without intentionality
&\cellcolor{green!40}\checkmark &\cellcolor{black!40}\xmark &\cellcolor{black!40}\xmark &\cellcolor{black!40}\xmark &\cellcolor{black!40}\xmark &\cellcolor{black!40}\xmark &\cellcolor{green!40}\checkmark\\
Properties of social deception &\cellcolor{green!40}\checkmark &\cellcolor{black!40}\xmark &\cellcolor{black!40}\xmark &\cellcolor{black!40}\xmark &\cellcolor{black!40}\xmark &\cellcolor{black!40}\xmark &\cellcolor{green!40}\checkmark\\
Susceptibility factors to OSD attacks &\cellcolor{green!40}\checkmark &\cellcolor{black!40}\xmark &\cellcolor{black!40}\xmark &\cellcolor{black!40}\xmark &\cellcolor{black!40}\xmark &\cellcolor{black!40}\xmark &\cellcolor{yellow!60} Limited\\
\hline
\multicolumn{8}{|c|}{\bf Security Threat Categorization/Classification}\\
\hline
Fake news &\cellcolor{green!40}\checkmark &\cellcolor{black!40}\xmark &\cellcolor{black!40}\xmark &\cellcolor{black!40}\xmark &\cellcolor{black!40}\xmark &\cellcolor{black!40}\xmark &\cellcolor{black!40}\xmark\\
Rumors &\cellcolor{green!40}\checkmark &\cellcolor{black!40}\xmark &\cellcolor{black!40}\xmark &\cellcolor{black!40}\xmark &\cellcolor{black!40}\xmark &\cellcolor{black!40}\xmark &\cellcolor{black!40}\xmark\\
Information manipulation &\cellcolor{green!40}\checkmark &\cellcolor{black!40}\xmark &\cellcolor{black!40}\xmark &\cellcolor{black!40}\xmark &\cellcolor{black!40}\xmark &\cellcolor{black!40}\xmark &\cellcolor{black!40}\xmark\\
Fake reviews &\cellcolor{green!40}\checkmark &\cellcolor{black!40}\xmark &\cellcolor{black!40}\xmark &\cellcolor{black!40}\xmark &\cellcolor{black!40}\xmark &\cellcolor{black!40}\xmark &\cellcolor{black!40}\xmark\\
Phishing &\cellcolor{green!40}\checkmark &\cellcolor{green!40}\checkmark &\cellcolor{black!40}\xmark &\cellcolor{green!40}\checkmark &\cellcolor{green!40}\checkmark &\cellcolor{green!40}\checkmark &\cellcolor{black!40}\xmark\\
Spamming &\cellcolor{green!40}\checkmark &\cellcolor{green!40}\checkmark &\cellcolor{black!40}\xmark &\cellcolor{green!40}\checkmark &\cellcolor{green!40}\checkmark &\cellcolor{green!40}\checkmark &\cellcolor{black!40}\xmark\\
Fake identity &\cellcolor{green!40}\checkmark &\cellcolor{green!40}\checkmark &\cellcolor{green!40}\checkmark &\cellcolor{green!40}\checkmark &\cellcolor{green!40}\checkmark &\cellcolor{green!40}\checkmark &\cellcolor{green!40}\checkmark\\
Compromised account &\cellcolor{green!40}\checkmark &\cellcolor{black!40}\xmark &\cellcolor{black!40}\xmark &\cellcolor{black!40}\xmark &\cellcolor{black!40}\xmark &\cellcolor{green!40}\checkmark &\cellcolor{green!40}\checkmark\\
Profile cloning attack &\cellcolor{green!40}\checkmark &\cellcolor{green!40}\checkmark &\cellcolor{black!40}\xmark &\cellcolor{green!40}\checkmark &\cellcolor{green!40}\checkmark &\cellcolor{black!40}\xmark  &\cellcolor{green!40}\checkmark\\
Crowdturfing &\cellcolor{green!40}\checkmark &\cellcolor{black!40}\xmark &\cellcolor{black!40}\xmark &\cellcolor{black!40}\xmark &\cellcolor{black!40}\xmark &\cellcolor{black!40}\xmark  &\cellcolor{black!40}\xmark\\
Human trafficking &\cellcolor{green!40}\checkmark &\cellcolor{black!40}\xmark &\cellcolor{black!40}\xmark &\cellcolor{black!40}\xmark &\cellcolor{black!40}\xmark &\cellcolor{black!40}\xmark  &\cellcolor{black!40}\xmark\\
Cyberbullying &\cellcolor{green!40}\checkmark &\cellcolor{green!40}\checkmark &\cellcolor{black!40}\xmark &\cellcolor{black!40}\xmark &\cellcolor{green!40}\checkmark &\cellcolor{black!40}\xmark  &\cellcolor{black!40}\xmark\\
Cyber-grooming &\cellcolor{green!40}\checkmark &\cellcolor{green!40}\checkmark &\cellcolor{black!40}\xmark &\cellcolor{black!40}\xmark &\cellcolor{green!40}\checkmark &\cellcolor{black!40}\xmark  &\cellcolor{black!40}\xmark\\
Cyberstalking &\cellcolor{green!40}\checkmark &\cellcolor{green!40}\checkmark &\cellcolor{black!40}\xmark &\cellcolor{black!40}\xmark &\cellcolor{black!40}\xmark &\cellcolor{black!40}\xmark  &\cellcolor{black!40}\xmark\\
\hline
\multicolumn{8}{|c|}{\bf Existing OSNs Security Solutions}\\
\hline
Security issues and challenge &\cellcolor{green!40}\checkmark &\cellcolor{green!40}\checkmark &\cellcolor{black!40}\xmark &\cellcolor{green!40}\checkmark &\cellcolor{black!40}\xmark &\cellcolor{yellow!60} Limited  &\cellcolor{green!40}\checkmark\\
Prevention &\cellcolor{green!40}\checkmark &\cellcolor{yellow!60} Limited &\cellcolor{black!40}\xmark &\cellcolor{yellow!60}Limited &\cellcolor{green!40}\checkmark &\cellcolor{black!40}\xmark  &\cellcolor{yellow!60} Limited\\
Detection &\cellcolor{green!40}\checkmark &\cellcolor{green!40}\checkmark &\cellcolor{green!40}\checkmark &\cellcolor{green!40}\checkmark &\cellcolor{green!40}\checkmark &\cellcolor{green!40}\checkmark  &\cellcolor{black!40}\xmark\\
Mitigation &\cellcolor{green!40}\checkmark &\cellcolor{black!40}\xmark &\cellcolor{black!40}\xmark &\cellcolor{black!40}\xmark &\cellcolor{black!40}\xmark &\cellcolor{green!40}\checkmark  &\cellcolor{black!40}\xmark\\
Security suggestions &\cellcolor{green!40}\checkmark &\cellcolor{green!40}\checkmark &\cellcolor{black!40}\xmark &\cellcolor{yellow!60}Limited &\cellcolor{green!40}\checkmark &\cellcolor{yellow!60} Limited  &\cellcolor{black!40}\xmark\\
\hline
\multicolumn{8}{|c|}{\bf Discussing Limitation, Pros and Cons of Detection}\\
\hline
Ethical Issues &\cellcolor{green!40}\checkmark &\cellcolor{black!40}\xmark &\cellcolor{black!40}\xmark &\cellcolor{black!40}\xmark &\cellcolor{black!40}\xmark &\cellcolor{black!40}\xmark  &\cellcolor{black!40}\xmark\\
Discussing Key Limitations &\cellcolor{green!40}\checkmark &\cellcolor{black!40}\xmark &\cellcolor{black!40}\xmark &\cellcolor{black!40}\xmark &\cellcolor{black!40}\xmark &\cellcolor{black!40}\xmark  &\cellcolor{black!40}\xmark\\
Pros and Cons of Techniques &\cellcolor{green!40}\checkmark &\cellcolor{black!40}\xmark &\cellcolor{black!40}\xmark &\cellcolor{black!40}\xmark &\cellcolor{black!40}\xmark &\cellcolor{black!40}\xmark &\cellcolor{black!40}\xmark\\
\hline
\end{tabular}
\vspace{-3mm}
\end{table}

\subsection{Comparison with Existing Survey Papers} \label{subsec:comparison-other-surveys}
As social deception leverages online social networks (OSNs) as platforms, there have been several survey papers~\cite{acemoglu2010spread, fire2014online, gao2011security, kayes2017privacy, kumar2018false, novak2012survey, Rathore17-is-survey, Wu16-bc, Wu2017-crowdturfing} discussing social network attacks and/or threats.  Due to the space constraint, we provided the detailed discussion of each existing survey paper in the appendix document (see Section A).

Based on the existing survey papers~\cite{acemoglu2010spread, fire2014online, gao2011security, kayes2017privacy, kumar2018false, novak2012survey, Rathore17-is-survey, Wu16-bc, Wu2017-crowdturfing}, we found that there is no comprehensive survey paper on online social deception (OSD) which sits between OSN threats and cybercrimes.  The most related work discussed above focused on security and privacy issues and their solutions in OSNs.  We demonstrated the key differences in scope and techniques between our survey paper and the existing OSN security and/or attack papers in Table~\ref{tab:comparison-survey-papers}.  We compared them based on a set of criteria in terms of security threat categories, existing security detection and suggestions, and discussion of limitations. Most previous studies analyzed various types of OSN threats and provided detection methods for specific types of security threats.  However, they usually discussed traditional types of security issues, which only partially overlaps our definitions of social deception threats.  We list the key contributions of our survey paper compared to existing survey papers in the following section.

\subsection{Key Contributions} \label{subsec:key-contributions}
Our survey paper has the following {\bf key contributions}: 
\begin{itemize}
\item To understand the fundamental meaning of social deception and its key characteristics, we comprehensively survey the multidisciplinary concepts and key properties of social deception. No previous survey paper has addressed all these concepts together to understand the fundamental meanings of social deception. 

\item We address a comprehensive set of OSD attacks by following the key properties of social deception discussed in Section~\ref{subsec:properties-deception}. We based our survey on five major categories of attacks: false information, luring, fake identity, crowdturfing, and human targeted attacks.  As shown in Table~\ref{tab:comparison-survey-papers}, no prior survey papers have embraced this comprehensive set of OSD attacks.  In addition, we outline the relationships between social network attacks, OSD attacks, and cybercrimes by describing how they are related to each other, what malicious behaviors are major attacks in each category, and what are the attack goals of OSD in terms of conventional CIA (confidentiality, integrity, and availability) security goals.  


\item To provide a more comprehensive understanding on a system-level defense framework including all three steps of defense, i.e., prevention, detection, and mitigation/recovery against intrusions (i.e., OSD attacks in this paper), we extensively survey the three types of defense mechanisms to fight against the OSD attacks based on a significant amount of references (i.e., 18 papers for prevention for 2008-2019, 31 papers for detection for 2011-2019, and 6 papers for mitigation/response for 2007-2018).  These comprehensive surveys of prevention, detection, and mitigation mechanisms are summarized in Tables A4 -- A6 of the appendix document.

\item We provide pros and cons of major defense approaches to combat OSD attacks and the overall trends of the state-of-the-art OSD defense techniques.  This gives a reader to understand which techniques are more relevant in a given context, which may be limited in some resources and/or requires a more feasible implementation plan.

\item We identify the common datasets and metrics that have been used to validate the performance of defense mechanisms combating the SDN attacks. From this comprehensive survey on datasets and metrics, we also provide useful directions of the OSD research to enhance the validation and verification methods, which have not been discussed in other existing survey papers on OSD. 
\item Based on the extensive survey provided in this work, we also comprehensively discussed key findings, insights and lessons learned, limitations, and future research directions.
\end{itemize}

\section{Concepts and Characteristics of Deception} \label{sec:deception-concept-causes}

The concept of deception is highly multidisciplinary and has been studied in various domains.  In this section, we discuss the root definitions of deception and the fundamental properties of deception which have been applied in launching OSD attacks in OSN platforms.

\subsection{Multidisciplinary Concept of Deception} \label{subsec:deception-concept}

Let us start by looking at the dictionary definition of deception~\cite{OED89}. Deception is defined as: ``To cause to believe what is false.''  However, the definition is too broad and many deception researchers raised doubts on the definition.  \\
\vspace{-4mm}
\begin{wrapfigure}{r}{0.4\textwidth}
\vspace{-2mm}
    \centering
    \includegraphics[scale=0.7]{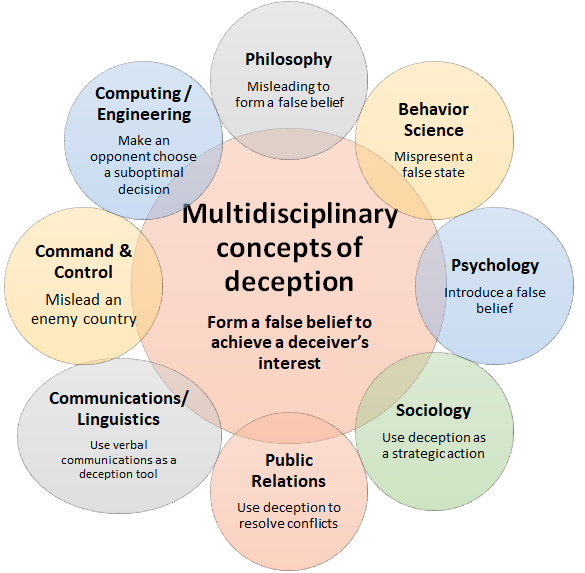}
    \caption{The key multidisciplinary concepts of deception.}
    \label{fig:deception-concepts}
\vspace{-4mm}
\end{wrapfigure} 
In the literature, the concepts of deception have been discussed with different perspectives under different disciplines, such as philosophy, behavioral science, psychology, sociology, public relations, communication/linguistics, command and control, and computing/engineering.  Due to the space constraint, we simply summarize the key concept of deception in different disciplines in Fig.~\ref{fig:deception-concepts}. We also provide a detailed discussion on the concepts of deception under the eight different disciplines in the appendix document (see Section B Multidisciplinary Concepts of Deception and Table A1).  

In addition, as the threat of phishing emails increases, an individual online user's susceptibility to phishing attacks is studied in terms of demographics~\cite{Lin19, Oliveira17-chi-susceptibility, Sheng10} or personality traits~\cite{Darwish12, Enos06, Halevi13, Halevi15, Modic11, Parrish09, Pattinson12}. We discuss the details of susceptibility to OSD attacks in Section F of the appendix document.  For easy grasping of the key multidisciplinary concepts of deception, we summarize the meanings and goals of deception under each domain in Table A1 of the appendix document. 

\vspace{-2mm}
\subsection{Properties of Deception} \label{subsec:properties-deception}
Via the in-depth literature review, we observe the following {\bf key properties of deception}: 
\begin{itemize}
\item {\em Misleading one's belief}: People may use deception intentionally or unintentionally (or mistakenly) with good or bad intent.  However, regardless of intent (good or bad or even without any intent), deception can mislead one's belief which is actually false. Since deception as an action induces confusion or false information (e.g., speaking or acting to induce a misbelief), false beliefs may be formed regardless of its intent or outcome.
\item {\em Impact by deception}: Confusion or misbelief introduced by deception brings an outcome which can be negative or positive based on its original intent and/or its proper execution. However, when deception with a certain intent is not properly executed as planned or is used mistakenly, the outcome as its impact may not be predictable, resulting in high uncertainty (e.g., uncertain outcome). Hence, if deception is intended, it should be planned with multiple scenarios to lower down the risk introduced by deception in terms of a deceiver's perspective.
\item {\em Success only by a deceivee's cooperation}: For deception to be successful, a deceivee should be deceived by the deception.  Even if deception is performed but the deceiver detects the deception, not being deceived, no impact of the deception can be introduced.
\item {\em Action as a strategy}: Deception can be used as a strategy to deal with situations with conflicts. The aim of the deception with intent is to mislead a target entity's belief and make the target choose a suboptimal (or poor) action that can be beneficial for the deceiver to achieve its goal.
\item {\em Signals as deception cues}: When deception is used, even if it can be very subtle, there exists some signals. Well-known deception strategies are to increase uncertainty (e.g., no signal increases uncertainty) or mislead one's belief (e.g., a false signal leads to false beliefs).  Although both deception techniques aim to make a deceiver choose a wrong decision, if deception by misleading with false signal is detected, this provides more information about a deceiver to a deceivee than providing no signal.
\item {\em Effect of intent}: Although deception is mostly understood as a negative action taken by an entity with bad intent, it can appear as misconception about situations or information. If the deceiver mistakenly uses deception (e.g., sending false information as it believes it is true at the time the information is found), after the truth is known, it can fix its stance/action. In addition, if the deception does not have bad intent behind it, but it impacts negatively in the current situation, the deceiver may reveal its intent to resolve any conflict derived from the deception. 
\end{itemize}
Investigating the key properties of deception is critical in developing defense mechanisms to combat OSD attacks as the features of deception-based attacks, distinguished from other common online social network attacks.  In this section, we discussed a variety of cues and susceptability traits of social deception behaviors across online and offline platforms. Thanks to the fast advances of social media and online social network technologies, many offline deception characteristics tend to be easily observed even in online deception behaviors. However, due to the limited real-time and/or interactions feeling people's presence in online platforms with the current state-of-the-art SNSs and social media technologies, some physiological or psychological cues may not be applicable in detecting online social deception. In addition, upon the detection of the deception, a deceiver can easily get out of the online situation while a deceivee can easily lose a track of the deceiver.  Now we look into various types of online social deception behaviors currently studied in the literature.

\section{Types of Online Social Deception} \label{sec:types-social-deception}

We categorize OSD attacks to mislead people's beliefs by the following strategy types: {\em false information}, {\em luring}, {\em fake identity}, {\em crowdturfing}, and {\em human targeted attacks} (summarized in Table A2 of the appendix document).  Further, we discuss how OSD attacks differ from other OSN attacks and what security goals each OSD attack aims to breach.

\vspace{-3mm}
\subsection{False Information} \label{subsec:false-information}
False information on the web and social media can be classified as {\em misinformation} and {\em disinformation}.  Misinformation can be considered as `deception without intent' which mistakenly misleads people's belief due to the false information propagated. Disinformation can be categorized as `deception with intent,' aiming to mislead people's beliefs.  False information can be also  categorized as {\em opinion-based} vs. {\em fact-based}.  The opinion-based false information propagates without ground truth.  On the other hand, the fact-based false information misleads people's beliefs due to the fraud from ground truth, such as hoaxes and fake news in social media~\cite{jiang2018linguistic}.

\citet{jiang2018linguistic} compared and summarized different definitions and ranges of misinformation, based on two critera, {\em veracity} and {\em intentionality}~\cite{shu2017fake}, as follows:
\vspace{-2mm}
\begin{itemize}
\item {\em Fake News}: Fake news caused by serious fabrications or large-scale hoaxes\cite{Rubin15} has spread wildly since the beginning of the 2016 US presidential election cycle. \citet{flintham2018falling} reported that two third of survey respondents accessed news via Facebook. Facebook and Twitter have banned thousands of pages and identified as the major culprit of generating and promoting misinformation~\cite{jiang2018linguistic}. Fact-checking from different sources is a means to determine the veracity of social media posts. \citet{vosoughi2018spread} found that fake news spread faster than truthful news. The time lag between fake news and fact-checking by fact-checking websites (for automatic fact-checking) is 10-20 hours~\cite{shao2016hoaxy}.
\item {\em Rumors}: \citet{vosoughi2017rumor} defined a rumor as an unverified assertion that starts from one or more sources and spreads over time from one user to another user in a network.  A rumor can be validated as true or false via real-time verification in Twitter or remain unresolved.
\item {\em Information Manipulation}: One of the causes of information manipulation is opportunistic disinformation~\cite{dhs18}.  This means false information is deliberately and often covertly spread (e.g.,  planting a rumor) in order to influence public opinions or obscure the truth.  Opportunistic disinformation falls into two categories: financially or politically incentivized. 
\item {\em Deceptive Online Comments/Review}: Malicious users write fake reviews, opinions, or comments in social media to mislead other users. Fake reviews can be generated automatically~\cite{yao2017automated}. 
\end{itemize}

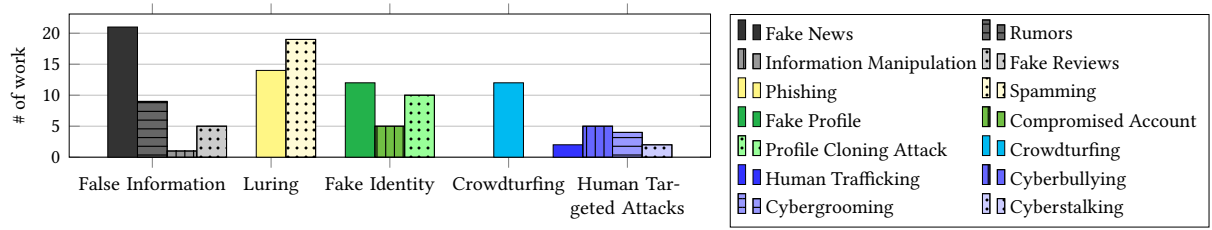
\begin{figure}[!t]
    \centering
\begin{tikzpicture}[font=\footnotesize, scale = 0.85]
\begin{axis}[
    ylabel={\# of work},
    ylabel near ticks,
    ymin=0,
    ytick={0,5,...,20},
    xtick={2,6,10,14,18},
    xticklabels={False Information,Luring,Fake Identity\enspace\enspace,Crowdturfing,Human Targeted Attacks},
    x tick label style={text width=2.4cm,align=center},
    ymajorgrids,
    ybar=-\pgfplotbarwidth,
    x=\pgfplotbarwidth,
    bar width=12pt,
    height=.26\textwidth,
    legend columns=2,
    legend cell align={left},
    legend style={
        cells={align=left},
        font=\footnotesize,
        legend pos=outer north east,
        }
    ]
\addplot [fill={black!80!}] coordinates {(1,21)};
\addplot [fill={black!60!}, postaction={pattern=horizontal lines}] coordinates {(2,9)};
\addplot [fill={black!40!}, postaction={pattern=vertical lines}] coordinates {(3,1)};
\addplot [fill={black!20!}, postaction={pattern=dots}] coordinates {(4,5)};
\addplot [fill={yellow!60!}] coordinates {(6,14)};
\addplot [fill={yellow!20!}, postaction={pattern=dots}] coordinates {(7,19)};
\addplot [fill={yellow!20!green!}] coordinates {(9,12)};
\addplot [fill={yellow!40!green!}, postaction={pattern=vertical lines}] coordinates {(10,5)};
\addplot [fill={green!40!}, postaction={pattern=dots}] coordinates {(11,10)};
\addplot [fill={cyan!80!}] coordinates {(14,12)};
\addplot [fill={blue!80!}] coordinates {(16,2)};
\addplot [fill={blue!60!}, postaction={pattern=vertical lines}] coordinates {(17,5)};
\addplot [fill={blue!40!}, postaction={pattern=horizontal lines}] coordinates {(18,4)};
\addplot [fill={blue!20!}, postaction={pattern=dots}] coordinates {(19,2)};
\legend{Fake News,Rumors,Information Manipulation, Fake Reviews, Phishing, Spamming, Fake Profile, Compromised Account, Profile Cloning Attack, Crowdturfing, Human Trafficking, Cyberbullying, Cybergrooming, Cyberstalking}
\end{axis}
\end{tikzpicture}
\caption{Research work count for five categories of deception: False Information, Luring, Fake Identity, Crowdturfing and Human Targeted Attacks and subcategories.}
\label{fig:workcount}
\vspace{-5mm}
\end{figure}

\vspace{-3mm}
\subsection{Luring}
Deception can be realized by luring people.  The most common luring techniques are as follows:
\vspace{-1mm}
\begin{itemize}
\item {\em Spamming}: Social media platform users can receive unsolicited messages (spam) that are ranging from advertising to phishing messages~\cite{Rathore17-is-survey}. The Attackers usually send spam messages in bulk to influence many normal users.

\item {\em Phishing}: Online phishing attacks, such as phishing webpages, are one type of cybercrimes that can lure users to reveal sensitive information and steal privacy data or financial information through social engineering techniques~\cite{ding2019keyword}.  Attackers exploit the financial credentials and other personal data in daily life in other fraudulent activities~\cite{abutair2019cbr}.  Those illegal activities can cause sever economic losses and threaten credibility and financial security of OSN users. Kaspersky Lab's quarterly Spam and phishing report~\cite{Vergelis19} showed that phishing attacks are increasing in Q1 2019. Attackers use social networks to reach their targets and even launched advertising campaigns using celebrities.  Scammers exploited high-profit media events to redirect users to their phishing links and scam websites, such as Apple product launch and holiday celebration. Banks are established as the top phishing targets.  Phishing links can be also found from emails to force users to update accounts or payment information. 
\end{itemize}

\vspace{-3mm}
\subsection{Fake Identity} \label{subsec:fake-identity}
This section discusses the following OSD attacks associated with fake identity:
\begin{itemize}
\item {\em Fake Profile} (a.k.a. {\em Sybil attack}): In OSNs, attackers create a huge amount of fake identities for their own benefits. For example, in Facebook, personal information, such as e-mail and physical addresses, date of birth, employment data were leaked. Identity theft can access photographs of the friends of the victims; in addition, fraudsters steal money~\cite{haddadi2010add}. 
\item {\em Profile Cloning}: Attackers secretly create a duplicate of existing user profile in the same or different social media platforms.  Since the cloned profile resembles the current profile, attackers can utilize the friend relationship and deceive and send friend requests to the contacts of the cloned user.  By constructing the trust relationship, the attacker can steal sensitive data from the existing user's friends.  Profile cloning exposed severe threats because attackers can commit even serious cybercrimes, such as cyberbullying, cyberstalking, and blackmail~\cite{Rathore17-is-survey}.
\item {\em Compromised accounts}: Legitimate user accounts can be hacked by attackers and then those accounts are compromised by attackers~\cite{egele2013compa}.  Unlike Sybil accounts, compromised accounts are originally maintained by real users with normal social network usage history and have established social connections.
\end{itemize}

\vspace{-5mm}
\subsection{\bf Crowdturfing}  \label{subsubsec:crowdturfing}
In this section, we discuss human attackers who are paid workers to achieve their employer's malicious intent, called {\em crowdturfing}. Crowdturfing refers to the behavior that participants of an astroturfing campaign are organized by crowdsourcing systems~\cite{wang2012serf}.  Crowturfing gathers crowdturfing workers and spreads fake information to mislead people's beliefs and/or opinions.  Crowdturfing activities in social media exploit social networking platforms (e.g., instant message groups, microblogs, blogs, and online forums) as the main information channel of the campaign~\cite{Wu2017-crowdturfing}.  The crowdturfing in social media is usually involved with spreading malicious URLs, forming astroturf campaigns, and manipulating. Crowdturfing workers spread information and posts from their social media accounts. This is hard to detect because their social media accounts are mixed with normal posts as a camouflage.  Campaign types in both Chinese crowdsourcing sites~\cite{wang2012serf} and Western sites~\cite{lee2014social} have been studied. Three classes of crowdturfers (i.e., professional users, casual users, and middlemen) are identified in Twitter networks. In addition, their profile, activity, and linguistic characteristics have been analyzed to detect workers~\cite{lee2013crowdturfers}.
\citet{wang2014man} studied adversarial attacks against machine learning (ML) models of detecting malicious crowdsourcing workers.  Two types of adversarial attacks were identified: evasion attacks (i.e., attackers change behavioral features) and poisoning attacks (i.e., administrators pollute training data). ML is the best classifier to detect crowdturfing activity. The powerful features are user interactions and tweet dynamics. Evasion attacks can be very powerful when attackers have total knowledge. Poison attacks can reduce the detection efficacy by injecting carefully crafted data.  

\vspace{-3mm}
\subsection{\bf Human Targeted Attacks}
Recently, OSD attacks are extended to directly hurt humans which are obviously considered as cybercrimes. We discuss the following human related targeted OSD attacks: {\em human trafficking}, {\em cyberbullying}, {\em cybergrooming}, and {\em cyberstalking}. Each OSD attack under this category is detailed as:
\begin{itemize}
\item {\em Human Trafficking}: Offline traditional human trafficking means traffickers kidnap the victims (mostly victims women and children) for trading. There are labor trafficking and sex trafficking but less than half of the victims are in the sex trade~\cite{Feingold05}.  Cybertrafficking is traffickers using cyber platforms to exploit a great number of victims and advertise service across geographic boundaries~\cite{latonero2011human}.  Cybertrafficking is defined as traffickers transport persons by using any electronic, cyber platforms (e.g., social media, Internet services, etc.) to `coerce, deceive, or consent' with the aim of `exploitation' ~\cite{Greiman13}.
\item {\em Cyberbullying}: This is one type of cybercrime attacks that commits the deliberate and repetitive online harassing of someone, especially adolescents~\cite{Rathore17-is-survey}. Cyberbullying causes serious fear and harms for the victims through the online platforms.
\item {\em Cybergrooming}: This is another type of cybercrime attacks that the adult criminals intend to have sexual abuse activities with a child and hunt for children victims and create emotional connection in online social media platforms~\cite{Rathore17-is-survey, Zambrano19-cybergrooming}. 
\item {\em Cyberstalking}: The malicious users and cybercriminals exploit the normal user's online information and harass them by cyberstalking~\cite{Rathore17-is-survey}.  Without proper information security protection, many personal information can be disclosed in social media platforms unintentionally.  From user's profiles, posting and connections, the sensitive information revealed may include phone number, home address, location, and schedules.
\end{itemize}

\vspace{-4mm}
\subsection{Relationship between Social Deception Attacks, Social Network Attacks and Cybercrimes} \label{subsec:distinction-social-deception-attack}

\begin{wrapfigure}{r}{0.5\textwidth}
\vspace{-2mm}
    \centering
    \includegraphics[width=0.5\textwidth]{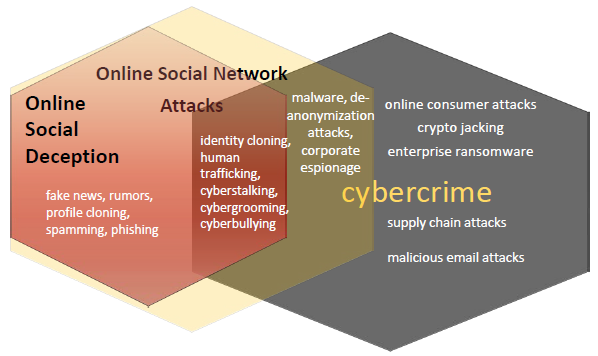}
    \caption{The relationships between OSN attacks, social deception, and  cybercrime.} 
\label{fig:broadness}
\vspace{-2mm}
\end{wrapfigure}

Social network attacks, including traditional threats, social threats and multimedia content threats, are the general security threats concerned in the literature~\cite{Rathore17-is-survey}.  Those security and privacy threats include all the detrimental activities with malicious intent. Social deception is part of social network attacks, as shown in Fig.~\ref{fig:broadness}, because social deception attacks can only be successful when the victims are being deceived from the attacker's perspective.  \\
\vspace{-4mm}
\vspace{-3mm}
\begin{wraptable}{r}{0.35\textwidth}
\vspace{-5mm}
\scriptsize
\centering
\caption{OSD attack types and their impact on security breach.}
\vspace{-3mm}
\label{tab:ons-attack-security-breach}
\begin{tabular}{|P{1.8cm}|P{2.2cm}|}
\hline
\bfseries Social Deception Attack &\bfseries Security Breach \\
\hline
Fake News  & Data Integrity  \\
\hline
Rumors  & Data Integrity \\
\hline
Information Manipulation &  Data Integrity\\
\hline
Fake Reviews
&  Data Integrity\\
\hline
Spamming & Account Confidentiality\\
\hline
Phishing & Account Confidentiality\\
\hline
Fake Profile & Account Integrity\\
\hline
Profile Cloning Attack & Authentication\\
\hline
Compromised Account & Account Integrity, Account Availability\\
\hline
Crowdturfing & Data Integrity, Network Integrity\\
\hline
Human trafficking & Confidentiality, Safety\\
\hline
Cyberbullying & Confidentiality, Safety\\
\hline
Cyber-grooming & Confidentiality, Safety\\
\hline
Cyberstalking & Confidentiality, Safety\\
\hline
\end{tabular}
\vspace{-6mm}
\end{wraptable} 
Three types of social network attacks are considered the online social deception (OSD) attacks: Unsolicited fake information attacks, identity attacks, crowdturfing, and human targeted attacks.  The specific types of attacks were described in Section~\ref{sec:types-social-deception}.  Some OSD attacks (e.g., personal and confidential information leakout, identity theft) have been treated as  cybercrimes~\cite{Nykodym05} since early 2000's. The advanced features of social network service technologies further have facilitated the significant increase of serious, sophisticated cybercrimes, such as human trafficking, online consumer fraud, identity cloning, hacking, child pornography, and/or online stalking~\cite{van2018determinants}.

Fig.~\ref{fig:broadness} illustrates the relations between OSN attacks, OSD attacks, and cybercrime.  Although cybercrime is considered most serious as cyberattacks, we can observe there are many attacks that overlap to each other.  OSD attacks overlap either OSN attacks or cybercrime or both.  Cybercrimes, such as consumer fraud, cryptojacking, enterprise ransomware, supply chain attacks, and malicious email attacks~\cite{Symantec19} fall in a separate group because these attacks are spread in the Internet, which is much broader than OSN platforms.  There are no explicit guidelines if certain OSN attacks or threats are illegal or if threats are illegal but their impact may not be direct. For example, when a user's data privacy (or integrity) is breached but no actually loss is found, it is hard to predict if there are future security concerns.  When the influences of OSN attacks are worse toward attack victims or organizations, the concept of social deception can define these security concerns.  

Although cybercriminals caused serious adverse effects to the society and individuals, 44\% of the victims reported to the police~\cite{Goucher10-victim}.  Victims' reporting is a beneficial practice to increase the awareness of the communities to defend against potential cybercrimes. Victims may report to not only the police, but also the corporation in an active dialogue environment, or share the victim stories to families and close friends~\cite{Goucher10-victim}.  Cybercriminal profiling is highly challenging, compared to profiles of traditional criminals; however, it is very beneficial to identify common characteristics of cybercriminals~\cite{Nykodym05}. Profiling can follow the procedure in the Behavioral Evidence Analysis~\cite{turvey2011criminal}.  Since most cybercrime victims are corporations and/or their customers, corporations can predict the potential insider criminals more intelligently with the help of cybercriminal profiling~\cite{Nykodym05}.

\vspace{-3mm}
\subsection{Security Goals Breach by Online Social Deception Attacks} \label{subsec:security-breach}  The CIA (confidentiality, Integrity and Availability) triad security goals play a major role in the information security practice.  With the growth of socio-technical security issues, the original CIA triad is expanded with more specialized aspects, such as authentication and non-repudiation   \cite{mell2011nist}.  However, they still have limitations in systems and data for the wider organizational and social aspects of security~\cite{samonas2014cia}.  
OSN security has three levels of security goals: network-level, account-level, and message-level. Achieving the CIA security goals can contribute to all social network security levels. We summarize how OSD attacks can breach security goals in Table~\ref{tab:ons-attack-security-breach}.  As the OSD research is closely related to many different disciplines studying human behaviors, a variety of deception cues have been studied in the literature. For the readers who may be interested in obtaining insights from those studies to develop tools to deal with OSD attacks based on the deception cues, we discussed various types of deception cues in the appendix document (see Section D Cues of Social Deception).  In addition, it is critical to investigating the victim profiles in terms of the predictors of potential victims' vulnerabilities to the OSD attacks for taking proactive actions to prevent them.  For those who are interested in the detailed survey on the susceptibility to the OSD, we also discussed it in the appendix document (see Section F Susceptibility to Online Social Deception).

\vspace{-2mm}
\section{Prevention Mechanisms of Online Social Deception} \label{sec:deception-prevention}

As the first set of defense mechanisms against OSD attacks, we discuss defense mechanisms to prevent OSD attacks in terms of using two types of techniques: {\em Data-driven prevention mechanisms} and {\em social honeypots} as discussed below.

\vspace{-2mm}
\subsection{Data-Driven Prevention Mechanisms} \label{subsec:data-driven-prevention-mechanisms}

Compared to their detection mechanisms, as discussed in Section~\ref{sec:deception-detection}, defense mechanisms to prevent OSD attacks have been less explored.  But in this paper, to make our survey complete, we discuss several types of prevention mechanisms that have been commonly used to deal with OSD attacks based on data-driven approaches, as follows:
\begin{itemize}
\item {\em Fake News Prevention}: \citet{saad2019fighting} proposed a blockchain-based system to fight against fake news by recording a transaction in blockchain when posting a news article and applying authentication consensus of the record. The result was indicated by an authentication indicator along with the post. In this design, when a user saw a post, authentication indicator showed the status of verification: successful, failed or pending. This mechanism achieved three goals of preventing fake news spread in the OSN: swift consensus was issued by the chaincode; the malicious user can be identified from the transaction record; and the false information posts can be deleted and a penalty can be applied to the fake news attackers. In general, the malicious attackers are the normal users but normal users do not have write access to the blockchain. Only the information source from a group of publishers or a group of social network are allowed to commit transactions to the blockchain. 
\item {\em Phishing Prevention}: \citet{florencio2007evaluating} proposed a phishing prevention method with a client reporting user password reuse activities in unknown websites and a server to make decisions and update the blocked list. The benefit is detecting phishing attacks reliably with low latency. \citet{gupta2011socio} proposed a defense model to classify web-pages on a collaborative platform PhishTank. This defense model uses a plug-in method into a browser to check blacklisting and blocking lists. 
\item {\em Identity Theft Prevention}: \citet{tsikerdekis2016identity} discussed a proactive approach of identity deception prevention using social network data.  Data in common contribution networks are used to establish a community's behavioral profile. Malicious accounts can be barred before joining a community based on the deviation of user behaviors from the community's profile. 
\item {\em Cyberbullying Prevention}:  \citet{dinakar2012common} proposed a dashboard reflective user interface in social network platforms for both cyberbullying attackers and victims.  The reflective user interface integrated notifications, action delay, and interactive education. Their user study revealed that the in-context dynamic help in the user interface is effective for the end-users.
\end{itemize}

\vspace{-2mm}
\noindent {\bf Pros and Cons}: These effective system design and real-time data analysis would be reliable to prevent online social deception. However, there is no real-world implementation of those proposed methods. Response delay issues may exist for the implementation. 

\vspace{-2mm}
\subsection{Social Honeypots} \label{subsec:social-honeypots}


Recently, there has been an idea that creating social network avatars (often called `good bots' in contrast to `bad bots') may be a good solution to identify malicious activities by highly intelligent, sophisticated attacks, such as advanced persistent attacks (APTs)~\cite{Virvilis14}. {\em Honeypots technology} is not new and has been popularly used in communication networks as a {\em defensive deception} to proactively deal with attackers by luring them to honeypots, preventing them from accessing a target~\cite{Cohen06}.  The existing approaches using {\em social honeypots} have mainly focused on detecting social spammers, socialbots~\cite{Zhu14-sh}, or malware~\cite{lee2010uncovering, lee2011seven, Paradise14, Paradise15, Paradise17, stringhini2010detecting, webb2008social} as a passive monitoring tool. These works use some profiles of attackers to detect them based on the features collected from the social honeypots placed as fake SNS accounts (e.g., Facebook or Twitter). But no real victim profiles have been used to develop the social honeypots. 

Although the original purpose of social honeypots was to proactively prevent attackers from accessing system/network resources, they have been used as a complement to detect various OSN attacks.  However, the original purpose of social honeypots lies in a proactive intrusion prevention mechanism.  In addition, although the social honeypots can be used as a detection tool for OSN or OSD attacks, their goal is an early detection or mitigation based on the proactive defense in nature.  Hence, we include social honeypots as prevention mechanisms of OSD attacks.   

For the social honeypots to be used as detection mechanisms, they are defined as information resources that monitor a spammer's behaviors and log their information (e.g., their profiles and contents in social networking communities)~\cite{lee2010uncovering}. This early study detected deceptive spam profiles in MySpace and Twitter by social honeypot deployment. Based on the spammer they attracted, a Support Vector Machine (SVM) spam classifier was trained to identify spammers and legitimate users.  An ML-based classifier was also developed to identify unknown spammers with high precision in two social network communities.

\citet{lee2011seven} detected content polluters in Twitter by designing Twitter-based social honeypots. The 60 social honeypot accounts follow other social honeypot accounts and post four types of tweets to each other. They investigated the harvested users to nine clusters via the Expectation-Maximization algorithm.  They did content polluters classification by Random Forest and improved the results by standard boosting and bagging and by different feature group combinations.

\citet{haddadi2010add} focused on privacy and fake profiles by characterizing fake profiles and reducing the threats of identity theft.  They set social honeypots using the fake identities of celebrities and ordinary people and analyzed the different behaviors (e.g., a number of friends, friends requests, and public/private messages) between those fake accounts.  \citet{stringhini2010detecting} studied 900 honey-profiles to detect spammers in three social network communities (e.g., MySpace, Facebook, and Twitter). They collected activity data for a long time (i.e., one year).  Their honey-profiles have geographic networks.  In addition, this work identified both spam profiles and spam campaigns based on the shared URL.

\citet{Virvilis14} described the common characteristics of advanced persistent threat (APT) and malicious insiders and discussed multiple deception techniques for early detection of sophisticated attackers, including creation of social network avatars in attack preparation phase (information gathering), along with fake DNS records and HTML comments.  \citet{Zhu14-sh} showed the analysis and simulation of infiltrating social honeybots defense into botnets of social networks.  The framework SODEXO (SOcial network Deception and EXploitation) had three components: Honeybot Deployment (HD), Honeybot Exploitation (HE), and Protection and Alert System (PAS).  HD set up a moderate number of honeybots in the social network.  HE modeled the dynamics and utility optimization of honeybots and botmaster by a Stackelberg game model.  The results showed that a small number of honeybots can significantly decrease the infected population (i.e., a botnet) in a large social network. 

\citet{Paradise14} and \citet{Paradise15} simulated defense account monitoring attack strategies in OSNs. The attackers sent friend requests to some community members chosen by different attacker strategies. In addition, the attackers may have full knowledge of the defence strategies.  The defender chose a set of accounts to monitor based on various criteria. They analyzed the acceptance rate, hit rate, a number of friends before hit, and monitoring cost between combinations of attackers and defenders.  The result showed that under the sophisticated attackers with the full knowledge of defence strategies, defense using PageRank and most connected profiles have the best detection with the minimum cost.

\citet{Paradise17} targeted at detecting the attackers in the reconnaissance stage of advanced persistent threat (APT). The social honeypot artificial profiles were assimilated into an organizational social network (Xing and LinkedIn) and received the friend requests to organization employees. The attacker profiles collected in the social honeypot were analyzed.  \citet{badri2016uncovering} collected fake Likers on Facebook by posting paid jobs using linkage and honeypot pages. They extracted the four types of profile and behavior features and trained classifiers to detect fake Likers. The temporal features were cost-efficient compared to the previous research. They also evaluated the robustness of their work by modifying features using individual attack model and coordinated attack model.  \citet{de2014paying} studied paying for `likes fraud' in Facebook and link the campaigns to honeypot pages to collect data. They analyzed and measured the page advertising and promotion activities.~\citet{nisrine2016security} discovered malicious profiles by social honeypot(s) and used both feature-based strategy and honeypot feature-based strategy to collect data. Combining honeypot features can increase the ML accuracy and recall compared to when traditional features are only used.  \citet{zhu2015fighting} defined ``active honeypots'' as active Twitter accounts, which capture more than 10 new spammers everyday, similar to the spammer network hubs.  They extracted 1,814 those accounts from the Twitter space and studied the properties and identification of active honeypots.  \citet{yang2014taste} conducted passive social honeypot to capture a spammer's preferences by designing social honeypots with various behaviors.  The design considered tweet behavior (i.e., tweet frequency, tweet keywords, and tweet topics), followed behaviors of famous people's accounts and application installation.  They analyzed which type of social honeypot has the highest capture rate and designed advanced social honeypot based on their results. They demonstrated that the advanced honeypot can capture spammers 26 times faster than the normal social honeypots.

\vspace{1mm}
\noindent {\bf Pros and Cons}: Social honeypots would be highly effective particularly when it is well deployed to attract targeted attackers.  However, developing social honeypots with fake accounts may introduce ethical issues because the use of the social honeypots itself is based on deceiving all other users as well.

\vspace{-2mm}
\section{Detection Mechanisms of Online Social Deception} \label{sec:deception-detection}

Most existing defense mechanisms to deal with OSD attacks focus on detecting those attacks.  We discuss those detection mechanisms based on three types: {\em user profile-based}, {\em message content-based}, and {\em network feature-based}. 

\vspace{-3mm}
\subsection{User Profile-based Deception Detection Mechanisms} \label{subsec:user-profile-detection}
Most profile cloning studies make use of the user profiles~\cite{kamhoua2017preventing, kontaxis2011detecting, shan2013enhancing}.  Identify cloned profiles, they all calculate profile similarities in different ways by using user profile attributes.  \citet{kontaxis2011detecting} proposed three components to detect profile cloning: an information distiller, a profile hunter, and a profile verifier. The profile verifier component calculated the profile similarity score between testing social profiles and the user's original profile.  Both the information field and profile pictures contributed to estimating the profile similarity. \citet{kamhoua2017preventing} detected user profiles across multiple OSNs in a supervised learning classifier.  The method consists of three steps: the profile information collection from a friend request, the friend list identity verification, and the report of possible colluders. The binary classifier was based on both the profile attributes similarity and friend list similarity.  \citet{shan2013enhancing} simulated profile cloning attacks by snowball sampling and iteration attack and then detected the attackers by a detector called `ChoneSpotter.'  The  context-free detection algorithm includes the profile information and friendship connections.  The input features include recently used IPs, a freind list, and the profile and profile similarity.  A cloned profile was determined by using the same IP prefix and the similarity over a certain threshold.

Some mechanisms detecting Sybil attacks, fake reviews and spamming extracted user profile features and user behavior/ activity features to detect malicious accounts~\cite{badri2016uncovering, cao2015detecting, cresci2017social, liang2015rumor, Paradise17, song2014discriminative, wang2013characterizing}.  \citet{badri2016uncovering} studied the feature engineering from the account of `fake likers.'  They considered profile features, such as the length of user introduction, the longevity of an account, and the number of friends.  Social activities represent a unique attribute observed in OSN and consist of the behavior features of an account, such as sending friend request, posting, retweeting, liking/disliking and social attention~\cite{badri2016uncovering}.  More specific features under each activity category can be further extracted, such as the acceptance of a friend request sent from~\cite{Paradise17} and the average time interval of posting from~\cite{song2014discriminative}.  \citet{wang2013characterizing} investigated several behavioral signatures for the output of crowdturfing campaigns and tasks.  \citet{cao2015detecting} studied the behavioral features to detect spam URLs in OSNs.  They used fifteen click and posting-based features in Random Forest classifiers and evaluated the top six features. 

\citet{cresci2017social} proposed a novel DNA-inspired social fingerprinting approach of behavioral modeling to detect spambot accounts.  Twitter account behaviors were encoded as a string of behavioral units (e.g., tweet, reply and retweet).  This new model can deal with the new type of spambots which can be easily missed by most traditional tools.  Social fingerprinting sequences are characterized by the longest common substring (LCS) curve.  Spambots are related to high LCS values by sharing suspicious long behavioral patterns. The LCS curve from behavioral model is used to detect more sophisticated types of crowdsourcing spammers. 

User profiles and activities are the key features to detect OSD attacks (e.g, advanced spammers or crowdturfing), along with other content-based and graph-based features~\cite{inuwa2018detection, lee2010uncovering, lee2011seven, lee2013crowdturfers, vosoughi2017rumor, wang2014man}.  Those hybrid detection examples will be discussed later in Section~\ref{subsec:hybrid-detection}.

\vspace{1mm}
\noindent {\bf Pros and Cons}: User profile information provides specific activity features and behaviors about each user.  However, some profile information is private; thus, collecting private information itself is the violation of a user's privacy right.  In addition, even if the information itself is open to the public, how to use the information should be agreed with the owner of the information.  Besides, collecting profile and behavioral data incurs high cost and/or time under privacy protection of the social media platforms.  
\vspace{-4mm}
\subsection{Message Content-based Deception Detection Mechanisms} \label{subsec:message-content-detection}

In Table A5 in the appendix document, we showed that the majority of social deception detection approaches have used content-based features because the text of user posts and reviews can be easily collected and analyzed using existing linguistic models.  The proliferation of social media and/or network applications allowed numerous types of raw and advanced content features available.  Topic modeling and sentiment-based features have been popularly utilized for the linguistic analysis of deceptive messages.  

\vspace{-3mm}
\subsubsection{\bf Topic Modeling-based Detection} \label{subsubsec:topic-modeling-detection}
Most of the work built topic distribution by Latent Dirichlet Allocation (LDA)~\cite{lau2014probabilistic, liu2016detecting, song2014discriminative, swe2018fake, wu2015false}.  If each user's posts are collected as a document, LDA generates the topic probability distribution of the user's document. \citet{liu2016detecting} extended the topic features to two new features.  A global outlier standard score (GOSS) indicates a user's interests in specific topics, compared to other users while a local outlier standard score (LOSS) indicates a user's interests in various topics.  By adding those two topic-based features to classifiers, the averaged F1-score shows better performance.  \citet{swe2018fake} built a keyword ``blacklist" to detect fake accounts by extracting topics from LDA and keywords from TF-IDF algorithms.  The blacklist contributed to 500 fake words.  The number and ratio of fake words and a few other content-based features were extracted for their classifier. The result using a ``blacklist" showed better accuracy than the traditional spam word list by reducing false positive rate. \citet{wu2015false} extracted the topic distribution of 18 topics for one message following the official Weibo topic categories.  The probability of 18 topics was used as one feature vector for the SVM classifier.

Two other work modified the LDA algorithm to detect cybercriminal accounts and spams.  \citet{lau2014probabilistic} developed a weakly supervised cybercriminal network mining method supported by a probability generative model and a novel context-sensitive Gibbs sampling algorithm (CSLDA).  The algorithm can extract the semantically rich representations of latent concepts to predict transactional and collaborative relationships (e.g., cybercriminal indicator) in publicly accessible messages posted on social media.  \citet{song2014discriminative} used Labeled Latent Dirichlet Allocation (L-LDA) to indicate the probability of co-occurrence.  The latent topics were normalized to topic-based features, which has distinct properties with TF-IDF generated word-based features. 

\citet{golbeck2018fake} detected two types of false article stories, which are fake news and satires by themes and word vectors.  Then they defined a theme by a new codebook with 7 theme types, such as conspiracy theory and hyperbolic criticism.  Multiple themes can be labelled to an article as a theme coding.  The proposed classifier worked better for articles under a certain type of theme.

\vspace{1mm}
\noindent {\bf Pros and Cons}: The topic features can be easily obtained.  However, the content-only features may not be able to capture other dynamic information such as user activites describing the interactions with other users (e.g., likes, acceptance of friend requests).  In addition, the topic model is highly sensitive to datasets; hence, depending on datasets, the effectiveness of topic models cannot be guaranteed. 

\vspace{-3mm}
\subsubsection{\bf Feature-based Deception Detection} \label{subsubsec:feature-detection}
Table A5 in the appendix document lists the feature set of each research work.  The commonly used features include raw features, such as word vector, word embedding, hashtags, links and URLs~\cite{markines2009social}.  Advanced features include deep content features, statistics, Linguistic Inquiry and Word Count (LIWC) and other metadata, such as location, source, or time~\cite{vandam2018cadet}.  Most ML-based models are supervised models. Among the supervised models, random forest, support vector machine (SVM), na\"{i}ve Bayes, logistic regression, and k-nearest neighbors are the most favorable classifiers for detection. Neural network models, such as Recurrent Neural Networks (RNN)~\cite{yao2017automated} and Convolutional Neural Network with Long Short-Term Memory (CNN-LSTM)~\cite{yang2019phishing}, are used for textural features. Temporal models, such as Dynamic Time Warping (DTW) and Hidden Markov Models (HMM)~\cite{Everett16, vosoughi2017rumor}, are discussed in rumor detection.  The boosting-based ensemble models are implemented for spammmer detection~\cite{inuwa2018detection, yang2019phishing}.  A few studies used semi-supervised models~\cite{inuwa2018detection, sedhai2017semi} when the labeled dataset is not available.

\citet{Everett16} studied the veracity of the automated online reviews regular users.  The text is generated by second-order Markov chain model.  The key findings include: (i) The negative crowd's opinion reviews are more believable to humans; (ii) Light-hearted topics are easier to deceive than the factual topics; and (iii) Automated text on adult content is the most deceptive. \citet{yao2017automated} investigated attacks of fake Yelp restaurant reviews generated by an RNN model and LSTM model.  The model considers the reviews themselves only, not including metadata as reviewers. Similarity feature, structural features, syntactic  features, semantic features, and LIWC features were used in SVM to compare the character-level distribution.  They found that information loss was incurred in the process of generating fake reviews from RNN models and the generated reviews can be detected against real reviews.  \citet{song2015crowdtarget} detected crowdturfing targets and retweets from crowdturfing websites and black-market sites. 

\vspace{1mm}
\noindent {\bf Pros and Cons}:  Feature-based models have high accuracy and low false positive rates.  The raw content features are easily obtainable although the extraction of sophisticated features are expensive. However, the temporal pattern of messages influence the detection rate and performance. The semantic analysis method may ignore the hidden messages and background knowledge. In addition, the model requires tuning many input parameters. 

\vspace{-2mm}
\subsubsection{\bf Sentiment-based Deception Detection} \label{subsubsec:sentiment-detection}
Sentiment of social media messages serves as extra features of message contents.  Sentiment provides emotional involvement, such as like, agree, or negation. It is calculated by lexicon analysis~\cite{bhatt2018combining, dinakar2012common, hu2014social, jiang2018linguistic, vosoughi2015human}.  One research aims at designing better lexicon~\cite{jiang2018linguistic}.  {\em ComLex} was introduced as a novel emotional and topical lexicon. This work analyzed the linguistic signals in user comments, regarding misinformation and fact-checking.  Specifically, it discussed the signals from user comments to misinformation posts, veracity of social media posts, or fact-checking effects. There are signals for positive fact-checking effect as well as signals (e.g., increased swear word usage) indicating potential ``backfire'' effects~\cite{nyhan2010corrections}, where attempts to intervene against misinformation only entrench the original false belief.

Sentiment features are often used along with TF-IDF word vectors. Supervised classifiers in current research utilize sentiment analysis to improve prediction.  \citet{bhatt2018combining} detected fake news stance from neural embedding, $n$-gram TF vector and sentiment difference between news headline-body TF vector pair. \citet{dinakar2012common} proposed a sentiment analysis to predict bullying, aiming at discovering goals and emotions behind the contents. Note that Ortony lexicon~\cite{ortony1987referential} maintains a list of positive and negative words describing the affect.  The lexicon of negative words was only added in the feature list to detect bully-related rude comments.

\vspace{1mm}
\noindent {\bf Pros and Cons}: Sentiment analysis includes more emotional and background information, in addition to the explicit content, which can increase the prediction accuracy, when compared to semantic-only methods.  However, the use of sentiment analysis cannot fully leverage the linguistic information in the contents where the lexicon is domain-specific.

\vspace{-3mm}
\subsection{Network Structure Feature-Based Detection} \label{subsec:network-feature-detection}
 
Several general network features were extracted in supervised learning methods, such as topology, node in-degree and out-degree, edge weight, and clustering coefficient~\cite{kumar2016disinformation, ratkiewicz2011truthy, vosoughi2017rumor}.  \citet{Wu16-bc} summarized false information spreader detection based on network structures. \citet{ratkiewicz2011truthy} built Truthy system to enable the detection of astroturfing on Twitter.  Their Truthy system extracted a whole set of basic network features for each meme and sent those features with a meme mood by sentiment analysis to supervised learning toolkit. \citet{kumar2016disinformation} developed four feature sets including network features to identify hoaxes in Wikipedia. The network features measure the relation between the references of the article in the Wikipedia hyperlink network.  The performance of features sets was evaluated in a random forest classifier.

Below we discuss algorithms and supervised learning methods specifically designed for the network structure, such as propagation-based models, graph optimization algorithms, and graph anomaly detection algorithms.

\vspace{-2mm}
\subsubsection{\bf Epidemic Models} \label{subsubsec:epidemic-models}
Epidemic model is a direct way to model and simulate the diffusion of disease~\cite{newman2002spread}.  Since the spread of disease in a certain population is similar to the propagation of false information in the social media communities, epidemic models have been often modified to quantify the extent of false information propagation~\cite{jin2013epidemiological}.  The epidemic models are agent-based models, where an individual node can be described as an agent.  Different types of agents are characterized by distinct states and behaviors, such as the agents Susceptible (S), Infectious (I), and Recovered (R) in the traditional SIR (Susceptible, Infectious, and Recovered) model~\cite{mussumeci2016modeling} in false information propagation.  In OSNs, agents in the SIR model represent a group of users in each state as follows: (i) {\em Susceptible} ($S$): Users who have not received information (e.g., rumor posts or fake news) yet but are susceptible to receive and believe it; (ii) {\em Infectious} ($I$): Users who received the information and can actively spread it; and (iii) {\em Recovered} ($R$): Users who received the information and refuse to spread it~\cite{zhao2011rumor}.  

\begin{figure}[!t]
\centering
\subfloat[SIR Model\label{subfig:sir}]{\includegraphics[height=1.2in]{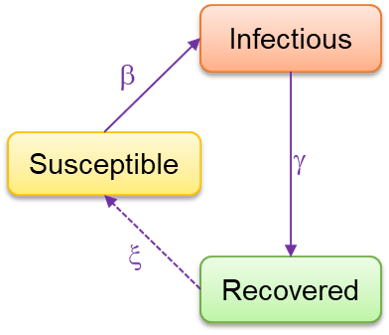}}
\subfloat[SIHR Model\label{subfig:sihr}]{\includegraphics[height=1.22in]{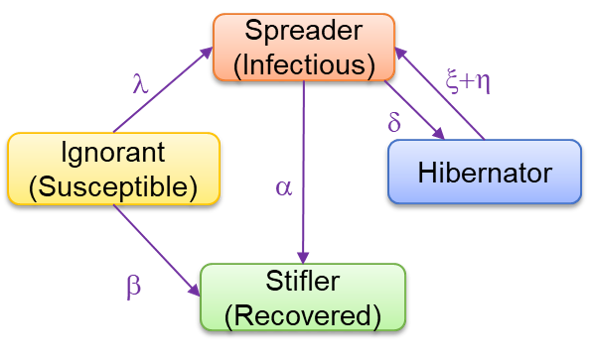}}
\subfloat[SEIZ Model\label{subfig:seiz}]{\includegraphics[height=1.2in]{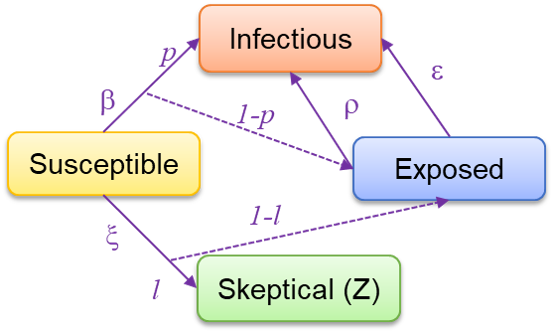}}

\caption{Three types of agent-based epidemic models. The solid line arrows are transitions from one state to another states with probabilites. The dotted line arrows are the transaction that may not exist at all times.  (a) SIR model: $\beta$ is infection rate, $\gamma$ is recovery rate, and $\xi$ is the rate of Recovered to Susceptible. (b) SIHR model: $\alpha$ is stifling rate, $\beta$ is refusing rate, $\gamma$ is spreading rate, $\delta$ is forgetting rate, $\eta$ is wakened remembering rate, and $\xi$ is spontaneous remembering rate. (c) SEIZ model: $\beta$ is infection rate, $\epsilon$ is self-adoption rate, $\phi$ is contact rate, and $\xi$ is skeptic rate. The details of $p$ and $l$ and the whole model were explained  in~\citet{jin2013epidemiological}.}
\label{fig:epidmic-models}
\vspace{-3mm}
\end{figure}

The state transitions are $S$ to $I$ by infection rate $\beta$, and $I$ to $R$ by recovery rate $\gamma$ depicted in Fig.~\ref{subfig:sir}.  The current false information propagation research has two tracks employing the epidemic models: (i) Adding more links and parameters to the traditional SIR model; or (ii) Building SEIZ model (Susceptible, Exposed, Infected, and Skeptic--Z; discussed below) to fit to the OSN data.

\noindent \textbf{\em SIR Model with Variations.}
Many variants of the basic SIR models  have been proposed in the current false information propagation research.  \citet{zhao2011rumor} added forgetting mechanisms to the SIR model for rumor spreading, so that the spreader ($I$) can be converted to stiflers ($R$).  Stiflers are defined similar to Recovered state.  They used the population size of $R$ to measure the impact of rumor.  They found that a forgetting mechanism can help reduce rumor influence and the rumor saturation threshold can be influenced by the average degree of nodes in the network. Another Hibernator state (i.e., users who refuse to spread rumor just because they forgot) was added to the SIHR (Susceptible, Infectious, Hibernator, and Recovered) model~\cite{zhao2012sihr} to measure forgetting rate $\alpha$ and remembering mechanism $\eta$. The new remembering mechanism was proved to delay the rumor termination time and reduce rumor maximum influence. The direct link from $S$ to $R$ was added by \cite{zhao2012sihr} and were extended by \cite{zhao2013sir}.  The update was that all users in state $S$ were finally converted to either $I$ or $R$ state if they had the chance to be exposed to spreaders ($I$).  Fig.~\ref{subfig:sir} and Fig.~\ref{subfig:sihr} describe the SIR and SIHR models, respectively.

\citet{cho2019uncertainty} extended the basic SIR model by replacing the transition between states to a decision based on the agent's belief on the extent of uncertainty in the agent's opinion.  The Subjective Logic opinion model is used to model an agent's opinion composition and update based on the extent of uncertainty.   The three states in the SIR are defined based on the degree of each dimension of an opinion which is defined by belief, disbelief, and uncertainty.  The opinion update involved interaction similarity between two agents, a conflict measure between belief and disbelief, and opinion decay upon no interactions between agents for opinion updates.  Based on the degree of uncertainty in a given opinion, an agent's opinion can move from any state to any other state.  This work investigated the effect of misinformation and disinformation in terms of how well false information can be effectively mitigated by propagating countering (true) information by selecting a good set of true informers.   


The evolutionary SIR model simulation has been used to model decision strategies in fake news attacks~\cite{kopp2018information}.  The state transitions in the SIR model was replaced by the decision model Iterated Prisoner's Dilemma (IPD).  The deception strategies can modify the prior knowledge of the agents by either adding uncertainty or changing false perceptions.  In their expensive simulation experiments, only a small population of fake news attackers can initiate the spread but the fitness of attackers was sensitive to the cost of deception. 

\noindent \textbf{\em SEIZ Model with Variations.}
\citet{jin2013epidemiological} captured diffusion of false and true news by the SEIZ epidemic model.  Instead of considering the Recovered state, they modeled a state of users being heard of the rumor but not spreading it (Skeptic, Z) and influenced users (E) posting the rumor with an exposure delay.  The SEIZ model was accurately capturing the diffusion patterns in real news and rumors events and was evaluated to be better than the simple SIS (Susceptible, Infectious, and Susceptible) model.  They also proposed a ratio $R_{SI}$, the transition rates entering $E$ from $S$ to the transition rates exiting $E$ to $I$, to differentiate rumor and real news events data.  \citet{isea2017new} extended the SEIZ model by modeling a forgetting rate of rumor posts. The forgetting rate is defined as a probability a user forgets the rumors across all the states.  Fig.~\ref{subfig:seiz} shows the key components of the SEIZ model and its process with the states and rates given from one state to another state.

\vspace{1mm}
\noindent \textbf{Pros and Cons:} Epidemic models provide a direct and straightforward mathematical model for the diffusion dynamics of the false information.  The agent density plot with time is a good way of observing the differences between the simulation and real values.  However, simulation tests face a common issue as the population size is unknown and stable, and initial variable values are unknown.  If the population size is as large as the real social media network, the computational cost cannot be ignored.  In addition, in the SIR model, the state change is controlled by probability; but this autonomous behavior ignores a user's intentions and belief.   To complement this, there have been some efforts~\cite{cho2019uncertainty, kopp2018information} focusing on modeling and evaluating the effect of subjective, uncertain opinion and trust of agents and the role of more agents in terms of false information diffusion.

\vspace{-2mm}
\subsubsection{\bf Credibility-based Models} \label{subsubsec:credibility-based models}
In OSNs, one of the methods detecting false information attackers, Sybil accounts and spammers is modeling the credibility score in the network~\cite{jin2014news,jin2016news, yu2008sybillimit}.  Existing works used various ways to represent the credibility score, such as reputation score, trust score and belief score.  Credibility in OSNs can be modeled by two methods: {\em classification-based} and {\em credibility propagation}.  A classification-based approach uses supervised learning algorithms~\cite{negm2018news}.  On the other hand, {\em the credibility propagation approach} constructs a network to propagate credibility values among users, tweet contents, events and activities~\cite{jin2014news}. Based on the credibility scores, ranking algorithms of users and posts can be conducted such as PageRank~\cite{akoglu2013opinion, chirita2005mailrank, ghosh2012understanding, yu2008sybillimit}. 

\citet{negm2018news} used 5Ws (i.e., who, what, when, where, and why) credibility to distinguish credible news and RSS files from news agencies to extract publication dates, headlines, contents, and locations to feed into different algorithms to calculate the credibility of a news agency.  The algorithms they compared are Term Frequency–Inverse Document Frequency (TF-IDF), TF-IDF with location, Latent Semantic Index (LSI), and TF with LSI and log entropy.  They concluded that TF-IDF and TF-IDF with location give the best results to calculate credibility. More recently, \citet{brian5W1H} leveraged the 5W1H extraction and news summarization techniques to propose the Inverted Pyramid Score (IPS) to distinguish structural differences between breaking and non-breaking news, with the long-term goal of contrasting reporting styles of mainstream and non-mainstream fake outlets.

\citet{jin2014news} have introduced a credibility propagation network for news content composed of three layers: message, sub-event, and event. The event layer talks about the main event the news covers, the sub-event layer relates events to the main event, and the message layer holds the content of the news article.  A graph optimization problem is formulated to calculate the credibility in this hierarchical network.  All the layers are content-based, and have direct relations with the credibility of the news. \citet{jin2016news} further proposed a verification method on credibility in a propagation model by using a topic modeling technique.  \citet{mitra2015credbank} constructed the CREDBANK corpus by tracking tweets, topics, events, and associated in-situ human credibility judgements to systematically study credibility of social media events tracked over real-time. They later leveraged this corpus to construct language and temporal models for credibility assessment \cite{mitra2017parsimonious,mitra2017credibility}. By identifying theoretically grounded linguistic dimensions, the authors presented a parsimonious model that maps language cues to perceived levels of credibility. For example, hedge words and positive emotion words were found to be associated with lower credibility. Additionally, by examining the temporal dynamics of the event reportages, they found that the amount of continued collective attention given to an event contain useful information about its associated levels of credibility \cite{mitra2017credibility}.

\citet{akoglu2010oddball} proposed an {\em OddBall} algorithm to detect anomalous behavior like malicious posts and fake donations.  They studied a sub-graph (egonets) of a target node with its neighbors.  They analyzed various scoring and ranking methods by using feature patterns in density, weights, principle eigenvalues, and ranks and compared their performance in different network topologies. 

\citet{kumar2018rev2} detected fake reviewers in user-to-item rating networks.  They developed a new trust system to rank users, products and ratings by fairness, goodness and reliability, respectively.  The intrinsic scores are calculated by combining network and behavior properties. Users have ratings with low reliability are more likely to be fake reviewers~\cite{kumar2018rev2}.  \citet{akoglu2013opinion} developed a FraudEagle algorithm to spot fraudsters as well as fake reviews in online review platforms.  There are two steps in the FraudEagle algorithm, namely, scoring users and reviews and grouping the analyzed results. For each review, the sentiment from true and false is only analyzed to assign the belief score.  The grouping step reviews top-ranked users in a subgraph by clustering and merging more evidence to reveal fraudsters.

\citet{ghosh2012understanding} developed a CollusionRank algorithm for detecting link farming type spammer attacks.  The influence scores were given to the users and web pages. By decreasing the influence score of the users connected to spammers, the follow-back behavior of social capitalists was discouraged.  \citet{yu2008sybillimit} developed a SybilLimit ranking algorithm for detecting Sybil attacks.  A Sybil node was identified by calculating the node's trust score.  \citet{chirita2005mailrank} developed a MailRank algorithm for detecting Sybil attacks in the email network. A sender is assessed by a global and personalized reputation score.  

\vspace{1mm}
\noindent \textbf{Pros and Cons:}  Credibility models can be applied in different stages and levels based on  contents, user behaviors, and posts/comments in highly heterogeneous networks.  In addition, a credibility model based on network features is agnostic to platforms and languages because the model only needs network features. However, how to accurately evaluate initial credibility values is not a trivial problem.  Considering credibility at multiple levels makes the computation more complex and expensive so it may not be preferred. Further, credibility may be subjective and cannot be ported across platforms and/or networks.  Lastly, a credibility model may not be able to detect sudden changes caused by instances which are not easily observable, thus impacting the accuracy of  credibility score assessment.

\vspace{-3mm} 
\subsubsection{\bf Cascades Features-based Models} \label{subsubsec:cascades-features-models}
Information network propagation patterns can be represented a cascading structure depicting the flow of OSD information flow that users time travelled through, posted, tweeted, and retweeted.  The cascading structure has two forms: {\em hop-based cascades} and {\em time-based cascades}~\cite{zhou2018fake}. The cascades features can be grouped into two approaches: (i) Calculating the similarity of cascades between true and false information; and (ii) Representing cascades using informative representation and features in a supervised learning model.

\noindent \textbf{\em Cascades Similarity.} \label{para:cascades-similarity} Cascades similarity is computed between fake news and true news.  Graph kernels~\cite{zhou2018fake} has been used as a common strategy for computing the cascades similarity. \citet{wu2015false} proposed a fake news detection method using a hybrid kernel function.  This graph kernel function calculates the similarity between different propagation trees. It also discussed about Radial Basis Function (RBF) kernel which calculates the distance between two vectors of traditional and semantic features.  The sentiment and doubt scores for user posts need to be verified for fakes news.  \citet{ma2017detect} proposed a top-down 
tree structure using recursive neural networks (RNNs) for false information detection.  The RNN learns the representation from tweets content, such as embedding various indicative signals hidden in the structure to improve rumors identification.

\noindent \textbf{\em Cascades Representation.} \label{para:cascades-representation}  Cascades representation pursues informative representation as features to distinguish fake news from true news.  For example, the number of nodes is a feature in a non-automated way.  Alternatively cascades representation can fit deep learning models~\cite{wu2018tracing}.  \citet{wu2018tracing} used LSTM-RNN to model propagation cascades of a message.  This work combines the propagation pathways with user embedding, which forms a heterogeneous network. A message is represented by a sequence of its spreaders.  Modularity maximization algorithm is used to cluster nodes with embedding vectors.  \citet{ma2018rumor} proposed propagation trees using Propagation Tree Kernel (PTK) for rumor detection. It can explore the suggested feature space when calculating the similarity between two objects.

\vspace{1mm}
\noindent {\bf Pros and Cons:} Similarity-based approaches consider the roles users play in false information propagation. Computing similarity between two cascades may require high computational complexity~\cite{zhou2018fake}.  Representation-based methods automatically represent news to be verified, but the depth of cascades may challenge such methods as it is equal to the depth of the neural network. All the approaches only provided experimental data to show their effectiveness.  However, it may not properly reflect real world settings. Training data is a time-consuming process and is often computationally expensive.

\vspace{-2mm}
\subsubsection{\bf Game Theoretic Models} 
\label{subsec:game-theory}
This explores the deception and defense by reward and penalty model in OSD attacks. In game theory, the actions and decisions of the players are mainly based on the reward and penalty of their previous activities and the other players' actions~\cite{Tadelis13}.  

\citet{kopp2018information} discussed a game theoretic false information propagation model as a deception model that simulates the propagation of fake news in the OSNs. They used three types of game theories: Greenberg's deception model~\cite{Greenberg82-jcr}, Li and Cruz's deception model~\cite{Li09-deception-game}, and hypergame theory~\cite{Bennett79-hyper}.  The Greenberg's deception model investigated the effect of deception on players' payoffs~\cite{Greenberg82-jcr}. \citet{kopp2018information} mapped false information to Greenberg's false signal model.  \citet{Li09-deception-game} used passive and active deception strategies by introducing noise and randomization, respectively, to increase uncertainty.  \citet{kopp2018information} used the deception game in \cite{Li09-deception-game} for consistently monitoring constraints and conditions, which affects game strategies. \citet{Bennett79-hyper} used hypergame theory to model a deception game where players have subjective perception and understandings of a complicated game. \citet{kopp2018information} also used \cite{Bennett79-hyper} to consider players' subjective belief which may introduce uncertainty as well.  The information theoretic model proposed by \citet{kopp2018information} found that attackers' deceptive behavior can be significantly mitigated when the cost of deception is fairly expensive.




\vspace{1mm}
\noindent \textbf{Pros and Cons:} Game theoretic approaches to model OSD attacks add extra features over and above other conventional network-structure based approaches by considering the cost and benefit of performing a deceptive behavior by users in OSNs.  Game theoretic deception detection is a promising approach that reflects human behaviors aiming to take an optimal action based on the expected outcome.  However, game theoretic approaches have been rarely adopted in modeling and analyzing online social deceptive behaviors when compared with data-driven deception detection approaches. Due to this reason,  the effectiveness of game theoretic deception detection approaches has not been fully investigated in the literature.  In addition, aligned with a conventional drawback in using game theory, a large number of deceptive actions may introduce a high solution complexity. In addition, uncertain, subjective beliefs of users should be carefully considered in terms of modeling incomplete information and/or imperfect information in game theory.  

\vspace{-2mm}
\subsubsection{\bf Blockchain-based Models} 
\citet{huckle2017fake} developed a tool called {\em Proventor} to prove the origin of the media. The Proventor is based on Blockchain storing provenance metadata for users to trust the authenticity of the metadata. Provenator can be used to validate news for news outlets like CNN and BBC where information and news is sometimes gathered from independent sources.  However, since Provenator uses Blockchain and cryptography, a small difference, such as one pixel difference between two images, can make the result vastly different, leading to generating numerous false alarms and human intervention for validation, which is labor-intensive.  In addition, managing the large ledger size in Blockchain is an issue as shared information in social media and news outlets grows exponentially.  \citet{mcevily2018incentivized} proposed a social media platform called Steem (i.e., a database)
based on Blockchain technology for building a community reward system. The reward system relies on users for consensus voting, reading content, and commenting. 

\vspace{1mm}
\noindent \textbf{Pros and Cons:} The original design of Blockchain has security benefits in terms of provenance, integrity and immutability.  The Blockchain system is a heterogeneous network that incorporate other stakeholders to detect and control online social deception activities.  In addition, it is resilient against OSD attacks. However, since both flagging accuracy and consensus verification rely on the contribution of crowd signals, it may break when too many users are malicious.  
For example, if a high number of attackers contribute to the crowd activities and even control the system, a user cannot access to write transactions.  In addition, the authorized party may be compromised by advanced attackers.

\vspace{-3mm}
\subsubsection{\bf Other Network Optimization Models} \label{subsubsec:network-optimization-models}  Several graph optimization algorithms were proposed in graph anomaly detection and community detection problems.  \citet{hu2013social} developed a matrix factorization-based algorithm to detect social spammers on Twitter.  Their framework utilized both content information and network information of an adjacency matrix and solved a non-smooth convex optimization problem.  Several approaches have been taken to detect link farming attacks via network structure-based algorithms.  \citet{araujo2014com2} detected temporal communities in cell networks and computer-traffic networks based on Tensor analysis.  \citet{jiang2014inferring} detected behavior patterns in OSNs where the spectral subspaces have different patterns and different lockstep behaviors.  In addition, \citet{jiang2014catchsync} identified synchronized behaviors from spammers.  \citet{kumar2014accurately} considered trolling as a social deception activity. They proposed a {\em decluttering algorithm} to break a network into smaller networks on which the detection algorithm can be run.   \citet{kumar2017army} considered {\em sockpuppets} as an OSD attack where users create multiple identities to manipulate a discussion.  They found that sockpuppets can be distinguished from normal users by having more clustered egonets.

\vspace{1mm}
\noindent {\bf Pros and Cons}: Graph-based features are more available compared to the user profile and/or user interaction features without violating privacy issues.  In addition, graph-based algorithms can be agnostic to any datasets with high applicability in diverse platforms.  However, collecting graph-based features, such as centrality measures, and solving graph optimization often require a high computational cost.  This hinders applicability to platforms that require real-time detection for streaming data.

\begin{figure}[!t]
    \centering
\begin{tikzpicture}[
  basic/.style  = {draw, text width=4cm, font=\sffamily, rectangle},
  root/.style   = {basic, rounded corners=2pt, thin, align=center, fill=green!60, text width=2.5cm, font=\relsize{0}},
  level 1/.style={basic, rounded corners=6pt, thin,align=center, fill=green!30, text width=6em, font=\relsize{-1}, sibling distance=68mm},
  level 2/.style={basic, rounded corners=6pt, thin,align=center, fill=yellow!60, text width=4.5em, font=\relsize{-1}, sibling distance=28mm},
  edge from parent/.style={->,draw},
  level 3/.style = {basic, thin, align=left, fill=pink!60, text width=4.2em, font=\relsize{-1}},
  level distance=3em,
  scale = 0.8]
\node[root] {OSD Defense}
  child {node[level 1] (x1) {Prevention}
    child {node[level 2] (c1) {Data-Driven}}
    child {node[level 2] (c2) {Social Honeypot}}
    }
  child {node[level 1] (x2) {Detection}
    child {node[level 2] (c3) {User Profile}}
    child {node[level 2] (c4) {Message Content}}
    child {node[level 2] (c5) {Network Structure}}
    }
  child {node[level 1] (x3) {Mitigation}};
\begin{scope}[every node/.style={level 3}]
\node [below of = c1, xshift=10pt, yshift=5pt] (c11) {Fake News};
\node [below of = c11, yshift=10pt] (c12) {Phishing};
\node [below of = c12, yshift=10pt] (c13) {Fake Profile};
\node [below of = c13, yshift=10pt] (c14) {Cyberbullying};

\node [below of = c2, xshift=10pt, yshift=4pt] (c21) {Spamming};
\node [below of = c21, yshift=10pt] (c22) {Fake Profile};
\node [below of = c22, yshift=10pt] (c23) {Socialbot};

\node [below of = c3, xshift=10pt, yshift=5pt] (c31) {Rumor};
\node [below of = c31, yshift=10pt] (c32) {Fake Review};
\node [below of = c32, yshift=10pt] (c33) {Spam};
\node [below of = c33, yshift=10pt] (c34) {Fake Profile};

\node [below of = c4, xshift=10pt, yshift=5pt] (c41) {Fake News};
\node [below of = c41, yshift=10pt] (c42) {Rumor};
\node [below of = c42, yshift=10pt] (c43) {Fake Review};
\node [below of = c43, yshift=10pt] (c44) {Phishing};
\node [below of = c44, yshift=10pt] (c45) {Spam};
\node [below of = c45, yshift=6pt] (c46) {Fake Account};
\node [below of = c46, yshift=2pt] (c47) {Compromised Account};
\node [below of = c47, yshift=6pt] (c48) {Crowdturfing};
\node [below of = c48, yshift=10pt] (c49) {Cyberbullying};
 
\node [below of = c5, xshift=10pt, yshift=5pt] (c51) {Fake News};
\node [below of = c51, yshift=10pt] (c52) {Rumor};
\node [below of = c52, yshift=10pt] (c53) {Fake Review};
\node [below of = c53, yshift=7pt] (c54) {False Information};
\node [below of = c54, yshift=6pt] (c55) {Spam};
\node [below of = c55, yshift=10pt] (c56) {Sybil Node};
\node [below of = c56, yshift=10pt] (c57) {Crowdturfing};

\node [below of = x3, xshift=10pt, yshift=-20pt] (x31) {Fake News};
\node [below of = x31, yshift=10pt] (x32) {Rumor};
\node [below of = x32, yshift=7pt] (x33) {Compromised Account};
\node [below of = x33, yshift=6pt] (x34) {Cyberbullying};
\end{scope}
\foreach \value in {1,...,4}
  \draw[->] (c1.195) |- (c1\value.west);
\foreach \value in {1,2,3}
  \draw[->] (c2.195) |- (c2\value.west);
\foreach \value in {1,...,4}
  \draw[->] (c3.195) |- (c3\value.west);
\foreach \value in {1,...,9}
  \draw[->] (c4.195) |- (c4\value.west);  
\foreach \value in {1,...,7}
  \draw[->] (c5.195) |- (c5\value.west);  
\foreach \value in {1,...,4}
  \draw[->] (x3.185) |- (x3\value.west);  
\end{tikzpicture}
\caption{Classification structure of our survey on OSD defense mechanisms.  The types and subtypes of OSD defense are illustrated in the tree structure.  Under each of the method subtype, the OSD attack types are summarized from the surveyed literature.}
\label{fig:defense-structure}
\end{figure}
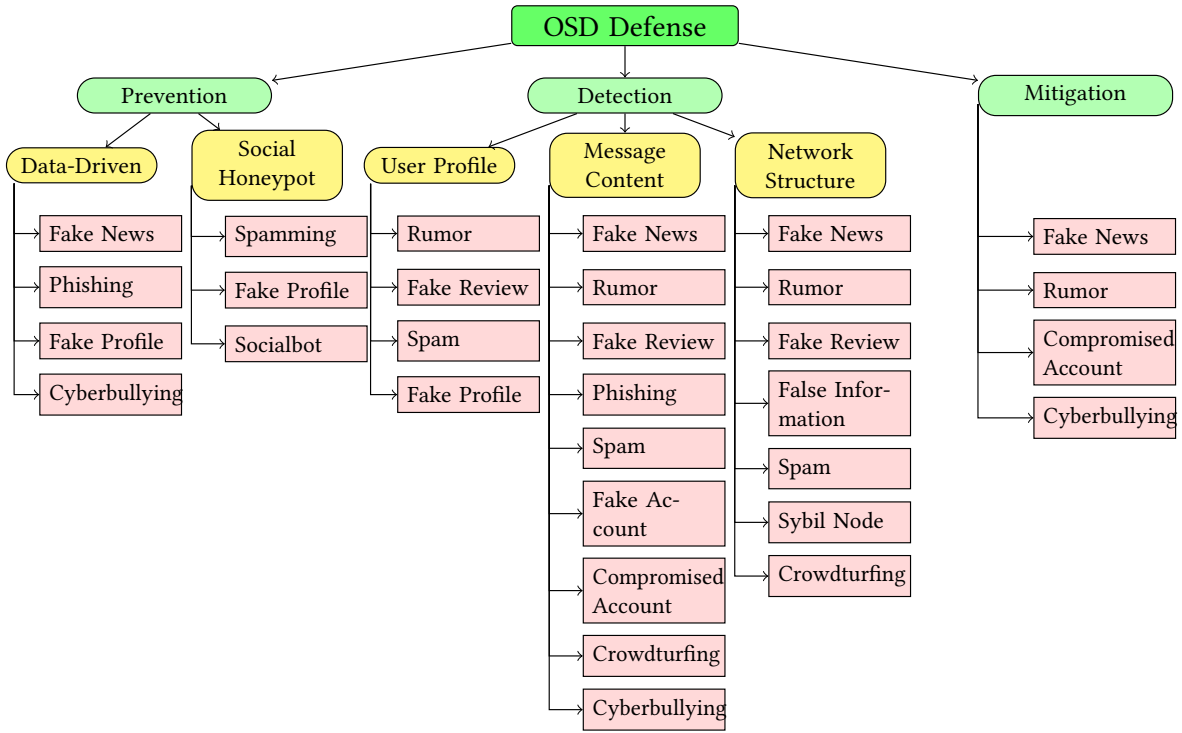

\vspace{-3mm}
\subsection{Hybrid Detection} \label{subsec:hybrid-detection}
Since ML/DL-based models can take an abundant amount of features, one can train a hybrid feature set combining the user profile, message content, and network features to detect OSD attacks. Unlike several existing survey papers which discussed only individual feature categories~\cite{kumar2018false, Wu2017-crowdturfing}, our discussion will focus on dealing with OSD attacks using hybrid features~\cite{inuwa2018detection, lee2010uncovering, lee2011seven, lee2013crowdturfers, vosoughi2017rumor, wang2014man}.

\citet{lee2013crowdturfers} detected crowdturfers 
from Twitter users.  A total of 92 features were divided into 4 groups: User demographics, user friendship networks, user activity (behavior-based features), and user content similarity including linguistic feature from LIWC dictionary.  \citet{vosoughi2017rumor} developed a tool called {\em Rumor Gauge} for automatically verifying rumors and predicting their veracity before they are verified by trusted channels.  Since rumors are temporal, time-series features are extracted as the rumor spreads.  A total of 17 features (e.g., linguistics, user involved, and propagation dynamics) were studied. They found that the fraction of low-to-high diffusion in the diffusion graph is the most predictive feature to represent the veracity of rumors.  The time-series features are processed in DTW and HMM models but DTW assumes all the time-series are independent and assigns equal weight to all 17 features.  The experiment evaluated the performance of the Rumor Gauge in terms of the accuracy of veracity prediction, contribution of each individual feature, and contribution of three groups of features and accuracy as a function of latency.

\vspace{1mm}
\noindent {\bf Pros and Cons}: Hybrid detection takes advantages of 
hybrid feature sets and can improve the accuracy in detecting rumors, spammers, and crowdturfings.  
A drawback is expensive feature engineering and acquisition. Furthermore, the training process is time-consuming with the complexity increasing as the feature size increases. 

\begin{figure}[!t]
    \centering
\begin{tikzpicture}[font=\scriptsize]
\begin{axis}[
    ybar,
    legend style={
        cells={anchor=west},
        legend pos=outer north east},
    bar width=.28cm,
    width=.92\textwidth,
    height=.25\textwidth,
    ymin=0,
    ymax=12,
    enlarge x limits=0.15,
    ylabel={\# dataset},
    ylabel near ticks,
    symbolic x coords={False Information, Luring, Fake Identity, Crowdturfing, Human Targeted Attacks},
    xtick=data,
    nodes near coords,
    nodes near coords align={vertical},
    ]
\addplot coordinates {(False Information,8) (Luring,10) (Fake Identity,6) (Crowdturfing,4) (Human Targeted Attacks,2)};
\addplot coordinates {(False Information,5) (Luring,3) (Fake Identity,0) (Crowdturfing,0) (Human Targeted Attacks,0)};
\addplot coordinates {(False Information,3) (Luring,1) (Fake Identity,1) (Crowdturfing,0) (Human Targeted Attacks,0)};
\addplot coordinates {(False Information,1) (Luring,3) (Fake Identity,4) (Crowdturfing,0) (Human Targeted Attacks,0)};
\addplot coordinates {(False Information,10) (Luring,10) (Fake Identity,4) (Crowdturfing,2) (Human Targeted Attacks,4)};
\legend{Twitter, Sina Weibo, Facebook, Synthetic, Others}
\end{axis}
\end{tikzpicture}
\caption{Dataset counts for the four categories of deception: False Information, Luring, Fake Identity, Crowdturfing, and Human Targeted Attacks where each category has several dataset sources from Twitter, Sina Weibo, Facebook, synthetic and other sources. The datasets are collected from all the approaches for the prevention, detection, and mitigation of OSD attacks. }
\label{fig:dataset-attack-type}
\vspace{-5mm}
\end{figure}
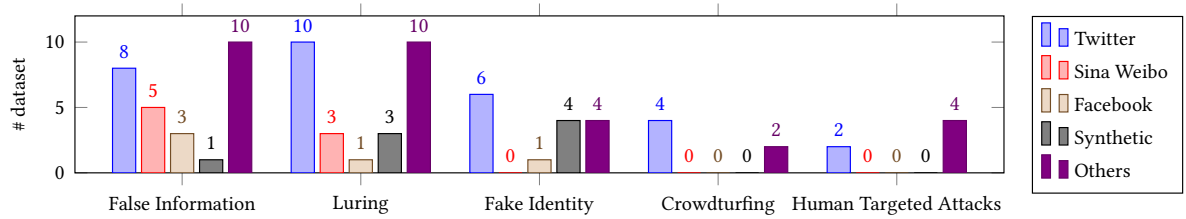
\vspace{-2mm}
\section{Response Mechanisms to Online Social Deception} \label{sec: recovery}
In this section, we survey existing mitigation or recovery mechanisms after OSD attacks are detected along with early detection mechanisms of OSD attacks~\cite{dinakar2012common, florencio2007evaluating, Wu16-bc}.  \citet{florencio2007evaluating} developed a mitigation strategy to deal with compromised accounts by detecting password reuse events and timely reporting it to financial institutions.  The aftermath actions are to take down identified phishing sites, restore the compromised accounts, and rescue users from bad decisions. \\
\vspace{-3mm}
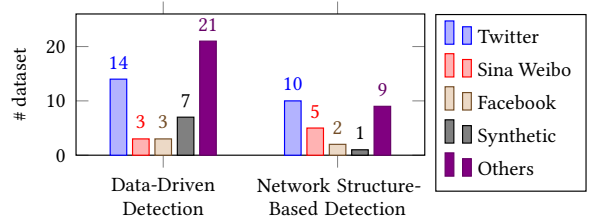
\begin{wrapfigure}{r}{0.5\textwidth}
\vspace{-6mm}
\centering
\begin{tikzpicture}[font=\scriptsize]
\begin{axis}[
    ybar,
    nodes near coords,
    nodes near coords align={vertical},
    legend style={
        cells={anchor=west},
        legend pos=outer north east},
    width=.45\textwidth,
    height=.25\textwidth,
    bar width=.2cm,
    ymin=0,
    ymax=26,
    enlarge x limits=0.5,
    symbolic x coords={ 0, Data-Driven Detection, Network Structure-Based Detection, 1},
    ylabel={\# dataset},
    ylabel near ticks,
    xtick=data,
    tick label style={font=\scriptsize},
    x tick label style={text width=2cm, align=center}, 
    scale=0.9]
    
\addplot coordinates {(Data-Driven Detection,14) (Network Structure-Based Detection,10)};
\addplot coordinates {(Data-Driven Detection,3) (Network Structure-Based Detection,5) };
\addplot coordinates {(Data-Driven Detection,3) (Network Structure-Based Detection,2)};
\addplot coordinates {(Data-Driven Detection,7) (Network Structure-Based Detection,1)};
\addplot coordinates {(Data-Driven Detection,21) (Network Structure-Based Detection,9)};
\legend{Twitter, Sina Weibo, Facebook, Synthetic, Others}
\end{axis}
\end{tikzpicture}
\vspace{-5mm}
\caption{Dataset count used in data-driven OSD detection techniques shown in Table A5 of the appendix document and network structure based OSD detection techniques shown in Table A6 of the appendix document: False Information, Luring, Fake Identity, Crowdturfing, and Human-Targeted Attacks and each category has several dataset sources from Twitter, Sina Weibo, Facebook, synthetic and other sources.}
\label{fig:dataset-source}
\vspace{-4mm}
\end{wrapfigure}
\citet{dinakar2012common} took a mitigation action to counter cyberbullying with two steps: (i) early detection; and (ii) reflective user interfaces that pop up notices and suggestions on user behaviors. Most efforts made to mitigate OSD attacks in OSNs mainly focused on reducing the effect of false information propagation.  \citet{Wu16-bc} summarized two misinformation intervention methods: (i) detecting and preventing misinformation from spreading in an early stage; and (ii) developing a competing campaign to fight against misinformation.  To limit the spread of fake news, a sample of fake news with maximal utility was identified in~\cite{tschiatschek2018fake}.  Within a certain constraint, this sample of fake news kept the largest number of users away from fake news posts.  Their algorithm was robust against a high amount of spammers. \citet{huckle2017fake} also made an effort to mitigate fake news spread based the validity proof of digital media data, such as a picture in the fake news. The blockchain technology was used to prove the origins of digital media data; however, this method cannot prove the authenticity of the whole news article.  \citet{kumar2018false} summarized misinformation mitigation by modeling true and false information. From the existing four different approaches, the authors concluded that these algorithms are effective in detecting the spread of rumor and their simulations can suggest rumor mitigation strategies.  \citet{okada2014sir} studied rumor diffusion by an SIR-extended information diffusion model and developed a mitigation mechanism to ask high influential users to spread correction diffusion.  The authors examined how false rumor diffuses and converges when help and/or correct information is given and how fast the convergence appears.

\vspace{1mm}
\noindent {\bf Pros and Cons}: 
Mitigation and recovery mechanisms relied heavily on early detection. The simulation model of spreading true information can mitigate the negative influence; but there is a lack of real-world deployment. Recovery in OSNs is more difficult than offline social networks.  Only one research~\cite{florencio2007evaluating} designed a system for account restoration.  More research efforts should be made to effectively mitigate the aftermath after early detection. 

Fig.~\ref{fig:defense-structure} summarizes the classification of OSD defense mechanisms including prevention, detection, and mitigation/response discussed in Sections~\ref{sec:deception-prevention},~\ref{sec:deception-detection} and~\ref{sec: recovery}. Existing works mostly focused on detection of OSD attacks we classified in Section~\ref{sec:types-social-deception}. Less attention is paid to prevention and mitigation with the focuses now mainly on false information, luring, and identity theft. There are still open questions to build trustworthy cyberspace against human-targeted attacks, especially for protecting children.  

\vspace{-2mm}
\section{Validation \& Verification} \label{sec:v-v}
\subsection{Datasets} \label{subsec:datasets}

We summarized all the datasets used in existing works in OSD prevention and detection in Tables~A4--A5 of the appendix document.  Most datasets are from large social media platforms, such as Twitter, Sina Weibo, Facebook, Youtube, and Reddit.  Twitter is the most frequently used data source probably because of the user friendly API for public users to download tweets in a certain time period.  Fig.~\ref{fig:dataset-attack-type} demonstrates the frequency distribution of each data source for four types of OSD attacks, namely, False Information, Luring, Fake Identity, Crowdturfing, and Human Targeted Attacks. Twitter, Weibo and Facebook platforms are drawn with synthetic datasets and datasets from all other sources. Datasets for false information attacks (e.g., rumors, fake news and fake reviews) and luring attacks (e.g., spamming and phishing) draw the most attention from researchers. Fig.~\ref{fig:dataset-source} illustrates the dataset platforms distribution for two types of OSD attack detection approaches, namely, data-driven detection and network structure-based detection.  Twitter is still the preferable data source.  It demonstrates the diversity of the sources of datasets used in the literature.  

\vspace{-2mm}
\subsubsection{\bf Datasets for Data-Driven Approaches} \label{subsubsec:data-data-driven}
Fig.~\ref{fig:dataset-source} shows the distribution of datasets used in data-driven approaches.  Twitter datasets are broadly used in all types of OSD attack detection mechanisms, such as spambot, malicious account, fake account, compromised account, rumors, and crowdturfing.  Other data sources include LinkedIn, YouTube, online forums Reddit, blacklisting websites, fact-checking websites, crowdturfing worker sites, and PhishiTank websites, depending on the type of OSD attacks. Several benchmark datasets are frequently used, such as a social hoenypot dataset~\cite{lee2011seven} in which the authors collected a lot of spammer accounts by using social honeypots deployed in Twitter networks for seven months. 

\vspace{-2mm}
\subsubsection{\bf Datasets for Network-Structure Approaches} \label{subsubsec:data-network-structure}

Fig.~\ref{fig:dataset-source} also shows the dataset distribution by sources in network structure-based detection. Twitter, Weibo, and Facebook are the top three individual data sources. The others include fact-checking websites, app store database, online forums, and rating platforms.  The datasets for network structure-based approaches can be divided into simulation research and detection research.  Synthetic datasets are more frequently used in simulation models, such as epidemic models and/or credibility/ranking-based models. 

\vspace{-2mm}
\subsection{Metrics} 
\label{subsec:metrics}

Most data-driven approaches have used metrics to estimate the detection accuracy of OSD attacks.  The following metrics have been considered in the literature: confusion matrix, precision, recall, F1 score or measure, accuracy, false positive rate (FPR), false negative rate (FNR), specificity, weighted cost, receiver operating characteristic (ROC) curve, area under the curve (AUC), discounted cumulative gain (DCG), Matthews correlation coefficient (MCC), Cohen's Kappa Value ($\kappa$), mean absolute error (MAE), 2-norm error, mean fraction of recovered agents per time unit (R), Spearman's Rank correlation coefficient, label ranking average precision (LRAP), and label ranking loss (LRL).  Due to the space constraint, we discuss each of this metric in the appendix document (Section G).

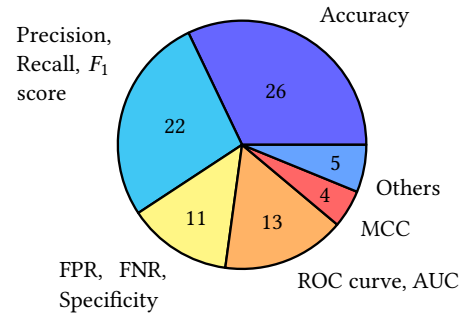
\begin{wrapfigure}{r}{0.4\textwidth}
\vspace{-5mm}
\centering
\begin{tikzpicture}[font=\footnotesize, scale=0.75]
 \pie [ sum = auto , after number = , radius =2]
    { 26/Accuracy, 22/\parbox{4em}{Precision, Recall, $F_1$ score}, 11/\parbox{4.5em}{FPR, FNR, Specificity}, 13/\text{ROC curve, AUC}, 4/MCC, 5/Others}
\end{tikzpicture}
\caption{Counts of research works by metrics.}
\label{fig:metric}
\vspace{-4mm}
\end{wrapfigure}

Fig.~\ref{fig:metric} illustrates the counts of papers (i.e., how many papers) that have used a particular group of metrics.
Since most of the current studies are to develop OSD attack detection mechanisms, the majority of the metrics is related to measuring detection accuracy.  Among all the detection metrics, Precision, Recall, $\mathrm{F}_1$ score, Accuracy are the most popular metrics used in the literature. FPR, FNR, Specificity, ROC, and AUC are also obtained based on the Confusion Matrix.  They are used to compare the performance of multiple classifiers. Algorithmic complexity of defense algorithms is rarely considered.

\vspace{-2mm}
\section{Ethical Issues of Social Deception} \label{sec:ethical-social-deception}

\citet{Paradise17} discussed legal and ethics considerations of social honeypots and artificial profiles.  \citet{dittrich2015ethics} provided an overview of ethics of social honeypot combined with the use of deception.  Social and behavioral research falls into the type of human subject research that is regulated by institutional review board (IRB).  The authors discussed privacy issues in using personal data, the use of deception in research, stackholder analysis, and ethical issues associated with deception. They also showed a case study that three early social honeypot studies~\cite{grier2010spam, lee2010uncovering, zhu2012sodexo} lacked the statement about issues of privacy, ethics, and IRB review. 
\citet{elovici2014ethical} provided guidelines for actions with ethical considerations.  They investigated the privacy and security concerns to obtain OSN data and the benefits to have reliable experimental research on OSNs. 

\citet{zhu2015fighting} discussed their ethical considerations on Twitter suspended some fake accounts in their hub account imitation study. They explained that their fake accounts are only for detecting spammers. 
For some activities against Twitter rules~\cite{Twitter19}, they limited their targets and posting frequencies to minimize their negative effects. \citet{de2014paying} also discussed ethics considerations and justification in their data collection activity using Fackbook honeypot page deployment. \citet{yang2014taste} brought up what if OSN normal operation is influenced by social honeypot deployment. They justified their tweets by not sending malicious tweets with @mention or URLs. Their actions only affect a few verified accounts and have little influences on other normal users.  Besides, \citet{matwyshyn2010ethics} advocated that security vulnerability research for security prevention and serving as social functions are neither unethical nor illegal.

There have been hot debates on legal and ethics considerations associated with developing security tools to deal with OSD attacks. Since human users are the key part of OSNs and the key entities to be protected in OSNs, we should be highly careful in developing defense tools against OSD attacks while keeping user privacy intact as the first priority.

\vspace{-2mm}
\section{Discussions: Insights \& Limitations} \label{sec:insight-limitations}

Based on the extensive survey conducted, we identify the following insights:
\begin{itemize}
\item {\bf Deception domains and intent}: Deception is defined across multidisciplinary domains with varying intent and detectability in type and extent.  Although social deception frequently is considered a negative connotation with low integrity and maliciousness, not necessarily all socially deceptive behaviors have bad intent. Rather, social deception can play a defensive social role for self-protection or self-presentation. 
\item {\bf OSD type category}: Like online social network (OSN) attacks and cybercrimes, online social deception (OSD) can be defined by deceptive intent. However, unlike OSN attacks or cybercrimes, a unique aspect of the OSD is that OSD is only possible when a deceivee cooperates with a deceiver. Hence, training and education of deceivees is critical for preventing OSD attacks.

\item {\bf Important social deception cues}: Traditional offline deception cues and vulnerabilities are from several domains: individual, cultural, linguistic, physiological and psychological.  The cues and vulnerabilities of OSD have variations compared to face-to-face communication.  For serious OSD attacks which mainly belong to cybercrimes, such as human targeted attacks (e.g., human trafficking, cyberbullying, cyberstalking, or cybergrooming), if OSD cues are effectively captured, there is a much higher chance to prevent and detect online social deception than offline social deception due to much less real-time interactions which trigger much less risky situations from the safety perspective.

\item {\bf Ethical design considerations of social honeypots}: A social honeypot is one of broadly studied OSD prevention/detection mechanism.  They are deployed to passively collect attackers account profiles.  However, since social honeypots deal with human users, there should be careful legal or ethical considerations in their design features.

\item {\bf OSD detection mechanisms}: Three dominant OSD detection approaches surveyed in this work are user-profile-based, message content-based, and network structure-based.  They each have pros and cons in different scenarios.  In particular, if a detection mechanism uses only network structure features to detect OSD attacks, it would better preserve user privacy but need to develop lightweight algorithms to efficiently calculate expensive network features, such as centrality values requiring knowledge of the entire network topology and high computation cost to estimate centrality values. To maximize the  synergy of all three approaches, hybrid approaches incorporating all are promising.
\item {\bf Metrics for performance evaluation}: As the majority of OSD defense mechanisms are explored to effectively detect OSD attacks, most works have used accuracy metrics to measure the performance of their proposed work.  A few of the metrics are based on correlations and ranks, which are mainly used to identify key signals to detect OSD attacks.
\end{itemize}
We also found the following {\bf limitations} of the existing OSD detection approaches:
\begin{itemize}
\item {\bf Lack of systematic, comprehensive defense strategies to combat OSD attacks}: Fighting against OSD attacks requires systematic, comprehensive, and active defense strategies covering prevention, detection, and mitigation/response. However, existing approaches have been heavily explored in detection strategies only.  In addition, some approaches are embracing multiple roles with a single mechanism.  For example, most current OSD mitigation approaches are based on the results from early detection.  Further, since a social honeypot collects attacker profiles, the analysis of social honeypots is used to design classifiers for both prevention and detection. 

\item {\bf Lack of experiments with real-time, dynamic datasets}: Current prevention and detection methods are based on simulation and/or real datasets, but only a few discussed effective training and detection using streaming data, such as Twitter API.  In addition, the high computational and time complexity for real-time detection remains an open issue.  

\item {\bf Insufficient proactive defense}: The inherent role of a social honeypot is proactively finding targeted attackers (i.e., a particular type of attackers).
This way allows a system to identify targeted OSD attackers and proactively take actions to prevent potential victims by the targeted OSD attackers, which may lead to cybercrimes. Although honeypots are used in communication networks as a proactive intrusion prevention mechanism, social honeypots are passively used in OSNs due to potential legal and ethical issues.  Without clarifying the legal/ethical design issues, the function and exploitation of social honeypots cannot be improved further to deal with highly intelligent attackers.  In particular, to deal with real human-based OSD attacks, such as crowdturfing by paid workers to conduct social deception activities, more active social honeypot designs should be allowed while preserving normal user privacy and ethical rights.

\item {\bf High complexity of features and models}: We substantially surveyed the features for data-driven detection methods in Sections~\ref{subsec:user-profile-detection} and~\ref{subsec:message-content-detection} and network/epidemic models for network structure feature-based methods in Section~\ref{subsec:network-feature-detection}. The complexity of extracting/evaluating features and the model optimization grows fast with the size of datasets. How to reduce the solution complexity and improve solution efficiency for OSD detection is still an open issue.
\item {\bf Lack of qualitative analysis for cues of OSD attacks}: Most OSD defense mechanisms have focused on dealing with attacks by machines (or bots).  However, for more serious OSD attacks (i.e., human targeted attacks), appropriate cues should be first carefully identified through qualitative analysis based on multidisciplinary research efforts with behavioral scientists.
\end{itemize}

\vspace{-2mm}
\section{Concluding Remarks}
\label{sec:conclusion}

From the comprehensive survey conducted in this work, we obtained the {\bf key findings to answer the research questions} raised in Section~\ref{subsec:goal-questions} as follows:
\begin{itemize}
\item {\bf Answer to RQ1}: The fundamental meanings and intent of social deception are commonly present in both offline and online social deception as we find surprisingly common trends/characteristics observed in socially deceptive behaviors.  The common goal is `misleading a potential deceivee for the benefit of a deceiver' by increasing the deceivee's misbelief or confusion.  In both online and offline platforms, social deception is successful only when the deceivee cooperates with actions taken by the deceiver.  Due to the unique characteristics of an online environment such as less real-time/face-to-face interactions without physical presence to each other, both the deceivee and deceivers can take advantages of them in terms of defense (i.e., prevention, detection, and response/mitigation) and attack (e.g., anonymous attacks or easily running away if something goes wrong).  

\item {\bf Answer to RQ2}:  More serious human targeted attacks (e.g., human trafficking, cyberstalking, cybergrooming, or cyberbullying) have emerged as new OSD attack types.  The seriousness has grown as online deception often leads to offline crimes, which become indeed the major concern of cybercrimes.  While human targeted attacks become a more serious social issue, there is a lack of cyber laws to respond to this serious social deception attack, easily leading to cybercrimes.

\item {\bf Answer to RQ3}: Many cues and susceptability traits of offline social deception behaviors are present in online social deception behaviors.  The examples include intentionality of social deception, its cues from linguistic, cultural, and/or technological contexts, and various susceptibility factors including demographics, cultural, and/or network structure feature-based traits. Moreover, due to the limited real-time and/or interactions feeling people's presence in online platforms, some cues such as physiological and/or psychological cues may be missed while they can be highly useful cues for detecting social deception.  However, as more advanced features of online platform-based interactions emerge, more physiological/psychological cues can be captured to improve deception detection (e.g., heart beats can be fed back to a detection mechanism).

\item {\bf Answer to RQ4}: Most defense mechanisms to combat OSD attacks only focused on detection, particularly in terms of data-driven approaches using machine/deep learning techniques. Prevention mechanisms are substantially limited and have often been considered along with detection mechanisms (e.g., social honeypots or data-driven approaches).  Response mechanisms after the detection of the OSD are even much less explored than prevention mechanisms.  
\item {\bf Answer to RQ5}: Popular datasets used in existing OSD research are from Twitter, Sina Weibo, and Facebook along with other synthetic datasets collected from simulation, as shown in Figs.~\ref{fig:dataset-attack-type} and~\ref{fig:dataset-source}.  In particular, to study human targeted attacks, there is a lack of datasets available because online human targeted deception data are based on individual chats or dyadic interactions.  In addition, most metrics are to measure detection accuracy of OSD attacks, which is natural to observe as most defense mechanisms mainly focus on detection.  Hence, there is a lack of efficiency metrics that can capture cost / complexity of the proposed defense techniques against OSD attacks.

\item {\bf Answer to RQ6}: The OSD research is inherently involved with human users and may introduce ethical issues. However, to conduct meaningful experiments, some real testbed-based validation/verification should be conducted to obtain high confidence in the developed technologies under realistic settings.  However, when deploying defense techniques in a real testbed (e.g., Facebook, Twitter, etc), the defense process may encounter inevitable deception towards normal, legitimate users. In addition, privacy is a big concern in cybersecurity and there is an inherent tradeoff between preserving users privacy and improving the quality of defense tools against OSD attacks.  To investigate serious OSD attacks, such as human targeted attacks, most interactions are peer-to-peer, such as dyadic conversations/chats, which is mostly unavailable. As a result, there is a lack of real datasets in studying highly serious human targeted attacks, such as human trafficking, cyberstalking, or cybergrooming attacks. Also, there is a lack of systematic legal and/or ethical logistics on how to proceed the OSD research with involvement of human users in real testbed settings.
\end{itemize}

We suggest the following {\bf future research directions} in the online social deception and its countermeasure research:
\begin{itemize}
\item {\bf Development of defense applications against online social deception considering multidimensional concepts of social deception}: Although various concepts, properties, and cues of social deception have been studied in diverse disciplines, the multidisciplinary nature of social deception has not been appropriately considered in developing defense mechanisms against online social deception (OSD) attacks.  In particular, deceivers and deceivees are both humans via online platforms. Without understanding the way deceivers and deceivees communicate and/or interact to each other, it is hard to detect deception easily.  Deception can be easily deployed on top of firm, trust relationships.  In order to distinguish deception from truthfulness, in-depth understanding of deception based on multidisciplinary research effort is a must for developing effective defense mechanisms against OSD attacks.

\item {\bf Distinction of benign deception from malicious deception}: In the cybersecurity domain, deception refers to a deceptive action with malicious intent. 
However, in a social network, many users may use OSD to promote self-presentation/protection for privacy protection. Therefore, if OSD is treated as a form of attacks, it can possibly result in a high false positive rate (i.e., detecting benign users as malicious users).  In order to prevent this, deception-specific online defense tools that can differentiate benign deception from malicious deception should be developed.

\item {\bf Culture-aware defense against OSD attacks}: Based on our survey, different cultural deception cues have been observed~\cite{Bond90, Heine11-deception-psychology, Lewis08, Sedikides95themultiply}.  Since deception cues are sensitive to cultural characteristics, culture-aware defense mechanisms should be developed to effectively deal with OSD attacks that consider unique cultural characteristics of a social network.

\item {\bf Detectability-aware and intent-aware defense against OSD attacks}: As discussed in Fig.~A2 of the appendix document, the spectrum of deception can span a wide range based on the extent of detectability and intent. Intelligent OSD attackers may establish trust relationships with potential victims and exploit the established trust to deceive the victims. This is especially observed in human targeted attacks such as human trafficking or cybergrooming, which is categorized as a serious cybercrime~\cite{Zambrano19-cybergrooming}.  Hence, developing detectability-aware and intent-aware cues against highly subtle hard-to-detect OSD attacks is a future research direction.

\item {\bf Security protection of adolescent online users in multiple roles}: Adolescents have high vulnerability to OSD attacks, as discussed in Section F of the appendix document.  Deceptions such as cyberbullying have exposed severe social, behavioral and security issues introduced due to collaboration in multiple roles by adolescents.  Educational and habitual guidelines, parental control, and/or security guard tools cannot protect potential deceivees.  Social media platforms need to enhance their effective OSD prevention mechanisms especially for young users.

\item {\bf Dynamic, updated defense mechanisms to obfuscate highly advanced attackers}: Recent studies showed that OSD attackers can build advanced social bots by analyzing the current detection models and fooling the existing models by leveraging adversarial machine learning (AML) techniques~\cite{kurakin2016adversarial}.  One countermeasure is to collect new datasets and retrain the classifiers. However, it is challenging to support updating the models with additional datasets.  The cost of repeatedly training the classifiers with the whole dataset is particularly high.  Another method is to identify unknown deception features based on linguistic, behavioral, and technological cues.

\item {\bf Defense against human attackers vs. social bots}: Human attackers are another type of advanced attackers where a real human is behind the social network platforms performing OSD attacks. They can bypass detection because the conversation is from real humans or the accounts are mimicking normal users.  There also exist crowdturfing workers who spread deceptive information in social media and get paid.  More research work is needed to investigate how to detect and differentiate social bots from human attackers.

\item {\bf Measurement of physiological and/or psychological cues to develop better prevention techniques against OSD attacks}:  Due to the unique characteristics of online platforms, some critical deception cues are missing and must be identified first, such as physiological and/or psychological cues. Measuring those cues can be critical in terms of improving prevention and early detection against OSD attacks.

\item {\bf More efforts are needed to explore prevention and response mechanisms to defend against OSD attacks}: In terms of the techniques used across all defense mechanisms, while machine/deep learning approaches are popularly used, game theoretic and/or network structure feature based approaches are still to be further explored to produce more mature approaches.  They have extra merits over data-driven approaches in that the game theoretic approach can predict an attacker's next move.  For prevention, although early detection as an OSD prevention strategy is receiving a high attention with growing amounts of recent works to fight against OSD attacks, there should be more prevention mechanisms that can provide more proactive defense such as identifying potential attacks even before the attacks occur. Response/mitigation after OSD detection, such as mitigation after false information spread or recovery after OSD attacks are launched, is little explored in the literature and calls for more efforts to further investigate effective mechanisms to minimize risk and aftermath effect after OSD detection.

\item {\bf Effective deception cues are needed to combat OSD attacks without violating user privacy}: Due to a lack of effective deception cues/datasets, it is difficult to conduct OSD research to defend against serious human targeted OSD attacks for validation and verification. A future direction is to develop techniques to capture clear deception cues without violating user privacy.

\item {\bf More efficiency metrics are needed to expedite the defense process}: Efficiency metrics for measuring algorithmic complexity of defense techniques have not been sufficiently used in existing approaches. More meaningful complexity/efficiency metrics should be considered in order to expedite the speed of prevention, detection, and recovery as a defense against OSD. 

\item {\bf Systematic legal and/or ethical guidelines are needed for conducting meaningful OSD research}: Since humans are the key factors in solving the problems associated with the OSD attacks, the research community and government need to provide clear guidelines on conducting OSD research without violating user privacy.  In communication networks, the research community appears to have reached some accord about using defensive deception techniques to defend against cyberattacks by emphasizing its benefits.  However, for cybersecurity research on OSN platforms likely involving human subjects, there is little research, let along a consensus, on what methodologies are allowed and what level of user privacy must be preserved before achieving the goal of defense effectiveness. 
\end{itemize}

\bibliographystyle{IEEETranSN}
\bibliography{sd}

\end{document}


\title{Appendices: Online Social Deception and Its Countermeasures for Trustworthy Cyberspace: A Survey}



\author{Zhen Guo}
\author{Jin-Hee Cho}
\author{Ing-Ray Chen}
\affiliation{%
  \institution{Computer Science, Virginia Tech}
\streetaddress{7054 Haycock Road}
  \city{Falls Church}
  \state{VA}
  \postcode{22043}
  \country{USA}
  }
\author{Srijan Sengupta}
\affiliation{%
  \institution{Statistics, Virginia Tech}
  \streetaddress{250 Drillfield Drive}
  \city{Blacksburg}
  \state{VA}
  \postcode{24061}
  \country{USA}
  }
\author{Michin Hong}
\affiliation{%
  \institution{School of Social Work, Indiana University}
  \streetaddress{902 West New York Street}
  \city{Indianapolis}
  \state{IN}
  \postcode{46202}
  \country{USA}
  }
\author{Tanushree Mitra}
\affiliation{%
  \institution{Computer Science, Virginia Tech}
\streetaddress{2202 Kraft Drive}
  \city{Blacksburg}
  \state{VA}
  \postcode{24060}
  \country{USA}
  }
\maketitle

\appendix


\begin{wrapfigure}{r}{0.45\textwidth}
    \centering
    \includegraphics[scale=0.7]{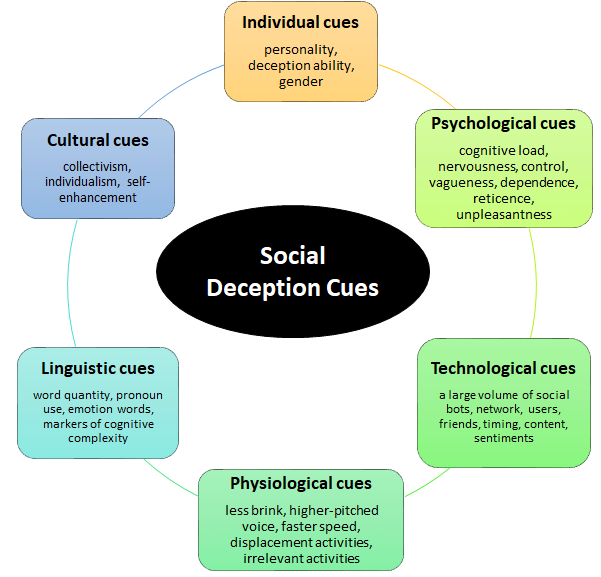}
    \caption{Multidimensional cues of social deception.}
    \label{fig:deception-cues}
\end{wrapfigure}


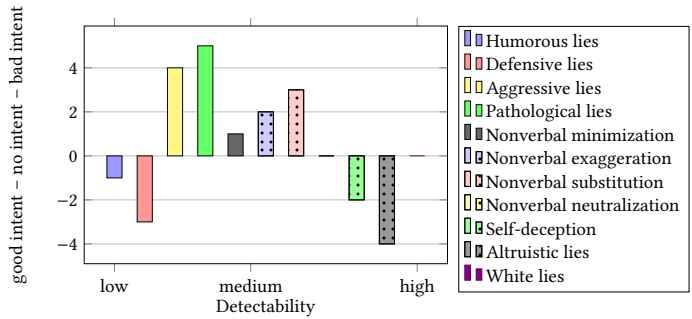
\begin{wrapfigure}{r}{0.55\textwidth}
    \centering
\begin{tikzpicture}[font=\normalsize, scale = 0.65]
\begin{axis}[
    ylabel={good intent -- no intent -- bad intent},
    ytick={-6,-4,-2,0,2,4,6},
    ymajorgrids,
    xtick={1,10,21},
    xticklabels={low, medium, high},
    ybar=-\pgfplotbarwidth,
    x=\pgfplotbarwidth,
    bar width=8pt,
    xlabel={Detectability},
    legend style={
        cells={anchor=west},
        legend pos=outer north east
        },
    ]
\addplot[fill={blue!40!}] coordinates {(1,-1)};
\addplot[fill={red!40!}] coordinates{(3,-3)};
\addplot[fill={yellow!60!}] coordinates{(5,4)};
\addplot[fill={green!60!}] coordinates{(7,5)};
\addplot[fill={black!60!}] coordinates{(9,1)};
\addplot[fill={blue!20!}, postaction={pattern=dots}] coordinates{(11,2)};
\addplot[fill={red!20!}, postaction={pattern=dots}] coordinates{(13,3)};
\addplot[fill={yellow!40!}, postaction={pattern=dots}] coordinates{(15,0)};
\addplot[fill={green!40!}, postaction={pattern=dots}] coordinates{(17,-2)};
\addplot[fill={black!40!}, postaction={pattern=dots}] coordinates{(19,-4)};
\addplot coordinates{(21,0)};
\legend{Humorous lies,Defensive lies,Aggressive lies,Pathological lies,Nonverbal minimization, Nonverbal exaggeration,Nonverbal substitution,Nonverbal neutralization,Self-deception,Altruistic lies,White lies}
\end{axis}
\end{tikzpicture}
\caption{The spectrum of deception based on the extent of detectability of deception (x-axis); and the extent of good/bad intent of deception and no intent (y-axis).}
\label{fig:detectability}
\end{wrapfigure}


\begin{landscape}
\begin{table}[t]
\small
\centering
\caption{Multidisciplinary Concepts of Deception.}
\label{tab:deception-concept}
\small
\begin{tabular}{|P{2cm}|p{9cm}|p{4cm}|p{4cm}|}
\hline
{\bf Discipline} & \multicolumn{1}{c|}{\bf Meaning of deception} & \multicolumn{1}{c|}{\bf Goal} & \multicolumn{1}{c|}{\bf Source} \\
\hline
{\bf Philosophy} & To deceive someone or something to intentionally and cause to have a false belief & To mislead an entity to form a false belief & \cite{Carson10, Gert15, Mahon16, Russow86} \\
\hline
{\bf Behavior Science} & Functional deception applied for an individual's behavior (i.e., a signal) to mislead the actions of others, 
leading to the misrepresentation of belief states & To mislead the actions of others; To misrepresent belief states &\cite{Hauser97, Langleben02, Spence01} \\
\hline
{\bf Psychology} & A behavior providing information to mislead subjects to some direction or explicit misrepresentation of a fact aiming to mislead  subjects & To mislead subjects to some direction, introducing a false belief toward a target or victim &\cite{Adair85, Hyman89, Nicks97} \\
\hline
{\bf Sociology} & A distinct human ability due to intentionality, use of language, and self-awareness &  To serve as relational or marketing strategy &\cite{Langer72, Meltzer03, Mitchell86} \\
\hline
{\bf Public Relations} & Internally, self-deception is a solution for an individual to eliminate the conflict from cognitive dissonance; Externally, self-deception is a way to avoid disastrous impact on an organization by attributing a problem (or guilty) to an individual or victim & To resolve internal or external crisis or conflicts & \cite{Clementson19, Hiebert03, Lee19-public-relations, Ortmann10, Seiffert-Brockmann17, Stone01} 
\\
\hline
{\bf Communica-tions / Linguistics} & Interpersonal deception theory (IDT) views deception as an interactive process between senders and receivers, exchanging non-verbal and verbal behaviors and interpreting their communicative meanings &  To manage deceivers' verbal communications to successfully deceive receivers & \cite{Buller96-testing, Buller96, Feng12, Hancock08, Zhou04-linguistics} \\
\hline
{\bf Command / Control} & Any planned maneuvers undertaken for revealing false information and hiding the truth to an enemy with the purpose of misleading the enemy and enticing the enemy to undertake the wrong operations & To confuse or mislead foreign country opponents &\cite{daniel2013strategic, Mitchell86, rothstein2013art, Whaley82} \\
\hline
{\bf Computing / Engineering} & Deception attacks have popularly performed by attackers as various forms such as phishing, social engineering attacks, fraud advertisements, stealthy attack, and so forth. Defensive deception is a planned action taken to mislead and/or confuse attackers and to thereby cause them to take (or not take) specific actions that aid computer-security defenses. & To lead a deceivee to respond suboptimally &\cite{almeshekah2016cyber, Darwish12, Enos06, Halevi13, Halevi15, Han18-deception-survey, Heartfield15, Lin19, Modic11, Oliveira17-chi-susceptibility, Parrish09, Pattinson12, Rathore17-is-survey, Rowe16, Sheng10, Yuill06} 
\\
\hline
{\bf Common Concept} & A successful attempt to make others believe something that is false, either by words or nonverbal behaviors, and either intentionally or unintentionally & To mislead others; To cause to believe what is false &\cite{OED89, Rowe16} \\
\hline
\end{tabular}
\end{table}
\end{landscape}

\begin{landscape}
\begin{table}[t]
    \small
    \centering
    \caption{Type, Description, and Intent of Online Social Deception Attacks}
    \label{tab:osd-type}
    \begin{tabular}{|P{2cm}|p{9cm}|p{6cm}|p{2cm}|}
    \hline
    \multicolumn{1}{|c|}{\bf OSD Type} &  \multicolumn{1}{c|}{\bf Description} & \multicolumn{1}{c|}{\bf Intent \& Potential Damage} & \multicolumn{1}{c|}{\bf Source} \\
    \hline
    \hline
    Fake news & News contradicts, fabricates or conflates the ground truth and spreads in OSN. & Credibility loss, economical and political misleading, controlling public opinions & \cite{jiang2018linguistic, shao2016hoaxy, vosoughi2018spread}\\
    \hline
    Rumors & An unverified assertion that starts from one or more sources and spreads over time from node to node in a network. & Misleading people's decision, panic in public, govenment credibility loss  &\cite{vosoughi2017rumor}\\
    \hline
    Information manipulation & False information deliberately and often covertly spread in order to influence public opinion or obscure the truth. & Advertising, campaigns &\cite{dhs18}\\
    \hline
    Fake reviews & Malicious users write fake reviews, opinions, or comments in social media to mislead other users. & Influencing user's option or decision, advertising, reputation loss &\cite{yao2017automated}\\
    \hline
    Phishing & Attackers trick users into revealing sensitive information related to work, financial credentials, or even personal  data to be used in fraudulent activities & Confidential personal data leakage, launch advertising campaigns, pornography &\cite{abutair2019cbr, ding2019keyword, Vergelis19} \\ 
    \hline
    Spamming & Attackers send unsolicited  messages (spam) in bulk to OSN users & Reputation loss, malicious advertising & \cite{Rathore17-is-survey}\\ 
    \hline
    Fake Profile  & Attackers create a huge  amount of fake identities for their own  benefits. & Personal information leakage, stealing money &\cite{haddadi2010add}\\ 
    \hline
    Compromised account & Attackers hacked legitimate user accounts that are created and used by their fair owners and later used for ill purposes. & Reputation loss, account loss, personal privacy leakage &\cite{egele2013compa, kayes2017privacy}\\ 
    \hline
    Profile cloning attack & Attacker clones a  pre-existing user profile either in the same  OSN or a different one. & Reputation loss, sensitive information leakage, account loss &\cite{Rathore17-is-survey}\\ 
    \hline
    Crowdturfing & Attackers are gathered by crowdsoursing system and speak fake and inaccurate information to mislead people & Speading malicious URLs, forming astroturf campaigns, manipulating opinions &\cite{lee2013crowdturfers, lee2014social, wang2012serf, wang2014man, Wu2017-crowdturfing}\\ 
    \hline
    Human trafficking & Traffickers use computers and networks to transport a great number of victims and advertise service across geographic boundaries for labor trade or sex trade & Sexual exploitation, modern slavery, forced labor or services, removal of organs &\cite{Feingold05, Greiman13, latonero2011human}\\
    \hline
    Cyberbullying & Cyberbullying is the deliberate and repetitive online harassing or harming of someone. & Reputation loss, cyber harassment, teen depression &\cite{Rathore17-is-survey}\\ 
    \hline
    Cyber-grooming & Cyber-grooming is when an adult tries to establish an online, emotional connection with a child in order to sexually abuse them. & Reputation loss, cyber harassment &\cite{Rathore17-is-survey}\\
    \hline
    Cyberstalking & Attackers exploit their personal information, such as their phone number, home address, location, and schedule, in their SNS user's profile & Reputation loss, personal data leakage, cyber harassment, safety loss &\cite{Rathore17-is-survey}\\
    \hline
    \end{tabular}
\end{table}
\end{landscape}
\begin{table}[t]
\footnotesize
\caption{Goal, intent, and security breach according to a different type of social deception.}
\label{tab:goal-intent-security-breach}
\centering
\begin{tabular}{|P{3.8cm}|P{5cm}|P{3.8cm}|}
\hline
\bfseries Goal of social deception &\bfseries Malicious vs. Non-malicious intent &\bfseries Breach of security goals \\
\hline
Parasitism & Malicious & Loss of integrity  \\
\hline
Cooperative deception & Malicious & Loss of integrity \\
\hline
Privacy protection & Non-malicious & Loss of confidentiality \\
\hline
Self-presentation
& Non-malicious & Loss of integrity \\
\hline
Self-deception & Non-malicious & Loss of integrity\\
\hline
\end{tabular}
\end{table}

\begin{landscape}
\begin{table}[!t]
\scriptsize 
\renewcommand{\arraystretch}{1.3}
\centering
\caption{Online Social Deception Prevention Mechanisms} 
\label{tab:prevention}
\begin{tabular}{|P{2cm}|p{5cm}|p{6cm}|p{4cm}|P{1cm}|}  \hline
 \bf Type &\multicolumn{1}{c|}{\bf Method} &\multicolumn{1}{c|}{\bf Features}  &  \multicolumn{1}{c|}{\bf Datasets} &\bf Ref. \\
 \hline\hline
 Fake news & Prototype of conventional Blockchain system & State tuple of unique transaction identifier, news payload, timestamp of news generation, hash of news payload, and user identifier & Synthetic data &\cite{saad2019fighting}
 \\
  \hline
 \multirow{2}{*}{Phishing} & A server-client system to report password reuse and update whitelist & Password, user ID, domain in the protected list & Phishing attacks data from third party vendor in three weeks &  \cite{florencio2007evaluating}
 \\
  \cline{2-5}
  & Blacklisting, heuristics and moderation based phishing prevention platform & Number of moderators, availability of moderators, unanimity of decision, network resources & Simulated data &\cite{gupta2011socio} 
  \\
  \hline
 Fake profile & Proactive sub-community behavioral profiles: support vector machine (SVM), random forest (RF), adaptive boosting & Closeness, betweenness, eigenvector centrality & Dataset \textit{giftcardexchange} from Reddit banned users &\cite{tsikerdekis2016identity} 
 \\
  \cline{2-5}
 & Weka toolkit and Decorate algorithm &Tweet similarity (TS) and 9 other content-based features; honeypot features: ratio of malicious accounts interacted (MAR), average daily new follower number of a social honeypot (DFN), social honeypots an account interacts with (AIN), and social honeypot follow back an account (AFB) & Seven types of Twitter accounts on the blue bird network: the Social Star, the Butterfly, the Distant Star, the Private Eye, the Cycler, the Listener and the Egghead & \cite{nisrine2016security}
 \\  
 \cline{2-5}
  & LogitBoost, RF, XGBoost; evaluate robustness by individual attack model and coordinated attack model & 4 sets: 3 profile features, 8 posting activity features, 2 page liking features, 3 social attention features. Temporal features: change rate of \# of liked pages and of category entropies in 30 days & Fake liker in Fiverr and Microworkers and legitimate liker in Facebook conference group & \cite{badri2016uncovering}
  \\
 \hline

 \multirow{4}{*}{Spammer} & Decorate, logistic regression (LR) and LibSVM & User profile features (longevity of account, average tweets per day, ratio of following/followers, percentage of bidirectional friends) and tweets features (number of URLs and @username in 20 most recent tweets) & MySpace and Twitter social honeypot deployment & \cite{lee2010uncovering}
 \\
  \cline{2-5} 
  & RF, standard boosting and bagging, feature grouping &Link payloads, user behavior over time, and followers/following network dynamics; 2 User demographics (longevity of account), 5 user friendship networks, 8 user content (average content similarity), user history (change rate of number of following) & 60 Twitter honeypot accounts, Twitter dataset of content polluters and legitimate users from \url{http://infolab.tamu.edu/data} &\cite{lee2011seven}
  \\
  \cline{2-5} 
 
  & Random forest & Following/followers ratio, URL ratio, message similarity, friend choice, number of messages sent, friend number & 900 MySpace, Facebook and Twitter honey-profiles &\cite{stringhini2010detecting} 
  \\
   \cline{2-5}
    &Random forest & Tweet behavior (tweet frequency, tweet keyword, tweet topics), follow behavior, and application usage &Tweeter accounts &\cite{yang2014taste} 
    \\
  \hline

\end{tabular}
\label{tribes}
\end{table}

\begin{table}[!t]
\renewcommand{\arraystretch}{1.3}
\scriptsize 
\centering
\caption*{Table~\ref{tab:prevention} (continued). Online Social Deception Prevention Mechanisms} 
\label{tab:prevention-2}
\begin{tabular}{|P{2cm}|p{5cm}|p{6cm}|p{4cm}|P{1cm}|} 
 \hline
 \bf Type &\multicolumn{1}{c|}{\bf Method} &\multicolumn{1}{c|}{\bf Features}  &  \multicolumn{1}{c|}{\bf Datasets} &\bf Ref. \\
 \hline\hline
 
  \multirow{2}{*}{Socialbot} & Account monitoring simulation, analysis of variance (ANOVA, $p=0.05$) & Attack strategy (no knowledge attacker, partial knowledge attacker, full knowledge attacker); defense strategy: random, most connected, eigenvector, PageRank, and cost eigenvector 1 and 2 & Stanford Large Network Data Collection, 50 communities in each of Friendster, LiveJournal, and Orkut &\cite{Paradise14,Paradise15} 
  \\
 \cline{2-5}
 & Statistical method & Profile: acceptance rate of friend requests sent, incoming friend request, insider's incoming friend requests, Discounted cumulative gain. Email: total received, spam and suspicious, from Xing, from LinkedIn, DCG score   & 7 Social Honeypot profiles in an European organization; messages in Xing, LindkedIn &\cite{Paradise17}
 \\
 \cline{2-5}
 & Equilibrium simulation & Honeybot deployment (HD); Honeybot exploitation (HE) and Protection and Alert System (PAS) & Generated network & \cite{zhu2012sodexo, Zhu14-sh} 
 \\
 \hline 
 Cyber-bullying & Dashboard reflective user interface: notifications, action delay, and interactive education & TF-IDF, Ortony lexicon for negative affect, list of profane words, part-of-speech tags, label-specific unigrams and bigrams & Two datasets from YouTube comments and Formspring with expert annotation &  \cite{dinakar2012common}
 \\  
  \hline 
 \end{tabular}
\label{tribes}
\end{table}

\begin{table}[!th]
\renewcommand{\arraystretch}{1.3}
\centering
\scriptsize 
\caption{Data-Driven Deception Detection Mechanisms}
\label{tab:data-detection-I}
\begin{tabular}{|P{2cm}|p{5cm}|p{6cm}|p{4cm}|P{1cm}|}  \hline
 {\bf Type} &\multicolumn{1}{c|}{\bf Method} &\multicolumn{1}{c|}{\bf Features}  &  \multicolumn{1}{c|}{\bf Datasets} &  {\bf Ref.} \\
 \hline
 \hline 
 Spam URL & Random forest & Behavioral factor of URL posting (posting count, posting standard deviation, posting intensity, posting user network) and click (rises+falls, spikes+troughs, peak difference, total clicks, average clicks, clicking days, max clicks, effective average clicks, click standard deviation, mean median ratio) & List labeled dataset, URL-category website URLBlacklist (\url{http://urlblacklist.com}) and manually labeled dataset &\cite{cao2015detecting}\\
  \hline
 Spambot & Longest common substring (LCS), supervised and unsupervised classifier &Twitter account behavior (type (tweet, reply, retweet) and content (entities in tweets)) as DNA string of characters, and LCS curve, LCS value & Two Twitter dataset & \cite{cresci2017social}  \\
\hline 
 Spam & SVM, adaboost, and random forests & Topic distribution (LDA) for each user, two new topic based features: Local Outlier Standard Score (LOSS) and Global Outlier Standard Score (GOSS) & Soical honeypot Twitter dataset from \cite{lee2010uncovering}, synthesized Weibo dataset &\cite{liu2016detecting}\\
 \cline{2-5}
 & Labeled latent Dirichlet allocation (L-LDA) model and SVM & Word-based (\textit{TF-IDF} scheme), topic-based (group of words that have a high probability of co-occurrence, normalized topic frequencies), and user-based features (average time interval of posting (ATI), and the average similarity (AS) of two adjacent comments)& YouTube social spam dataset &\cite{song2014discriminative}\\
\cline{2-5}  
 & Random forest, C4.5 decision tree, Bayes network, naive Bayes, k-nearest neighbor, and support vector machine &12 lightweight features: user-based (account\_age, no\_of
followers, no\_of followings, no\_userfavourites, no\_lists, and
no\_tweets) and tweet-based (no\_retweets, no\_hashtags,
no\_usermentions, no\_urls, no\_chars, and no\_digits) and feature discretization & Four datasets from Twitter's streaming API with spam to non-spam ratios and continuous sampling method, ground truth from commercial tool &  \cite{chen2015performance}\\
 \cline{2-5}  
 & Na\"{i}ve Bayes, logistic regression, RF and semi-supervised spam detection & Hashtag, content, user and domain features & HSpam14 dataset of 15 days of tweets &\cite{sedhai2017semi} \\
 \cline{2-5}  
 & Semi-supervised clue fusion algorithm, boosting-based fusion & Content, behavior (variance of posting times during a period, night activity, regularity of posting), relationship (ratio of follower to followee, average number of neighbors' messages/ followers), and interaction features (average number of comments per message, average number of repost, mentions fraction) & Data crawler from Weibo API &\cite{chen2018semi} \\
 \hline
 Phishing & CNN-LSTM algorithm, XGBoost &URL deep features, URL statistical features, webpage code features, webpage text features & Two historical data were crawled from PhishTank website &\cite{yang2019phishing} \\
 \cline{2-5}
 & Search-engine based, heuristic-based and logistic regression & Phishing vocabulary similarities, 37 ULR lexical features (information entropy, confused string, length) & Data source from PhishiTank, Yahoo, URLB and DMOZ &\cite{ding2019keyword} \\
 \hline 
 Fake comments & Markov chain model on topic-crowd opinion pairs & Second-order Markov chain probabilities; five topics (Information, Science, Entertainment, Humour, and Adult) and three Crowd opinions (Positive, Negative, and Neutral) & User comments from Reddit; test dataset with 300 automated texts &\cite{Everett16} \\
\cline{2-5}  
 & ComLex: word embedding and unsupervised spectral clustering. linear regression, SVM and nearest neighbors (NN) & ComLex linguistic signals from user comments, keep emojis, special tokens \textit{snopesref} or \textit{politifactref}, 100-dimention vector; EmoLex; LIWC & 5303 social media posts from Politifact and Snopes with 2614374 user comments from Facebook, Twitter, and YouTube &\cite{jiang2018linguistic} \\
  \hline 

\end{tabular}
\label{tribes}
\end{table}

\begin{table}[!t]
\renewcommand{\arraystretch}{1.3}
\centering
\scriptsize 
\caption*{Table~\ref{tab:data-detection-I} (continued). Data-Driven Deception Detection Mechanisms}
\label{tab:data-detection2}
\begin{tabular}{|P{2cm}|p{5cm}|p{6cm}|p{4cm}|P{1cm}|}  \hline
 {\bf Type} & \multicolumn{1}{c|}{\bf Method} &\multicolumn{1}{c|}{\bf Features}  & \multicolumn{1}{c|}{\bf Datasets}  &  \multicolumn{1}{c|}{\bf Ref.}  \\
 \hline
  \hline 
 Rumor & Logistic regression, SVM, naive Bayes, and decision tree & Five new features from rumor publisher's behavior (verified user or not, number of followers, average number of followees per day, average number of posts per day, and number of possible microblog sources) with existing behavior-based features as follower's comments and reposting & Published rumor data from Sina Weibo & \cite{liang2015rumor} \\
 \cline{2-5}
  & Temporal model: dynamic time warping (DTW), and hiddern Markove models (HMM); non-temporal model: SVM and  logistic regression & Time-series features: 4 linguistics (ratio of tweets containing negation, average formality \& sophistication of tweets, opinion \& insight (LIWC), inferring \& tentative tweets), 6 user involved (controversiality, originality, credibility, influence, role, and engagement), and 7 temporal propagation dynamics features, feature contribution is ranked by Chi-squre test & 209 manually annotated rumors from \url{Snopes.com} and \url{FactCheck.com} and Twitter historical API & \cite{vosoughi2017rumor} \\
 \cline{2-5}
 & One-class support vector machine, support vector data description, $k$-nearest neighbors, principle component analysis (PCA), $k$-means, autoencoder & Content features (13 syntactic and 13 semantic), 18 user profile features, and 8 meta-message features  & Two published Twitter datasets & \cite{fard2019computational} \\
 \hline 
  
 Fake news & Deep RNN model & Neural embedding, n-gram TF vector, and external features including polarity, lexiton based sentiment difference & FNC-1 challenge from Emergent dataset~\cite{ferreira2016emergent} &\cite{bhatt2018combining}\\
 \cline{2-5}
 & Na\"{i}ve Bayes Multinomial algorithm & Basic word usage pattern (word vector); seven themes (code book) between fake news and satire & Created: \url{https://github.com/jgolbeck/fakenews} &\cite{golbeck2018fake} \\
\hline
 Profile cloning & A binary classifier calculating the attributes and friend list similarities from different OSNs & Profile attributes similarities, friend list similarity, friend request information, friends lists & Synthetic dataset of 2000 people's profiles &  \cite{kamhoua2017preventing}\\
 \cline{2-5}
  & Three components: Information Distiller, Profile Hunter, and Profile Verifier & User-identifying terms, profile-records, profile similarity & LinkedIn automated profile creation &  \cite{kontaxis2011detecting}\\
 \cline{2-5}
  & CloneSpotter: A real-time context-free detection algorithm & Recently used IPs, friend list, profile, profile similarity & Synthesized accounts in Renren network &  \cite{shan2013enhancing}\\
  \hline   
 Fake account & Decorate emsemble classifier &Blacklist: 50 top LDA topic words, 500 fake word from TF-IDF. 14 content-based features: fake word ratio, mean time between tweets, extreme idle duration between tweets &1KS-10KN dataset and social honeypot dataset on Twitter& \cite{swe2018fake}\\
 \hline 
 Spam account & Gradient boosting, RF, extremly randomized trees (ExtraTrees), maximum-entropy (MaxEnt), multi-layer perceptron (MLP), SVM & Lightweight features: user profile features (user name, screen name, location and description), account features (account age and verification flag), pairwise engage-with features (user activities) and engaged-by features (indirect from users) & Honeypot dataset from \cite{lee2011seven} and manually annotated dataset &\cite{inuwa2018detection} \\
 \hline 
\end{tabular}
\end{table}

\begin{table}[!t]
\renewcommand{\arraystretch}{1.3}
\centering
\scriptsize 
\caption*{Table~\ref{tab:data-detection-I} (continued). Data-Driven Deception Detection Mechanisms}
\label{tab:data-detection2}
\begin{tabular}{|P{2cm}|p{5cm}|p{6cm}|p{4cm}|P{1cm}|}  \hline
 {\bf Type} & \multicolumn{1}{c|}{\bf Method} &\multicolumn{1}{c|}{\bf Features}  & \multicolumn{1}{c|}{\bf Datasets}  &  \multicolumn{1}{c|}{\bf Ref.}  \\
 \hline
  \hline 

 Malicious / compromised account & Weka & 21 account features from contents and account activity behaviors and Petri net based features & Twitter data & \cite{sahoo2019hybrid} \\
 \cline{2-5}
 & SVM & Semantic representation of clickstreams & Synthetic data banksim and paysim &\cite{weller2018compromised}\\
 \cline{2-5}
 
 & $k$-NN & $n$-gram built baseline to identify writing style and identify users; keep updating the baseline with new posts & Twitter dataset of 1000 users: \url{https://wiki.illinois.edu//wiki/display/forward/Dataset-UDI-TwitterCrawl-Aug2012#Dataset-UDI-TwitterCrawl-Aug2012-4.Creation}&  \cite{barbon2017authorship}\\
 \cline{2-5}
 & CADET: nonlinear autoencoders & Feature embeddings from tweets’ content, source, location, and timing & Twitter data of posted geotagged tweets&\cite{vandam2018cadet} \\
 \hline 
  Crowd-turfing & Random forest, decision tree, SVM, Bayesian probability models & 9 user profile fields (FFRatio, reciprocity, user tweets per day, account age, ratio of tweets with URLs and mentions), 8 user interactions (comments and retweets), 5 tweeting clients (devices), and temporal behavior (12 tweet burstiness and 1 entropy regularity)  & Two baseline datasets: authenticated dataset and active users from Sina Weibo accounts from three-year crowdturfing campaigns;  & \cite{wang2014man}\\
\cline{2-5}  
  & CrowdTarget to detect crowdturfing targets: Ada Boost, Gaussian naive Bayes, $k$-nearest neighbors & Retweet-based features from crowdturfing targets: retweet time distribution (mean, standard deviation, skewness, and kurtosis); ratio of the most dominant application; number of unreachable retweeters; ratio of number of received clicks to the number of retweets for tweets containing URLs & Twitter, crowdturfing sites, and five black-market sites, e.g. \verb+retweets.pro+ and \verb+socialshop.co+ & \cite{song2015crowdtarget}\\
  \cline{2-5}
  & Character-level RNN, SVM linguistic classifier & Temperature, a parameter used in softmax function; character-level probability distribution $P(X_{t+1}=x_{t+1}|x_{1,...,t})$; 1 similarity feature, 4 structural features, 6 syntactic features, 4 semantic features, 62 LIWC features & 2017 Yelp Challenge Dataset \url{https://www.yelp.com/dataset/challenge} and Attack dataset, replacing fake reviews with RNN generated reviews &  \cite{yao2017automated}\\
  \cline{2-5}
  & Random forest, Na\"{i}ve Bayes, logistic regression, SVM & Four groups 92 features: user demographics 5 features,  user friendship networks 4 features, user activity 12 features (behavioral), user content 3 features including 68 LIWC dictionary & Random sample 10,000 twitter users &  \cite{lee2013crowdturfers}\\   
  \hline 
 Cybercrime account & Content-sensitive Gibbs sampling (CSLDA) & Semantic labels (transactional or collaborative), Laplacian semantic ranking score & 2 cybercrime related corpora from Twitter and online forums &\cite{lau2014probabilistic} \\
 
 \hline 
 Cyber-bullying & JRip, J48, SVM from Weka tool & TF-IDF, Ortony lexicon for negative affect, list of profane words, part-of-speech tags, label-specific unigrams and bigrams & Two datasets from YouTube (comments) and Formspring (young people) with expert annotation &\cite{dinakar2012common}\\  
\hline   
\end{tabular}
\end{table}

\begin{table}[!t]
\renewcommand{\arraystretch}{1.3}
\centering
\scriptsize 
\caption{Network Structure-Driven Deception Detection Mechanisms} \label{tab:network-detection-I}
\begin{tabular}{|P{2cm}|p{5cm}|p{6cm}|p{4cm}|P{1cm}|}  \hline
 {\bf Type} & \multicolumn{1}{c|}{\bf Method} & \multicolumn{1}{c|}{\bf Features}  & \multicolumn{1}{c|}{\bf Datasets} & {\bf Ref.} \\
 \hline\hline
 Fake news & Opinion model with subjective logic, epidemic model & Opinion consisting of belief, disbelief and uncertainty, agent features (prior belief and centrality degree), agent types (disinformers, misinformers and true informers) & Facebook dataset with 1033 nodes and 26747 edges  &\cite{cho2019uncertainty} \\
  \cline{2-5}
  & Hierarchical propagation model  & Three-layer credibility network of event, sub-events and messages & Microblog datasets SW-2013 and SW-MH370 &\cite{jin2014news} \\
   \cline{2-5}
& Credibility propagation network & News verification by mining conflicting viewpoints  & Built upon Sina Weibo &\cite{jin2016news} \\
   \cline{2-5}
 & Independent cascading model and Bayesian inference & User's flagging activity, user's observed activity, utility of blocking a news & Social circles Facebook graph of 4039 users and 88234 edges & \cite{tschiatschek2018fake} \\
\cline{2-5}  
 & Cascades representation, LSTM-RNN sequence classifier & Embedding of users & Twitter API used for certain topics and the dataset consisting of 68892 news with 288591 posts and 121211 users &\cite{wu2018tracing} \\
  \hline  
 Fake reviews & FraudEagle algorithm: scoring by loopy belief propagation (LBP) and grouping by cross-association clustering & User nodes and product nodes with signed links and prior belief & New SoftWare Marketplace (SWM) dataset from an app store database &\cite{akoglu2013opinion} \\  
 \cline{2-5}  
& Random forest & Linguistic traits, activities, reply network structure  & Disqus communities datasets &\cite{kumar2017army} \\
 \cline{2-5}  

 & REV2, unsupervised and supervised RF prediction & Quality/trust scores from fairness, goodness and reliability for users, ratings, and products & Flipkart, Bitcoin OTC, Bitcoin Alpha, Epinions, and Amazon dataset &\cite{kumar2018rev2} \\
 \hline
 Rumor & SEIZ model & 4 topics and 5 geographic regions, state Exposed (E) users taking some time to post, state Skeptic (Z) users who heard a news but decided not to retweet, retweet cascades & 8 Twitter dataset for real news and rumors &\cite{jin2013epidemiological} \\
   \cline{2-5}
 & Correct information diffusion in SIR-extended diffusion model & SIR agents, condition to become R, diffusion of corrected information and new situations & Tweets collected from two weeks before and after Great East Japan Earthquake & \cite{okada2014sir} \\
   \cline{2-5}
 & Propagation structure via tree kernel  & Cascades similarity, SVM-time series, decision tree based ranking, random forest & Twitter15 and Twitter16 data & \cite{ma2017detect} \\
   \cline{2-5}
 & Cascades representation, LSTM-RNN & Bottom-up tree, top-down tree & Twitter15 and Twitter16 data &\cite{ma2018rumor} \\
  \cline{2-5}
  & Temporal model: Dynamic Time Warping (DTW), and Hiddern Markove Models (HMM); non-temporal model: SVM and logistic regression & 8 temporal propagation dynamics features (time-inferred diffusion, fraction of low-to-high diffusion, fraction of nodes in largest connected component, average depth to breadth ratio, ratio of new users, ratio of original tweets, fraction of tweets containing outside links, fraction of isolated nodes) & 209 manually annotated rumors from \url{Snopes.com} and \url{FactCheck.com} and Twitter historical API & \cite{vosoughi2017rumor} \\
 \cline{2-5}  
 & Cascades similarity, hybrid SVM with graph kernel and RBF & 23 features from propagation tree and 8 new features (topic type LDA, search engine, user type, ave sentiment, ave doubt, ave surprise, ave emotion, and re-post time) & Sina Weibo data of 2601 false rumors from the official management center & \cite{wu2015false} \\
\hline  
\end{tabular}
\label{tribes}
\end{table} 

\begin{table}[!t]
\renewcommand{\arraystretch}{1.3}
\centering
\scriptsize
\caption*{Table~\ref{tab:network-detection-I} (continued). Network Structure-Driven Deception Mechanisms} \label{tab:network-detection2}
\begin{tabular}{|P{2cm}|p{5cm}|p{6cm}|p{4cm}|P{1cm}|} \hline
 {\bf Type} & \multicolumn{1}{c|}{\bf Method} & \multicolumn{1}{c|}{\bf Features} & \multicolumn{1}{c|}{\bf Datasets} & {\bf Ref.} \\
 \hline
 \hline 
 Misinfor-mation & Profit minimization of misinformation (PMM), influence maximization & Total activity profit of edges & Three real-world datasets from existing work: soc-wiki-Vote, p2p-Gnutella08 and ca-HepTh & \cite{chen2019minimizing} \\
 \hline
 Astroturf-ing & Klatsch data model, Gephi toolkit & Topology of the largest connected component, number of nodes and edges, mean degree and strength of nodes, mean edge weight, clustering coefficient of largest connected coefficient, standard deviation and skew of in-degree/out-degree & Dataset from Tweets of political keywords, Yahoo Meme, Google Buzz &\cite{ratkiewicz2011truthy} \\   \hline    
 Spam & Malicious relevance score propagation algorithm, criminal account inference algorithm & Social relationship graph: criminal supporters community (social butterflies, social promoters, dummies), social relationships and semantic coordination & Twitter half million account (2060 spammer, 5924 criminal supporters) & \cite{yang2012analyzing} \\
 \cline{2-5} 
 & User ranking algorithm Collusionrank to penalize users who connected to spammers & Node rank, followers, indegree, outdegree, indegree/outdegree ratio, social capitalists & Twitter dataset & \cite{ghosh2012understanding} \\
 \cline{2-5} 
 & SSDM: directed Laplacian formulation to model social networks, SVM and elastic net (EN) & Network information by adjacency matrix & Crawled Twitter dataset from Twitter search API & \cite{hu2013social}\\
 \cline{2-5} 
 & Lockstep propagation algorithm & Adjacency matrix, ``block," ``ray" and ``pearl" subspace, lockstep score & Weibo network with 100 million nodes, synthetic data &\cite{jiang2014inferring} \\
 \cline{2-5} 
 & CatchSync algorithm & Topology & Twitter and Weibo dataset, synthetic data &\cite{jiang2014catchsync} \\
\cline{2-5} 
& Troll identification algorithm (TIA), five de-cluttering operations & Centrality measures: Freaks, Fans Minus Freaks (FMF), PageRank (PR), Signed Spectral Ranking (SSR), Negative Ranking (NR), Singed Eigenvector Centrality (SEC), Modified HITS (M-HITS), Bias and Deserve (BAD) & Slashdot Zoo dataset &\cite{kumar2014accurately} \\
 \cline{2-5} 
  & Incremental tensor analysis & Tensor decomposition and deflation & Phone-call network data, computer-traffic network & \cite{araujo2014com2} \\
 \hline    
 Sybil & MailRank algorithm: basic and personalized scores & Email interactions link analysis, sender rating by social reputation score based on PageRank score & Synthetic dataset &\cite{chirita2005mailrank} \\
 \cline{2-5} 
  & A near-optimal defense algorithm SybilLimit with node trust ranking & Undirected social network with nodes and trust relations, each node consisting of a suspect and a verifier & Three crawled datasets from Friendster, LiveJournal, DBLP and Kleinberg &\cite{yu2008sybillimit} \\
  \hline 
 Abnormal nodes & OddBall algorithm, scalable and unsupervised & Features for egonets: density, weights, ranks and eigenvalues & Bipartite network: Auth2Conf from DBLP, Don2Com and Com2Cand from donations of political candidates &\cite{akoglu2010oddball}\\
  \hline
\end{tabular}
\label{tribes}
\end{table} 

\end{landscape}


\section{Comparison with Existing Survey Papers}

In this section, we provide more detailed discussion on each of the existing survey papers in Section 1.3 of the main paper.

\citet{fire2014online} mainly discussed social network threats targeted at young children in terms of phishing, spamming, fake identity, profile cloning attacks, cyberbullying, and cyber-grooming.  \citet{Rathore17-is-survey} surveyed social network attacks with a special emphasis on multi-media security and privacy.  Since fake news is an emerging deception attack in OSNs, a recent effort by \citet{kumar2018false} discussed the details of fake news detection methods.  Although the existing works stated above~\cite{fire2014online, kumar2018false, Rathore17-is-survey} proposed mechanisms to mitigate specific social deception threats, they focused on discussing prevention methods and practical security suggestions.  An interesting observation is that no work has discussed ethical issues in developing techniques to deal with OSN threats/attacks.  Besides, we observed a lack of understanding on the pros and cons of each detection or mitigation technique that combat online social deception attacks.

\citet{Rathore17-is-survey} conducted a comprehensive survey on social network security. They classified social network security threats in three categories, including multimedia content threats, traditional threats, and social threats with 21 types of threats/attacks.  The authors mainly discussed multimedia content threats, along with their definitions, impact, and security response methods, including detection methods for each type of threat.  They also compared various security attacks in terms of the nature of attack (attack source), attack difficulty, risk to data privacy/integrity, and attack impact on users.  In the end, they proposed a framework to measure and optimize the security of social network services (SNSs).  \citet{gao2011security} discussed the four types of social network attacks, which include privacy breaches, viral marketing attacks, network structural attacks, and malware attacks. The authors compared various attacks, including information leak, de-anonymizing, phishing, Sybil, malware, and spamming, and discussed countermeasure defense mechanisms against them.

\citet{novak2012survey} focused on OSN security and data privacy issues.  They discussed how to protect user data from attacks, social network inference (e.g., user attributes, location hubs, and link prediction) and the research in anonymizing social network data.  \citet{kayes2017privacy} reviewed the taxonomies of privacy and security attacks and their solutions in OSNs.  The authors categorized the attacks based on OSN's stakeholders (on users and their OSNs) and entities (i.e., human, computer programs, or organizations) performing the attacks. They discussed attacks on users' information and how to counter leakages and linkages.  However, the attacks discussed as social deception are common social network attacks, such as Sybil attacks, compromised accounts and/or spams.  The defense techniques to mitigate each attack type were discussed as ways to detect and resist against those attacks.

\citet{fire2014online} discussed key OSN threats and solutions against them.  The authors outlined OSN threats based on the four groups with an additional focus on attacks against children and teenagers.  The threats are 5 classic threats, 9 modern threats, combination threats and 3 threats targeting children. The discussed defense solutions were techniques provided by OSN operators, commercial companies, and academic researchers and discussed the protection ability of various solutions. In the end, they provided recommendations for OSN users to protect their security and privacy when using social networks.  \citet{kumar2018false} discussed the characteristics and detection of false information on Web and social media, with two knowledge-based types: {\em opinion-based methods with ground truth} (e.g., fake reviews), and {\em fact-based methods without ground truth} (e.g., hoaxes and rumors).  They described how false information can perform successful deception attacks, and their impacts on the speed of false information propagation and characteristics for each type.  Based on the specific characteristics, the authors discussed the detection algorithms for each type utilizing different features and propagation models in terms of the analysis of classification, key actors, impacts, features, and measurements. In addition, they discussed the detection algorithms for opinion-based and fact-based detection mechanisms, respectively.

\citet{Wu16-bc} summarized misinformation in social media, focusing more on the unintentional-spread misinformation, such as meme, spam, rumors, and fake news.  It discussed information diffusion models and network structure, misinformation detection and spreader detection, misinformation intervention, and detailed evaluation datasets and metrics.  The diffusion models are SIR (Susceptible-Infected-Recovered/Removed), Tipping Point, Independent Cascade, and Linear Threshold model.  In the diffusion process, user types can be categorized as {\em forceful individuals}~\cite{acemoglu2010spread}, which refer to users not affected upon belief exchange.  \citet{Wu2017-crowdturfing} described detecting crowdturfing in social media.  The authors summarized the history of astroturfing campaign and crowdturfing.  The methods to investigate crowdturfing is mining and profiling social media users as attackers and modeling information diffusion in social media.  Finally, crowdturfing detection can be performed in content-based, behavior-based, and diffusion-based approaches in the state-of-the-art research.  However, this work~\cite{Wu2017-crowdturfing} limited its scope only to crowdturfing. Hence, we didn't include it 
for the comparison of our survey paper with other counterpart survey papers.  \citet{tsikerdekis2014online} analyzed the motivations and techniques of online deception in social media platforms. They categorized social media by the extent of media richness and self-disclosure. Due to the user connection and content sharing nature of social media, online deception techniques can involve multiple roles, such as content, sender, and communication channel. They also provided an insightful discussion of challenges in prevention and detection of online deception. However, this work didn't discuss any attack behaviors concerned as in our paper.

\section{Multidisciplinary Concepts of Deception}

\subsection{\bf Philosophy} Many philosophers raised issues towards the general definition of deception which can happen mistakenly or intentionally. Some philosophers admitted the possibility of `inadvertent or mistaken deceiving'~\cite{Carson10}; however, others disagreed with it and emphasized the aspect of `misleading a belief' by either inadvertently or mistakenly~\cite{Gert15, Russow86}. In addition, as the distinction between `lying' and `deceiving,' `deceiving' must go with the achievement or success by the act of deceiving, such as a `false belief'~\cite{Mahon16}.  
The core aspects of deception in Philosophy lies in an {\em intentional} act to {\em mislead} an entity to believe a {\em false belief}.  


\vspace{-2mm}
\subsection{\bf Behavioral Science} Behavior scientists~\footnote{We consider biologists, ecologists, neuroscientists, and medical scientists as `behavioral scientists' in this work.} investigated the concept of deception and its process in the behaviors of animals and/or humans. Two main concepts of deception are: (1) {\em Functional deception} for an individual's behavior (i.e., a signal) to mislead the actions of others; and (2) {\em intentional deception} referring to intentional states, such as beliefs and/or desires, guide an individual's behavior, leading to the misrepresentation of belief states~\cite{Hauser97}.  The common deception study in this field is investigating the effect of lying or deception on memory and/or the part of a brain activated, associated with the concept of functional deception~\cite{Langleben02, Spence01}.

\vspace{-2mm}
\subsection{\bf Psychology} Psychologists defined deception as a behavior providing information to mislead subjects to some direction~\cite{Adair85} or explicit misrepresentation of a fact aiming to mislead subjects~\cite{Hyman89, Nicks97}. 
The major psychological deception study focused on identifying cues as committing a crime~\cite{Granhag08}, psychological symptoms for self-deception~\cite{Chance15, Heine11-deception-psychology}, individual differences and/or cues to deception~\cite{Riggo83}, verbal or non-verbal communication cues~\cite{Zuckerman81}. 

\vspace{-2mm}
\subsection{\bf Sociology} 
Since 1970's, deception has been studied as a distinct human ability due to intentionality, use of language, and self-awareness~\cite{Langer72}; however, since the 1980's, deceptive behaviors were observed among animals and have been studied by comparing human and non-human deceptive behaviors in terms of their unique characteristics~\cite{Meltzer03, Mitchell86}.  Sociological deception research has mainly studied the effect of deception in various social context on both positive and negative aspects~\cite{Meltzer03}, or deception as a relational, or marketing strategy~\cite{Patwardhan09}.


\vspace{-2mm}
\subsection{\bf Public Relations} In this domain, the concept of self-deception has been studied as a strategic solution to resolve internal or external crisis~\cite{Seiffert-Brockmann17}.  The external role of self-deception is described as a way to avoid disastrous impact on an organization~\cite{Ortmann10} by attributing a problem (or guilty) to an individual or victim.  This is called `parasitism' forming a false belief towards the victim for responsibilities.  In this domain, deception is mainly studied as a strategy to deal with conflicts in public (or international) relations~\cite{Hiebert03} or the relationships between public practitioners and journalists~\cite{Clementson19, Lee19-public-relations}.

\vspace{-2mm}
\subsection{\bf Communications or Linguistics} In this domain, deception research often aimed to identify either verbal or non-verbal indicators for deceptive communications.  Interpersonal deception theory (IDT) views deception as an interactive process between senders and receivers, exchanging non-verbal and verbal behaviors and interpreting their communicative meanings.  IDT further explains that deceivers strategically manage their verbal communications to successfully deceive receivers~\cite{Buller96-testing, Buller96}.  Experimental studies showed that deceivers  produced more words, fewer self-oriented (e.g., I, me, my) and more sense-based words (e.g., seeing, touching) than truth-tellers~\cite{Hancock08}.  As computer-mediated systems have been popularly used, systems for automated linguistics-based cues to deception have been developed considering text-based asynchronous communications~\cite{Zhou04-linguistics}. 

\vspace{-2mm}
\subsection{\bf Command and Control} In the military domain, deception refers to any planned maneuvers undertaken for revealing false information and hiding the truth to an enemy with the purpose of misleading the enemy and enticing the enemy to undertake the wrong operations~\cite{daniel2013strategic, Mitchell86, Whaley82}. The theory of military deception is to spread the deceived messages to the enemy masses through many channels and finally deceive the enemy's nerve centers and decision makers, which is to the opposite of psychological warfare~\cite{Mitchell86}. Military deception is used as a strategy and the strategic levels of political and military interaction are higher than the tactical level~\cite{rothstein2013art, daniel2013strategic}. 

\vspace{-2mm}
\subsection{\bf Computing and Engineering} In this domain, deception has been employed by both attackers and defenders.  Deceptive behaviors have popularly exhibited by cyber attackers in various forms, such as phishing, social engineering attacks, fraud advertisements, stealthy attack, and so forth~\cite{Heartfield15, Rathore17-is-survey}.  On the other hand, by leveraging the concept of deception, various types of defensive deception techniques have been proposed. \citet{almeshekah2016cyber} enhanced the definition of cyberdeception by \citet{Yuill06} as ``planned actions taken to mislead and/or confuse attackers and to thereby cause them to take (or not take) specific actions that aid computer-security defenses.''  A common idea lies in that deception is a way to cause an entity to form false beliefs~\cite{Rowe16}.  Hence, successful deception leads a {\em deceivee} to respond suboptimally, which benefits the {\em deceiver}. \citet{Han18-deception-survey} conducted an extensive survey on defensive deception techniques, in terms of classification, current applications, modeling, and deployment techniques aiming to solve cybersecurity problems.  In this work, we study deception from an attacker's perspective as part of OSN attacks.  Table~\ref{tab:deception-concept} summarizes the meaning and goal of deception studied in different disciplines.

\section{Type, Description, and Intent of Online Social Deception Attacks}

See Table~\ref{tab:osd-type}.

\vspace{-2mm}
\section{Cues of Social Deception} \label{sec:cues-osd}
Since many cues predicting offline social deception have been discussed, we discuss them in this section in order to link the cues of offline social deception and those of online social deception. We discuss deception cues in terms of individual, cultural, linguistic, physiological, psychological, and technological traits below. From this discussion, we aim to deliver insights on how the estimates of those deception cues can provide the key predictors of detecting online social deception.

\subsection{Individual Deception Cues} \label{subsec:individual-cues}

\citet{Riggo83} studied correlations between individual types and behavioral patterns and found individuals vary systematically in displaying certain behavioral cues (e.g., dominance, a social skills measure) are correlated with facial animation behavior.  Certain types of individuals can control the display of cues to increase the likelihood of deception.  For example, dominant, extroverted, and exhibitionistic individuals tend to restrain the display of nervous behaviors while deceiving others. In addition, a significant relationship is found between exhibitionistic personality (e.g., a good actor-like personality) and successful deception ability in different experimental settings.  In the correlation study of personality and deception ability, they also found that males appear to be good liars especially when their faces can be seen by the victims~\cite{Riggo83}.  \citet{kraut1980behavioral} studied demographic characteristics and behaviors as deception cues for custom inspectors. They found that the occupational status and age were the top predictors.

\subsection{Cultural Deception Cues} \label{subsec:cultural-cues}

Cultural variations are used to explain deception cues in the literature.  \citet{Lewis08} showed that individuals from collectivistic cultures were more apt to employ deception in business negotiation than those from individualistic cultures.  \citet{Heine11-deception-psychology} discussed self-enhancement in Western people where self-enhancement refers to a motivation that can make a person feel positive about himself/herself with a high self-esteem~\cite{Sedikides95themultiply}. They found that East Asians have different motivations and effects because of the cultural variations.  \citet{Bond90} had an experiment of cultural influences to a set of deception behaviors using Jordanians and Americans as subjects. The results showed in the lying settings, Jordanians displayed more behavioral cues than Americans in terms of eye contact and filled pauses.

\vspace{-2mm}
\subsection{Linguistic Deception Cues}  \label{subsec:linguistic-cue}

Linguistic and/or communicative cues exhibiting deception in communications have been studied.  Linguistic profiles are studied in deceptive communication, choice and use of languages, and linguistic patterns in deceptive messages~\cite{Buller96, Buller96-testing}.
The following four categories of linguistic deception cues have been discussed: 

\begin{itemize}
\item {\em Word quantity}~\cite{Hancock08,newman2003lying}: This cue was found inconsistently across studies.  The fewer or more quantities depended on what were examined.  In deceptive dyadic communications in an online synchronous text-based setting, the experiments showed that deceptive communication produced more words. 
\item {\em Pronoun use}~\cite{Brinke2012CryMA}: Deceptive communication used fewer of first-person singular, and more of third-person pronouns.
\item {\em Emotion words}~\cite{pennebaker2003psychological}: More sense-based words (e.g., seeing, touching, smelling, feeling) were found in deceptive communications.
\item {\em Markers of cognitive complexity}~\cite{Hancock08, newman2003lying, pennebaker2003psychological, Vrij04-catch-liars}: Describing what was done requires much more complex cognitive process than describing what was not done~\cite{pennebaker2003psychological}.  In this sense, deception is associated with language requiring lower cognitive complexity. Deception uses fewer distinction markers, such as exclusive words, negations (e.g., no, never, or but, except, either, not)~\cite{Hancock08, newman2003lying}. Using exclusive words requires the person to distinguish what is in a category from what is not.
\end{itemize}

\subsection{Physiological Deception Cues}
\label{subsec:physiological-deception-cues}
Physiological or behavioral cues are the emotions in deceiving that liars are expressing because they are indicators of guilt~\cite{DePaulo03}. In the studies of behavioral cues to deception~\cite{DePaulo03} and physiological cues to identifying deception~\cite{Vrij04-catch-liars},   
liars may have at least one of emotions, content complexity, and attempted control phenomena.  The behavioral cues have the following categories:
\begin{itemize}
\item {\em Less blink}~\cite{Vrij04-catch-liars, Vrij06}: The strongest evidence of behavioral change in deception is a decrease in eye blinking.  \citet{wallbott1991stress} showed that increased cognitive load rather than nervousness results in a decrease in eye blinking.
\item {\em Higher-pitched voices and faster speech}~\cite{DePaulo03}: A natural outcome of suppressing behaviors is to be more rigid.  An example of behavioral failure is to regulate expressive behaviors, such as a voice tone but result in higher tones.
\item {\em Displacement activities}~\cite{Troisi02}: Displacement activities consist of most self-directed behaviors in situations with high social tension.  Increased displacement behavior correlates with a subject's anxiety and conscious deception.  
\item {\em Irrelevant activities}~\cite{Vrij06}: These symptoms include decreased hand and finger movements for males and tensing up the body because lying is a cognitive demanding to prevent the conflicts and disagreement of lies. 
\end{itemize}
\vspace{-2mm}
\subsection{Psychological Deception Cues} \label{subsec:psychological-cues}

Psychological or cognitive cues include nonverbal anxiety responses that are consciously revealed in the intentional deception~\cite{Knapp74}. \citet{Mitchell86} described the mental process of deceptions from a social cognitive perspective based on children verbal deception and nonverbal deception in sports. \citet{Knapp74} used controlled lab settings to determine the characteristics of intentional deception with verbal and nonverbal cues.  They revealed the following psychological cues:
\begin{itemize}
\item {\em Cognitive load}~\cite{Trivers10, Vrij04-catch-liars, Vrij06}: This is the most important deception cue~\cite{Trivers10}. Self-deception is effective to hide the deception and reduce its immediate cognitive costs. The innovative deception detection is a cognitive-load approach~\cite{Vrij06}, which focused more on the cues of `thinking-hard' rather than attending to the behavioral cues of liars, assuming that lying is cognitive demanding. In practice, the interview strategy can be improved by increasing the interviewees' cognitive demand from a parallel task.
\item {\em Nervousness}~\cite{DePaulo03, Trivers10, Vrij04-catch-liars}:  Nervousness arises from a mixture of emotions of guilty, fear of being caught and excitement~\cite{Vrij04-catch-liars}. Nervousness is regarded as a weak indicator of predicting deception~\cite{Trivers10}. The cues of nervousness are gaze aversion and fidgeting but~\citet{Vrij04-catch-liars} argued they are not always related to deception in situations of low severity negative consequences of being caught. They further pointed out that nervousness could be decreased if the frequent suspects were familiar with the police interviews.
\item {\em Control}~\cite{Trivers10, Vrij04-catch-liars}: In order to avoid the signs of deceiving and suppress the nervous behaviors, deceivers tend to exert control to obtain an impression of being an honest person~\cite{Trivers10}. There are two possible unexpected consequences of attempted control. One could be inadequate control of behavior~\cite{Vrij04-catch-liars}, which is another cue of deception and the other is `overcontrol'~\cite{Trivers10}, which is a side effect of impression of planned and rehearsed behaviors.  Physiological cues are very closely related to psychological cues, for example, as control also shows displacement activities (e.g., exhibiting inappropriate behavior upon confusion or conflict).
\end{itemize}
\citet{Trivers10} emphasized nervousness, control and cognitive load as three key deception cues. In addition, other anxiety responses are discussed~\cite{Knapp74}.  Deceivers tend to exhibit cognitive cues, such as more uncertainty, vagueness, nervousness, reticence, dependence, and/or unpleasantness as a negative effect.

\vspace{-2mm}
\subsection{Technological Deception Cues}
\label{subsec:technological-deception-cues}
\citet{ferrara2016rise} discussed the impact and detection of social bots which are the outcome of abusing new technologies.  Social bots with malicious intents caused several levels of damage to society. Early bots mainly automatically posted content and can be spotted by the cues of a high volume of content generation.  Several social honeypot studies attracted social bots followers by carefully designed bots and analyzed the technology cues of social bots.  However, sophisticated social bots are becoming more intelligent and tend to mimic human-like behaviors, making it hard to detect the social bots.  The advanced detection strategy leveraged the technological cues from social graph structure, such as densely connected communities, and behavioral patterns.  The proposed behavioral signature contains classes of features including network, user, friends, timing, content, and sentiment~\cite{ferrara2016rise}.

\vspace{-2mm}
\subsection{Relationships Between Deception Cues of Offline and Online Platforms} \label{subsec:online-offline-deception-cues}

Fig.~\ref{fig:deception-cues} summarizes the key social deception cues discussed so far.  Via the in-depth survey of deception cues, we identified the commonalities and differences between online and offline deceptive behaviors as below.

\vspace{-2mm}
\subsubsection{\bf Commonalities between Online and Offline Deceptive Behaviors} Deception usually spreads via communication between deceivers and deceivees.  The online media platforms support chat-based communications and even synchronous similar to the traditional face-to-face chatting or interviews~\cite{tsikerdekis2014online}.  Interpersonal deception theory~\cite{Buller96} discusses several verbal and non-verbal deception cues for traditional offline communications.  Most of the verbal deception cues (e.g., linguistic cues) are relevant to both offline and online deception~\cite{derrick2013detecting}.  Messages and posts are the main source of online information so that the linguistic cues are most useful cues for online deception~\cite{zhou2008following}.  These days online platforms also provide face-to-face chatting.  Although it is limited to some extent, some physiological cues and/or body movement can be captured.  

\vspace{-2mm}
\subsubsection{\bf Differences between Online and Offline Deceptive Behaviors}  Although face-to-face social media platforms make people feel much closer to each other by delivering body movement and/or facial expressions, feeling some physiological cues and/or subtle behavioral changes may not be captured like face-to-face interactions~\cite{tsikerdekis2014online}.  In addition, typing behavior, including response time and the number of edits, for online chatting were studied as typing cues of online deception~\cite{derrick2013detecting}, which is not often observed in offline interactions.  In addition, online behaviors are known different from offline behaviors in their motivations and attitudes~\cite{Denker18-online-coms}.  In this sense, more research should be done to identify unique cues of online human deceptive behaviors. 

\section{Types of Deception and Related Taxonomies}

\subsection{Types of Deception} \label{sec:deception-causes}
In this section, we first categorize deception based on {\em intentionality} to explain whether a deceiver has an intent to deceive a target deceivee or not.  Intentional deception~\cite{Englehardt94} can be further classified into malicious deception and non-malicious deception.

\subsubsection{\bf Intentional Deception} This deception consists of deception with malicious intent and with non-malicious intent for a deceiver's interest. 

{\bf The goals of malicious deception} can be as follows:
\begin{itemize}
\item {\em Financial benefit}:  Many deceptive behaviors has its purpose to obtain a monetary benefit. Corporate deception may have financial benefits in a short term.  \citet{Englehardt94} described a fortune 500 company covered the cheating of huge expenses from one of its senior executives.  Another example is a deceptive advertising, such as the advertising from Volkswagen's 2015 emission scandal~\cite{Seiffert-Brockmann17}. For the financial benefits, Volkswagen deceived the consumers in its sustainability reports in 2013 and 2014.  Financial benefit is also a common reason of an individual's online deceptive behavior. For example, a spammer can be paid from clicking advertisements by attracting online traffic to the specific sites~\cite{Nextgate13}.  Malicious users spread phishing links during a holiday period to collect payment details from victims~\cite{Vergelis19}.

\item {\em Manipulation of public opinions}: In social media, social and political bots play a role in influencing public opinions~\cite{forelle2015political}. Malicious bots spread spam and phishing links.  Politicians and governments worldwide have been using such bots to manipulate public opinions.  For example, \citet{forelle2015political} reported that most active bots are those used by Venezuela's radical opposition.

\item {\em Self-deception}: Self-deception is part of a unconscious self-deception process.  The cost of lying is heavy so liars need to pay heavy cognitive costs to maintain the consistence of the lies to avoid any imperfections.  The cognitive costs of conscious deception is higher than the unconscious deception~\cite{Trivers10}.  Self-deception is part of an unconscious process which can reduce the immediate cognitive costs of deception and prevent showing nonverbal cues of guilt~\cite{Chance15}.  Lies covered up by self-deception are harder to be detected.

\item {\em Cooperative deception}:  Cooperation is a strategy of balancing costs and benefits and maintaining stakeholder relationships in the deception or cooperation interactions with opponents~\cite{Trivers10}, often used in public relations.  The research~\cite{charron1989relations, Clementson19} reported this relationship between public relations practitioners and journalists in a crisis communication experiment and studied how deception plays a role. In this deception, victims often willingly contribute to successful deception by ignoring the evidence against false information or by convincing themselves that deception does not exist~\cite{Mitchell86}. 
\item {\em Parasitism}~\cite{Seiffert-Brockmann17}: This refers to `false framing of responsibility' which can be easily used as a strategy to solve complicated issues without introducing long-term investigations that may cause structural changes.  As discussed earlier in the concept of deception in public relations, parasitism is used as a strategy to create a false belief towards the victim for a deceiver to avoid any consequence of a certain conflict.  
\end{itemize}

{\bf The goals of non-malicious deception} are commonly discussed as follows:
\vspace{-1mm}
\begin{itemize}
\item {\em Privacy protection}: Deception can be used as a defense for the privacy protection at the organization-level or individual-level. The organization-level refers to the protection of internal information and control of information flow.  This is also called {\em defensive deception}.  There are a few methods for the individual-level privacy protection in cyberspace.  Some privacy techniques add a noise to a user's data for protection against attackers~\cite{pawlick2019game} because the data can be modified before being published.  Another defensive deception is to hide a user by using outside users. Using mix networks for anonymity is an example for messages privacy protection~\cite{pawlick2019game}. 
\item {\em Self-presentation}: People use fake presentation to present themselves as certain roles or intents~\cite{schlenker1982social}. Self-presentation is an activity to impress others for both liars and truth tellers. Self-presentation is one way of understanding nonverbal communication~\cite{DePaulo03}.  Deceptive self-presentation can make people more reliable or protect themselves from getting hurt by disagreements. Self-presentation can be used as prediction cues to deception~\cite{DePaulo03}. 
\item {\em Self-deception}: This is to hide true information reflecting conscious mind unconsciously~\cite{Trivers10}, with the two main benefits of not being detected easily and reducing immediate cognitive costs.  \citet{Chance15} discussed a social benefit of self-deceptive confidence, leading to higher social status~\cite{kunda1990case}.  In addition, \citet{Chance15} examined psychological benefits of self-deception in terms of optimistic opinions and/or high motivations towards themselves and other people.  In addition, self-deception is also used as strategies to deal with public relations~\cite{Seiffert-Brockmann17} to resolve conflict situations. 
\end{itemize}

\subsubsection{\bf Unintentional Deception} This type of deception is usually used without explicit intent.  One can use deception without certain intent mistakenly or unconsciously.  This deception is understood as an action with the following reasons:

\begin{itemize}
\item {\em Lack of knowledge on fact}: A false belief can be conveyed by unintentional half-truths, such as exaggerations, omissions or obscured information~\cite{Seiffert-Brockmann17}.  In this situation, deception is caused by a lack of full knowledge.  Deception includes both verbal form (e.g., lying to lure a deceivee or hide truth) and non-verbal forms (e.g., hiding information such as secrecy) for misleading the deceivee's belief/action via biased information processing~\cite{mele2001self}.

\item {\em Following social norms or rules}: Deception can be driven by the need of fake conformity by adjusting internal regulations complying with external social norms expectations (e.g., submissive attitudes are preferred in some Asian cultures).  Inside the organization, actions caused by either individuals or groups should follow the external social norm expectations~\cite{Seiffert-Brockmann17}.
\item {\em Resolving cognitive dissonance by using self-deception}: Self-deception can imply an unintentional state of mind, potentially resulting in motivational bias~\cite{mele2009have}, 
thereby contributing to errors in behavioral judgements or decisions~\cite{montibeller2015cognitive}.  Self-deception hides facts in the unconscious mind in order to solve the conflicts of truth and people's internal beliefs~\cite{Trivers10}.  Individuals tend to use self-deception to cope with cognitive dissonance in their mind~\cite{Seiffert-Brockmann17}.
\end{itemize}

\vspace{-3mm}
\subsection{Taxonomies and Spectrum of Deception} \label{subsec:deception-taxonomies-concepts}

This section discusses the related concepts and spectrum of deception. Deception can be defined and explained by a set of related terminologies in which those concepts should be defined and compared.  Deception exists in our daily life in both verbal and nonverbal forms.  Deception ranges a wide spectrum with varying intent and detectability (i.e., the extent of deception being detected). 

\vspace{-2mm}
\subsubsection{\bf Key Taxonomies of Deception}
Here are the concepts related to deception in social sciences. Each of them has a specific meaning and represents some properties of deception.  Most common concepts are defined in the dictionary and discussed in the cybersecurity literature~\cite{Chance15, DePaulo03, OED89, Rowe16, Seiffert-Brockmann17}. 
\begin{itemize}
\item {\em Deceivee}~\cite{Rowe16}: The victim of a deception.
\item {\em Deceiver}~\cite{Rowe16}: The perpetrator of a deception.
\item {\em Susceptibility}~\cite{OED89}: Likelihood to be deceived.
\item {\em Exploitation}~\cite{OED89}: The use of resources and benefit from them (e.g., damage to systems) by attackers.  
\item {\em Self-deception}~\cite{Chance15}: A conscious false belief held with a conflicting unconscious true belief.
\item {\em Trust}~\cite{OED89}: Reliance on the confidentiality and integrity from other sources and with confidence. Earning high trust from a deceivee can be easily exploited by a deceiver.
\item {\em Lying}~\cite{DePaulo03,Rowe16}: Deliberate verbal deceptions. People often lie in pursuit of material gain, personal convenience, or escaping from punishment. 
\item {\em White lying}~\cite{Seiffert-Brockmann17}: Normal standards for the lighthearted type of deception. 
\item {\em Belief}~\cite{OED89}: A truth in somebody's mind, truth basis.
\item {\em Misbelief}~\cite{OED89}: A misplaced belief (i.e., mistakenly believing in false information)
\item {\em Perception}~\cite{OED89}: The state of being aware of something through the senses.
\end{itemize}

\vspace{-2mm}
\subsubsection{\bf Spectrum of Deception} In daily life and social networks, deception spans a spectrum of verbal and non-verbal behaviors. This section lists a few of the various deceptions based on~ \cite{Ekman09, Rowe16, smith1953compleat}. 
\begin{itemize}
\item {\em White lies}~\cite{Rowe16}: Harmless lies to avoid hurting other's feelings and smooth relationships.
\item {\em Humorous lies}~\cite{smith1953compleat}: Jokes that are obvious lies, such as practical jokes.
\item {\em Altruistic lies}~\cite{Rowe16}: Good lies for protecting others, such as for preventing children from worrying.
\item {\em Defensive lies}~\cite{Rowe16}: Lies to protect the deceiver, such as lies to get rid of repeated telemarketers.
\item {\em Aggressive lies}~\cite{Rowe16}: Lies to deceive others for the benefit of the deceivers.
\item {\em Pathological lies}~\cite{Rowe16}: Lies by a deceiver with psychological disorder. 
\item {\em Nonverbal minimization}~\cite{Ekman09}: Understating an important case in nonverbal camouflage.
\item {\em Nonverbal exaggeration}~\cite{Ekman09}: Overstating an important case to hide others.
\item {\em Nonverbal neutralization}~\cite{Ekman09}: Intentionally hiding normal emotions when inquired about emotional things.
\item {\em Nonverbal substitution}~\cite{Rowe16}: Intentionally changing a sensitive concept with a less sensitive one.
\item {\em Self-deception}~\cite{Rowe16}: Pushing of a reality into the subconsciousness.
\end{itemize}

Fig.~\ref{fig:detectability} represents the spectrum of deception from the lowest detectability to the highest detectability and from lowest bad intent (good intent) to no intent and to highest bad intent.  In general, the deception with lower detectability are more with good intent, such as altruistic lies and white lies. Nonverbal deception are usually with bad intent and can be detected by professionals. Those behaviors can also be used as cues to detect lies. The deceptions with neutral intent can also be easily detected.

\section{Susceptibility to Online Social Deception} \label{sec:susceptibility-online-deception}

Attackers aim to achieve their attack goals as efficient as possible with minimum cost.  To this end, the attackers may target highly susceptible people to the OSD attacks.  In this section, we discuss the following key susceptability traits to the OSD attacks: {\em demographic}, {\em personality}, {\em cultural}, {\em social and economic}, and {\em network structure feature-based factors}.  This survey will provide insights on how to protect people with different levels of susceptability to the OSD attacks.
\vspace{-2mm}
\subsection{Demographic Susceptible Factors} \label{subsec:demographics-susceptibility}

The commonly used demographic factors related to susceptibility to OSD attacks are as follows: 
\begin{itemize}
\item {\em Age}: \citet{Sheng10} conducted an extensive survey with 1,001 online survey respondents and found young age groups between 18 and 25 are more susceptible to phishing than other age groups. \citet{Berson02} and \citet{Wolak10} discussed a new medium for the victimization of children and sexual exploitation in cyberspace. They studied demographics, online habits, online interaction patterns, and high-risk youths and risky online behaviors to design prevention mechanism. 
\item {\em Gender}: \citet{Sheng10} observed in their experiment that women are more susceptible to phishing than men. \citet{Oliveira17-chi-susceptibility} and \citet{Lin19} further examined phishing susceptibility in terms of age groups based on the 21-day observations on their online activities.  They found old women as the most vulnerable groups. 
\item {\em Education}: \citet{Sheng10} showed in their experiment that educational materials were 40\% more effective in reducing phishing susceptibility. 
\item {\em Perception and Knowledge}: In online seller deception, most consumer victims cannot detect the fraud from sellers~\cite{Grazioli00}.  In behavioral science research, perceived capabilities to detect a risk~\cite{Wright10} and knowledge about risk~\cite{malhotra2004internet} have been often examined as a predictor to avoid such risk.  However, prior research provided different findings on the effect of perception and knowledge on social deception.  \citet{Wright10} found no relationship between perceived self-efficacy/risk and deception detection, concluding that no significant effect of such perception is found on social deception.  \citet{Xun08} modeled user-phishing interactions in which the model allows to investigate how a user's perception can be a predictor of detecting phishing attacks.
\end{itemize}

\vspace{-2mm}
\subsection{Personality Traits-based Susceptible Factors} \label{subsec:personality-susceptibility}

An individual's personality traits are studied to investigate their impact on susceptibility to scams or phishing attacks~\cite{Darwish12, Enos06, Halevi13, Halevi15, Modic11, Parrish09, Pattinson12} using the Big Five personality traits model~\cite{Tupes92}.  However, due to the sample bias and lack of subjects covering a wide range of personality traits, the findings are not generalizable.  In order to overcome the issues of limited sampling, \citet{Cho16} developed a mathematical model based on Stochastic Petri Nets to investigate the effect of user personality traits on phishing susceptibility.  They used Big Five personality model to consider a user's perceived trust and risk towards phishing attacks, which are the key factors to phishing susceptibility.  However, their work is purely based on a probability model without using human subjects which lacks empirical validation. \citet{Ding15} classified phishing emails in terms of their corresponding target victims based on personality traits.  To this end, the authors constructed a dictionary based on semantic similarity of prospective words corresponding to the personality traits.  However, this work didn't validate their model based on the empirical study to measure the actual user susceptibility. 

As many deceptive attacks are widespread and online activities become the part of everyday life in most people, an human individual's susceptibility to online deception or attacks has been studied in the recent research in computing and computer science fields.  \citet{Albladi18} studied a user's susceptibility to social engineering attack by proposing a user-centric framework considering socio-psychological, habitual, socio-emotional, and perceptual user attributes.

\vspace{-2mm}
\subsection{Cultural Susceptible Factors} \label{subsec:cultural-susceptibility}
Culture has been defined in various forms.  However, it has not been proved if there exists a clear relationship between culture and its effect on individual's communication and/or behaviors~\cite{cho2009multinational}.  A well-known classification of cultural values is Hofstede's two cultural dimensions~\cite{hofstede2011dimensionalizing}: individualism vs. collectivism. In the individualistic culture, individuals are loosely tied to one another and a sense of `I' and an individual's `privacy' are valued. On the other hand, in the collectivistic culture,  individuals are tightly connected emphasizing `we-ness' and `belongings' to each other.  Since culture has been studied as a key factor impacting trust in a society where trust affects deceptive behavior, existing studies also have looked at how culture influences deception.  For example, individuals from collectivistic culture (e.g., East Asia countries) were more apt to employ deception~\cite{Lewis08} or to be deceptive in business negotiation than those from individualistic culture (e.g., Western countries).  However, these studies were conducted in offline worlds. 

\vspace{-2mm}
\subsection{Social and Economic Susceptible Factors} \label{subsec:social-economic-susceptibility}
Social media becomes one of prevalent communication tools, being unavoidable part of our modern life across different socio-demographic groups. Vulnerable status in a socio-economic ladder in the off-line world seems to be transferable to the online world.  For example, low education and/or income may influence the level of knowledge and awareness about online social deception (or phishing) or related threat~\cite{kalmus2011motives, Tembe14-phishing}.  However, previous studies discussed in Section~\ref{subsec:demographics-susceptibility} showed a lack of empirical evidence due to inconsistent findings on the relationships between individual characteristics related to social and economic status~\cite{kalmus2011motives}. 

\vspace{-2mm}
\subsection{Network Structure Feature-based Susceptible Factors} \label{subsec:network-feature-susceptibility}

Network structure features have been used to predict the extent of susceptibility to social bots in online social networks.  \citet{wagner2012social} found that a user's out-degree is identified as a key network feature social bots can target as their victim.  In addition, susceptible users tend to be more active (e.g., retweet, mention, follow or reply) in the Twitter network and interact with more users, but their communication is mainly for conversational purpose rather than informational purpose. Susceptible users tend to use more social words and show more affection.  Similarly, in Facebook, susceptible users tend to more engage in posting activities with less restrictive privacy settings, naturally resulting in higher vulnerability to privacy threats~\cite{Halevi13}.  Social isolation (loneliness) and risk-taking online behaviors are the indirect factors of vulnerable people, such as victims of  cybergrooming~\cite{whittle2013review,Williams17}.  \citet{Albladi18} analyzed various user characteristics, such as a level of involvement, for vulnerability of social engineering attacks. 

Engagement in social media is one of the most prominent attributes contributing to high susceptibility to social deception.  Habitual use of social media measured by the size of social network and time spent in social media increases the likelihood of being victims for social attacks in OSNs~\cite{Vishwanath15-individual-deception}.  Highly active social network users can be more favorable targets for attackers as they have more exposures to social media and accomplish their attacks through the active users' networks~\cite{Albladi18}. Individuals highly engaged in social media can be inattentive and automatically respond to online communications, being more susceptible to social deception. In addition, cybercrime research found that the more use of social media is significantly associated with a higher level of risks for sexual exploitation~\cite{Berson02,Wolak10} and cyberbullying~\cite{diomidous2016social}.

\section{Metrics}

This section gives the detailed explanation of metrics surveyed in this work, related to Section 9.2 of the main paper.

\begin{itemize}
\item {\em Confusion Matrix}~\cite{barbon2017authorship, cao2015detecting, chen2015performance, cresci2017social, ding2019keyword, Everett16, fard2019computational, golbeck2018fake, hu2013social, jin2014news, jin2016news, kamhoua2017preventing, kontaxis2011detecting, kumar2018rev2, lee2010uncovering, lee2011seven, liang2015rumor, liu2016detecting, ma2018rumor, nisrine2016security, sahoo2019hybrid, sedhai2017semi, song2015crowdtarget, song2014discriminative, stringhini2010detecting, swe2018fake, tsikerdekis2016identity, vosoughi2017rumor, wang2014man, yang2014taste}: The confusion matrix is made of True Positive (TP), False Positive (FP), True Negative (TN), and False Negative (FN). They are the basic components for other accuracy metrics, such as precision and recall.
\item {\em Precision}~\cite{barbon2017authorship, cao2015detecting, chen2015performance, cresci2017social, ding2019keyword, fard2019computational, golbeck2018fake, hu2013social, inuwa2018detection, jin2014news, jin2016news, kamhoua2017preventing, kumar2018rev2, lee2010uncovering, liang2015rumor, liu2016detecting, nisrine2016security, sahoo2019hybrid, sedhai2017semi, song2014discriminative, tsikerdekis2016identity, vandam2018cadet, wu2015false, wu2018tracing, yao2017automated}: This metric simply estimates the true positives over positives detected including true positives and false positives by:
\begin{equation}
\mathrm{Precision} = \frac{TP}{TP+FP}
\end{equation}

\item {\em Recall}~\cite{cao2015detecting, chen2015performance, cresci2017social, ding2019keyword, fard2019computational, golbeck2018fake, hu2013social, inuwa2018detection, jin2014news, jin2016news, kamhoua2017preventing, lee2010uncovering, liang2015rumor, liu2016detecting, nisrine2016security, sahoo2019hybrid, song2014discriminative, tsikerdekis2016identity, wu2015false, wu2018tracing, yao2017automated}:  This metric captures the true positives over the actual positives include true positives and false negatives.  This metric is estimated by:
\begin{equation}
\mathrm{Recall} = \frac{TP}{TP+FN} 
\end{equation}

\item {\em $\mathbf{F}_1$ Score or Measure}~\cite{cao2015detecting, chen2015performance, cresci2017social, dinakar2012common, ding2019keyword, fard2019computational, golbeck2018fake, hu2013social, inuwa2018detection, jin2014news, jin2016news, lee2010uncovering, lee2011seven, lee2013crowdturfers, liang2015rumor, sahoo2019hybrid,  tsikerdekis2016identity, wu2015false, wu2018tracing}: This metric is an indicator of the accuracy of detection based on both precision and recall.  It is measured by:
\begin{equation}
\mathrm{F}_1 = \frac{2 \times \mathrm{Precision} \times \mathrm{Recall}}{\mathrm{Precision}+\mathrm{Recall}} 
\end{equation}
\item {\em Accuracy}~\cite{badri2016uncovering, barbon2017authorship,  bhatt2018combining, cresci2017social, dinakar2012common, ding2019keyword, Everett16, ghosh2012understanding, inuwa2018detection, jiang2014catchsync, jiang2018linguistic, jin2014news, jin2016news, lee2010uncovering, lee2011seven, lee2013crowdturfers, ma2017detect, ma2018rumor, nisrine2016security, sedhai2017semi, song2014discriminative, tsikerdekis2016identity, vosoughi2017rumor, wu2015false, wu2018tracing, yang2019phishing}:  This metric measures correct detection for true positives and true negatives.  However, when the datasets are not balanced such as too large true positives with too small true negatives or vice-versa, this metric may mislead.  It is given by:
\begin{equation}
\mathrm{Accuracy} = \frac{TP+TN}{TP+TN+FP+FN} 
\end{equation}
There is also a weighted accuracy score~\cite{bhatt2018combining} with different weights on labels. Accuracy can also be used to evaluate the contribution of each features or feature sets~\cite{inuwa2018detection, sedhai2017semi, vosoughi2017rumor, wu2015false}. 

\item {\em False Positive Rate (FPR)}~\cite{badri2016uncovering, chen2015performance, Everett16, lee2013crowdturfers, sahoo2019hybrid, song2015crowdtarget, swe2018fake, tsikerdekis2016identity, wang2014man, yang2019phishing}: This metric is to measure misdetection in terms of false alarms among the ones detected as positives and computed by:
\begin{equation}
\mathrm{FPR} = \frac{FP}{FP+TN} 
\end{equation}

\item {\em False Negative Rate (FNR)}~\cite{badri2016uncovering, barbon2017authorship, lee2013crowdturfers, wang2014man, yang2019phishing}:  This metric captures how many positives are missed and is estimated by:
\begin{equation}
\mathrm{FNR} = \frac{FN}{TP+FN} 
\end{equation}

\item {\em Specificity}~\cite{barbon2017authorship, chen2015performance, cresci2017social, sahoo2019hybrid, song2015crowdtarget}:  This metric measures the extent of correctly detecting negatives over the actual number of negatives and is obtained by:
\begin{equation}
\mathrm{Specificity} = \frac{TN}{TN+FP}  = 1-FPR
\end{equation}

\item {\em Weighted Cost ($\mathrm{W}_{cost}$)} ~\cite{yang2019phishing}: In phishing detection, since the ratio of legitimate websites to phishing website is high, a legitimate website misclassified to a phishing one (FPR) has severe effects than the reverse (FNR). The weighted cost is used to balance the performance of FPR and FNR.  $\mathrm{W}_{cost}$ is estimated by:    
\begin{equation}
\mathrm{W}_{cost} = \mathrm{FNR}+\lambda \times \mathrm{FPR}, \enspace\lambda > 1.
\end{equation}
where $\lambda$ is the weight of FPR. Higher values of $\lambda$ means larger influence of FPR value.

\item {\em Receiver Operating Characteristic (ROC) Curve}~\cite{barbon2017authorship, inuwa2018detection, lau2014probabilistic, song2015crowdtarget, song2014discriminative, wang2014man}: ROC curve draws a plot of classifier's true positive rate (TPR) against FPR at various detection threshold scenarios.  This curve is used to measure and compare stability between several classifier models. 

\item {\em Area Under the Curve (AUC)}~\cite{barbon2017authorship, cao2015detecting, chen2018semi, golbeck2018fake, inuwa2018detection, kumar2018rev2, lau2014probabilistic, lee2011seven, lee2013crowdturfers, sahoo2019hybrid, song2015crowdtarget}: AUC is calculated by the the area under the ROC curve. It measures the probability of a classifier to correctly identify a true-positive data. Since AUC is insensitive to imbalance between classes, it can be better than \textit{Accuracy} in evaluating imbalanced dataset. AUC is another metric of classifier stability and classification quality for different settings.

\item {\em Discounted Cumulative Gain (DCG)}~\cite{Paradise17}: DCG measures the effectiveness of an algorithm, an alternative measure to AUC. A higher DCG is indictive of an early identification of suspicious cases and estimated by: 
\begin{equation}
\mathrm{DCG} = r[1]+\sum^n_{i=2}{\frac{r[i]}{\log_2 i}},   
\end{equation}
where $r[i]$ is 1 if the $i^{th}$ friend request was defined as suspicious or 0 if the $i^{th}$ friend request was defined as legitimate, and $n$ is the number of total incoming requests that require further investigation~\cite{Paradise17}. 

\item {\em Matthews Correlation Coefficient (MCC)}~\cite{cresci2017social, golbeck2018fake, sahoo2019hybrid, tsikerdekis2016identity}:  $\mathrm{MCC}$ measures the correlation between predicted class and real class of users. This metric is considered as the unbiased version of {\em $\mathrm{F_1}$-measure} and given by: 
\begin{equation}
\small 
\mathrm{MCC} = \frac{TP \times (TN-FP) \times FN}{\sqrt{(TP+FN)(TP+FP)(TN+FP)(TN+FN)}},
\end{equation}
\normalsize
where $\mathrm{MCC}\approx 1$ means high prediction accuracy. $\mathrm{MCC}\approx 0$ means the prediction is no better than random guessing.  $\emph{MCC}\approx -1$ means that the prediction is in disagreement with the real class.

\item {\em Cohen's Kappa Value ($\kappa$)}~\cite{dinakar2012common}:  This metric is a measure of reliability for two classifiers or raters, which considers true positive agreement by chance. 
Cohen's Kappa Value is used when \textit{Accuracy} alone is insufficient to evaluate model reliability~\cite{dinakar2012common}. Cohen's Kappa is calculated as:
\begin{equation}
\mathrm{\kappa} = \frac{P_o-P_e}{1-P_e}  
\end{equation}
where $P_o$ is the observed agreement in classification, the same as \textit{Accuracy}, and $P_e$ is the hypothetical probability of agreement by chance. High Cohen's Kappa Value ($0.8\leq\kappa\leq 1$) indicates good reliability~\cite{carletta1996assessing}. 

\item {\em Mean Absolute Error (MAE)}~\cite{jin2013epidemiological, Wu16-bc}: Many detection algorithms for OSD attacks use MAE to estimate their detection accuracy.  In addition, this metric is used to measure the simulation  fitting error of an epidemic model by calculating the absolute values of errors at each time points. 
\begin{equation}
\mathrm{MAE} = \frac{1}{|U|}\sum_{i\in U}{|p_i - l_i|}, 
\end{equation}
where $U$ is a user set, $p_i$ is a prediction result, $l_i$ is a true label, and $i$ is a data index.

\item {\em 2-norm Error}~\cite{jin2013epidemiological}: This measures the simulation fitting error of an epidemic model as one of the performance measures of model fitting and optimization. A good model would reduce this error through iterations.  This metric is estimated by:
\begin{equation}
\text{2-norm Error} = \frac{\parallel \mathrm{I}(t)-\mathrm{Tweets}(t)\parallel^2}{\parallel \mathrm{Tweets}(t)\parallel^2}, 
\end{equation}
where $\mathrm{I}(t)$ is the number of users (agent I) that spread the rumor tweet at time $t$. $\mathrm{Tweets}(t)$ is the number of tweets at time $t$ from the real data.
    
\item {\em Mean Fraction of Recovered Agents Per Time Unit ($\mathrm{R}$)} \cite{cho2019uncertainty}: This is a specific case of the statistics and plot metric.  Instead of plotting the count of each agent at each time point, the average fraction of recovered agents during the total session time $T$ is calculated.
\begin{equation}
\mathrm{R} = \frac{\sum^T_{t=1}R(t)}{T}, 
\end{equation}
where $R(t)$ is the number of agents recovered from false information (i.e., not believing in false information) and $T$ is the total simulation time.
    

\item {\em Spearman's Rank Correlation Coefficient ($\rho$)}
~\cite{Everett16, jiang2018linguistic}:  This metric measures the rank correlation between the predicted labels and the ground truth and is obtained by:
\begin{equation}
\rho = 1 - \frac{6\sum d^2_i}{n(n^2-1)},   
\end{equation}
where $n$ ranks are distinct integers and $d_i$ is the difference of two ranks between an element.  $\rho$ ranges in $[-1, 1]$ as a real number where $0$ refers to random guess while $1$ indicates positive correlation~\cite{Wikipedia19-spearman}.

\item {\em Label Ranking Average Precision (LRAP)} ~\cite{jiang2018linguistic}: This measures the ability to give more accurate prediction for each post message, with a prefect prediction of $1$. LRAP is measured by:
\begin{equation}
\mathrm{LRAP} = \frac{1}{n}\sum^{n-1}_{i=0}\frac{1}{\parallel y_i\parallel_0}\sum_{j,y_{ij}=1}\frac{|L_{ij}|}{\mathrm{rank}_{ij}},   
\end{equation}
where $n$ is number of data points, $y_i$ is the vector of ground truth labels of the $i$th data point, $\parallel \cdot \parallel_0$ is number of non-zero elements in a vector, $y_{ij}$ is the binary label of $j$th label from ground truth vector $y_i$,  $|L_{ij}|$ is number of positive labels for a given data point $i$, and $\mathrm{rank}_{ij}$ is the rank of predicted label ($p_{ij}$) in predicted label vector ($p_i$) for a given $i$~\cite{Scikit-learn19-lrap}. 

\item {\em Label Ranking Loss (LRL)}~\cite{jiang2018linguistic}:  This metric estimates the number of times that irrelevant labels are ranked higher than
relevant labels.  Due to its large volume of complex description, the interested readers can refer to~\cite{tsoumakas2009mining} for more details.
\end{itemize}

\bibliographystyle{IEEETranSN}
\bibliography{sd}